\documentclass[twocolumn,aps,prx,reprint,longbibliography,superscriptaddress,nofootinbib]{revtex4-2}

\usepackage{graphicx}
\usepackage{dcolumn}
\usepackage{bm}
\usepackage{hyperref}
\hypersetup{
    colorlinks = true,
    linkcolor = blue,
    urlcolor = gray,
    citecolor = magenta
}

\usepackage{amsfonts}
\usepackage{amsmath, amssymb}

\usepackage[table, dvipsnames]{xcolor}
\usepackage{pifont}
\usepackage{placeins}
\usepackage{stackengine}

\usepackage{dblfloatfix}
\usepackage{lipsum}

\usepackage{colortbl}
\usepackage{makecell}

\makeatletter

    \def\CT@@do@color{%
      \global\let\CT@do@color\relax
            \@tempdima\wd\z@
            \advance\@tempdima\@tempdimb
            \advance\@tempdima\@tempdimc
    \advance\@tempdimb\tabcolsep
    \advance\@tempdimc\tabcolsep
    \advance\@tempdima2\tabcolsep
            \kern-\@tempdimb
            \leaders\vrule
                    \hskip\@tempdima\@plus  1fill
            \kern-\@tempdimc
            \hskip-\wd\z@ \@plus -1fill }
    \makeatother

\definecolor{dgray}{gray}{0.9}
\definecolor{mygray}{gray}{0.97}

\usepackage{changepage}
\usepackage{tabularx}
\usepackage{multirow}
\usepackage{array}

\usepackage[]{mdframed}

\newcolumntype{C}{>{\centering\arraybackslash}p{8.3cm}}
\newcolumntype{K}{>{\centering\arraybackslash}p{6.5cm}}

\newcolumntype{A}[1]{>{\centering\arraybackslash\hspace{0pt}}m{#1}}

\newcommand\tstrut{\rule{0pt}{2.5ex}}
\newcommand\bstrut{\rule[-1.5ex]{0pt}{0pt}}
\usepackage{makecell}
\usepackage{soul}

\begin{document}

\preprint{APS/123-QED}

\title{Collinear Altermagnets and their Landau Theories}

\author{Hana Schiff}
\email{hschiff@uci.edu}
\affiliation{Department of Physics and Astronomy, University of California, Irvine, California 92697, USA}

\author{Paul McClarty}
\email{paul.mcclarty@cea.fr}
\affiliation{Laboratoire Léon Brillouin, CEA, CNRS, Université Paris-Saclay, CEA Saclay, 91191 Gif-sur-Yvette, France.}

\author{Jeffrey G. Rau}
\email{jrau@uwindsor.ca}
\affiliation{Department of Physics, University of Windsor, 401 Sunset Avenue, Windsor, Ontario, N9B 3P4, Canada}

\author{Judit Romhányi}
\email{jromhany@uci.edu}
\affiliation{Department of Physics and Astronomy, University of California, Irvine, California 92697, USA}

\date{\today}

\begin{abstract}
Altermagnets exhibit spontaneously spin-split electronic bands in the zero spin-orbit coupling (SOC) limit arising from the presence of collinear compensated magnetic order. The distinctive magneto-crystalline symmetries of altermagnets ensure that these spin splittings have a characteristic anisotropy in crystal momentum space. These systems have attracted a great deal of interest due to their potential for applications in spintronics. In this paper, we provide a general Landau theory that encompasses all three-dimensional altermagnets where the magnetic order does not enlarge the unit cell. We identify all crystal structures that admit altermagnetism and then reduce these to a relatively small set of distinct possible Landau theories governing such systems. In the zero SOC limit, we determine the possible local multipolar orders that are tied to the spin splitting of the band structure. We make precise the connection between altermagnetism as defined at zero SOC (``ideal" altermagnets) and the effects of weak SOC. In particular, we examine which response functions allowed by symmetry when SOC is present are guaranteed by the spin-orbit free theory, and spell out the distinctive properties of altermagnets in comparison with conventional collinear antiferromagnets. Finally, we show how these ideas can be applied by considering a number of altermagnetic candidate materials.
\end{abstract}

\maketitle

\tableofcontents
\clearpage
\section{\label{sec:intro}Introduction}

Understanding the interplay of heat, charge, and spin transport in magnetic materials has proven to be an important theme in modern condensed matter physics. Falling broadly under the umbrella of spintronics research, a plethora of phenomena have been uncovered that have motivated and guided the development of new devices to manipulate these currents. Early work focused on uncompensated magnetic metals that have a net magnetic moment, as these offer a straightforward means to induce spin-polarized currents~\cite{SpinCurrent_Dyakonov1971, spinHall_Sinova2015}. For such currents to be robust, as is necessary for devices to be useful, materials with weak spin-orbit coupling are preferable. As an alternative, more recent research has explored \emph{compensated} magnets, where the net moment is zero. These systems offer the potential to achieve THz switching speeds due to the larger underlying exchange scale~\cite{afmSpintronics_Hou2019, Smejkal2022a}. However, generating spin currents is a challenge
due to the compensated order.

Recently, it has been recognized that intrinsic spin-splitting -- characteristic of uncompensated magnets -- is possible even in compensated collinear antiferromagnets at {\it zero} spin-orbit coupling. In many cases, this allows for straightforward spin current generation~\cite{Okugawa2018,Naka2019,HayamiMultipole1,Smejkal2020,Naka2021,Smejkal2022a}. This class of magnets has sublattices with magnetic moments pointing in opposite directions, that are related not by translation or inversion, but instead by a spatial symmetry involving a rotation or reflection. From a fundamental point of view, these insights amount to the appearance of new physics $-$ often called altermagnetism $-$ in the remarkably simple setting of two sublattice collinear antiferromagnetism with spin isotropy in the interactions, a context traditionally associated with spin degenerate bands. 

These altermagnets are sharply distinguished from conventional ferromagnets or antiferromagnets in the idealized limit of zero spin-orbit coupling. In this limit (``ideal altermagnetism"), the characteristic pattern of spin splitting is symmetry enforced by the additional spin rotation symmetries that appear in the absence of spin-orbit coupling. While the weak spin-orbit coupling present in real materials breaks these symmetries, the dominant magnetic energy scale, derived from the idealized limit, controls many of the properties of real altermagnets and is crucial for understanding their behavior.

In ideal altermagnets, these spin symmetries impose a compensated collinear antiferromagnetic order (magnetization $\mathbf{M}\!=\!0$) and preserve spin as a good quantum number while lifting the spin-degeneracy often associated with N\'eel antiferromagnets~\cite{Mazin2021,Smejkal2022a,MazinPunchline2022}. Overall compensation is preserved through the symmetry-imposed constraint that constant energy surfaces in momentum space (and thus occupied electronic bands) display alternating spin patterns~\cite{Noda2016,Okugawa2018,Ahn2019,Hayami2019,Naka2021,Mazin2021,Smejkal2022b,Smejkal2022a,HayamiMultipole1,Radaelli2024TensorialApproach}. These spin splittings are even under inversion regardless of whether the crystal is centrosymmetric~\cite{Ahn2019, Hayami2019, Smejkal2022b, Smejkal2022a, HayamiMultipole1, Radaelli2024TensorialApproach} and can follow $d$-wave, $g$-wave or $i$-wave form factors. These anisotropic spin-splitting
patterns are directly tied to their ability to produce spin currents~\cite{Gonzalez2021,Ma2021_CSVL} and are related to underlying secondary multipolar order parameters~\cite{McClarty2024}.

In the presence of weak SOC, some altermagnets produce a large anomalous Hall response that does \emph{not} arise from canting of their magnetic moments (i.e. weak ferromagnetism)~\cite{Smejkal2020,Naka2020,Feng2022,Naka2022,Han2024_ElSwitchingNwAHE}. Other altermagnets exhibit a wide range of novel responses brought to light in Refs.~\cite{HayamiMultipole1, Smejkal2020,MnF2_Yuan2020} including the thermal Hall effect~\cite{MnF2_Hoyer2024_THE}, piezomagnetism~\cite{MnTe_Aoyama2023_Piezomag}, and anisotropic magnetoresistance~\cite{MnTe_Reichlova2017_MagAniso, MnTe_Kriegner2016_MagRes}, and topological transitions~\cite{Fernandes2024_Zeeman} among others, leading to a great deal of interest in the unconventional transport properties arising from altermagnetism. A crucial recent development has been direct experimental imaging of the altermagnetic spin splitting in candidate altermagnets MnTe and CrSb using photoemission spectroscopy both with and without spin polarization ~\cite{CrSbARPES2024, MnTe_Reichlova_ARPES}. As the definition of altermagnetism is grounded in symmetry, it has implications for all magnetic degrees of freedom meaning that a characteristic pattern of spin splitting of electronic bands should coincide with an analogous chirality splitting pattern in the spin-wave spectrum. Evidence of such a splitting of the magnon bands has been found in MnTe using inelastic neutron scattering~\cite{MnTe_Liu2024_MagnonNeutrons}, but has not been observed in the insulating candidate MnF$_2$~\cite{morano2024absence}.  

Much of the theoretical activity in this field has been focused on making detailed predictions of the properties of particular candidate altermagnetic materials using {\it ab initio} calculations of electronic band structures. However, soon after the discovery of altermagnetism, it was recognized that identification of candidate materials could be made on symmetry grounds under the assumption of weak SOC~\cite{Smejkal2022a}. For this reason, lists of candidate altermagnetic materials have been compiled by identifying materials possessing the characteristic magnetocrystalline symmetries from larger databases of magnetic materials~\cite{Guo23_AMCandidateMaterials}. It was further realized that the ideal limit brings enhanced magnetic symmetries and that these are intimately tied to the key features of altermagnetism~\cite{Smejkal2020,Smejkal2022b}. Understanding what properties of altermagnets are consequences of these higher symmetries, and which are not, is thus an important question. Further, understanding which features survive the introduction of weak spin-orbit coupling and whether
those features are unique to materials descended from ideal altermagnets is also essential in strengthening our understanding of the class of materials.


In this paper, we provide a general Landau theory of altermagnetism grounded in the enhanced symmetries enjoyed by these systems. By examining the ideal limit -- controlled by spin symmetries -- and the physical setting of finite SOC -- controlled by ordinary magnetic symmetries -- we are able to spell out many of the properties of these systems independent of the details of the electronic structure and also understand the extent to which properties of real materials are determined by the idealized limit. Landau theory is the method of choice to study properties common to the whole class of altermagnets because it allows one to be precise about the symmetry breaking and characteristic order parameters of these systems and to unify these with their observable features. 

The paper is organized as follows. In Section~\ref{sec:overview} we introduce altermagnets through a simple framework that encodes their characteristic symmetries. We then formulate a criterion that identifies altermagnets based on the transformation properties of the staggered magnetization under the space group of the crystal. This criterion is powerful enough to provide a complete classification of altermagnets based on symmetry. We carry out this classification for all altermagnets where the magnetic unit cell matches the crystal unit cell (i.e. $\mathbf{Q}\!=\!\mathbf{0}$ order). We include both centrosymmetric and non-centrosymmetric crystal structures in our analysis.  These results are summarized in a look-up table (presented in Table~\ref{tab:wpGN}) containing those Wyckoff positions for each space group that are altermagnetic if an appropriate collinear antiferromagnetic order is imposed at those sites. This result provides a tool to comprehensively study all altermagnets of a given crystal symmetry once the crystal symmetries and magnetic structure are known and is conducive to broad material searches for altermagnetic candidates.

Section~\ref{subsec:AMLTOverview} reviews the Landau theory of altermagnetism at zero SOC and in particular the fact that the characteristic spin splitting can be inferred from the nature of a multipolar order parameter that is fixed by the N\'{e}el order symmetry. Then using the classification scheme from the previous section, we show that any altermagnet at zero SOC can be described by one of $54$ possible Landau theories that we completely specify, including the associated multipolar order parameters and spin splitting anisotropies. This unifies all zero SOC altermagnets into a simple scheme that can be applied to any material candidate.

Section~\ref{sec:finiteSOC} extends this analysis to Landau theories in the more realistic case of altermagnetism at finite SOC. The novelty of this (otherwise standard) analysis lies in determining the special features arising from the particular magnetocrystalline symmetries of altermagnets, and contrasting these with the ideal limit and with properties of conventional antiferromagnets. 
Specifically, building on the multipolar order parameter of the SO-free Landau theory, we can identify symmetry-allowed characteristic observables, such as the components of transport tensors listed in Table~\ref{tab:properties}. Importantly, many of the characteristic responses that we identify at finite SOC are directly implied by the features of the ideal altermagnetic state.

In Section~\ref{sec:examples}, we demonstrate the efficacy of our method through a number of examples belonging to different point groups. Our examples include CrF$_2$, La$_2$CuO$_4$, MnF$_2$, and Fe$_2$O$_3$. Throughout the text, we use MnTe as a demonstrative example.

These discussions make reference to various comprehensive tables listed towards the end of the paper that include: the classification of altermagnets, the tower of multipolar couplings in the ideal limit together with explicit expressions for the lowest order multipole, and tables of allowed couplings to the N\'{e}el vector at finite SOC.

The paper is intended to be accessible to a general audience with at least a cursory familiarity with group theory. More technical discussions of various points may be found in the Appendices. For example, in the main text, we do not rely heavily on the formalism of spin-space groups though these are the complete symmetries of the broken symmetry phase of altermagnets in the zero SOC limit. In Appendices~\ref{app:circumventingSPGs} and \ref{app:noMagGroups}, we explain why we are able to avoid dealing with these groups for the purposes of our analyses.  

\section{\label{sec:overview}Altermagnets from their symmetries}

We begin this section with a general review of altermagnetism, translating the essential ideas into the language of representation theory. We then use this reformulation to perform a complete classification of crystal symmetries
that are compatible with altermagnetism. We note that our analysis concerns altermagnetism arising from staggered dipolar order, and therefore does not encompass scenarios in which orbital ordering \cite{Leeb_2024_OrbitalAM,Giuli_2025_OrbitalAM_NoMoment}, or ferromultipolar order with zero dipolar moment \cite{JaeschkeUbiergo_2025_AtomicAM,Giuli_2025_OrbitalAM_NoMoment}, is responsible for altermagnetic spin-splitting.

Altermagnets are compensated collinear magnets with intrinsic spin-split band structures \emph{at zero spin-orbit coupling}\footnote{Generalizations of the concepts to non-collinear compensated magnets have been proposed~\cite{McClarty2024,Cheong_2024,hu2024spinhalledelsteineffects}}. The key is to identify magneto-crystalline symmetries that do not protect spin degeneracy. This can be done in the simplest case, at zero spin-orbit coupling, by first requiring collinearity of the magnetic structure so that there is a global $\bf{U}(1)$ rotational symmetry in the magnetic degrees of freedom. We further require that the magnetic sublattices are related neither by inversion ($I$) nor lattice translation ($t_{\mathbf{R}}$). 

Ideal altermagnets, due to their lack of SOC, have symmetries that transform only their spin degrees of freedom~\cite{Brinkman1966,Brinkman1966b,spinPointLitvin,spinGroupsLO,corticelli2022, Smejkal2022a,yang2023symmetry,jiang2023enumeration,ren2023enumeration,xiao2023spin}. For collinear spin arrangements, these include all spin-space rotations about the moment direction, and all reflection planes containing this axis. These spin-space mirror symmetries impose a constraint on the bands requiring $\varepsilon_{\mathbf{s}}(\mathbf{k}) = \varepsilon_{\mathbf{s}}(-\mathbf{k})$ where $\mathbf{s}$ is spin-component along the collinear axis. This can be seen by expressing the spin-space mirrors as $\tau2^{s}_{\perp},$ where $\tau$ denotes time reversal (the spin-inversion element) and $2^{s}_{\perp}$ denotes a $\pi$ spin-space rotation perpendicular to the mutual spin axis~\cite{spinGroupsLO, spinPointLitvin, Liu,schiff2023}. This element preserves the spin orientation while flipping the momentum.

When $t_{\mathbf{R}}$ relates opposite spin sublattices, $\tau t_{\mathbf{R}}$ is a symmetry of the magnetic state, and thus $\varepsilon_{\mathbf{s}}(\mathbf{k}) = \varepsilon_{-\mathbf{s}}(-\mathbf{k})$. Combined with the effect of the spin-space mirrors, $\tau 2_{\perp\mathbf{n}}^{s}$, the bands then must be spin degenerate, $\varepsilon_{\mathbf{s}}(\mathbf{k}) = \varepsilon_{-\mathbf{s}}(\mathbf{k})$. When $I$ connects the magnetic sublattices in centrosymmetric systems, the collinear state is invariant under $\tau I$. Immediately, this symmetry also implies spin-degenerate bands, $\varepsilon_{\mathbf{s}}(\mathbf{k}) = \varepsilon_{-\mathbf{s}}(\mathbf{k})$.  

Without $\tau I$ or $\tau t_{\mathbf{R}}$ as symmetries of the magnetically ordered system, there is no constraint enforcing spin degeneracy throughout the entire Brillouin zone. As a result, opposite spin bands are generically split, though they may remain degenerate along high symmetry lines or points. Compensation of an ideal altermagnet must then be enforced by a different symmetry relating the opposite spin sublattices, an element that is neither $I$ nor $t_{\mathbf{R}}.$

The aforementioned symmetry constraints for ideal altermagnets can be encoded in the transformation properties of the Néel vector, $\mathbf{N}$. Under transformations acting only on the lattice (and not the spins, allowed by the lack of SOC), $\mathbf{N}$ may at most change sign. 
Therefore, $\mathbf{N}$ must transform as a one-dimensional, real representation, with the action of each lattice symmetry being $+1$ or $-1$~\cite{Smejkal2022a,Smejkal2022b,McClarty2024}. 

In the simplest case, $\mathbf{N}$ is assumed to be invariant under translations, implying that $t_{\mathbf{R}}$ is represented by $+1$, and the magnetic unit cell thus coincides with the crystallographic one ($\mathbf{Q}=\bf{0}$ AFM order). The more general case where the magnetic unit cell is enlarged is discussed in Ref.~[\onlinecite{supercellAltermagnets2024}]. For $\mathbf{Q}=\bf{0}$ order at least, it is sufficient to analyze transformation properties of $\mathbf{N}$ under the point group of the lattice. Where there is inversion symmetry in the magnetic structure, the invariance of $\mathbf{N}$ under $I$ means that in altermagnets $I$ is also represented by $+1$ (and thus is in an inversion-even irrep of the point group). Any such irrep, aside from the ``trivial" irrep (where all elements are represented by $+1$, corresponding to ferromagnetic order) is a potentially valid representation of the symmetries of an altermagnetic order parameter.

Based on these transformation properties of $\mathbf{N}$, one can frame the search for altermagnetic orders as identifying structures in which $\mathbf{N}$ transforms as a nontrivial 1D inversion-even irrep of the crystal point group. Practically, this can be accomplished by constructing a collinear antiferromagnetic (AFM) order on each Wyckoff position (WP) in each space group and isolating the cases where the corresponding irrep $\Gamma_{\mathbf{N}}$ under which $\mathbf{N}$ transforms obeys the symmetry constraints described above. A detailed procedure for accomplishing this for an arbitrary space group and WP is provided in Appendix~\ref{app:WPalg}, and the complete set of WPs compatible with altermagnetism is given in Table~\ref{tab:wpGN}. As there is a growing need for materials identification and design~\cite{Wei_2024_ChemAMPaper}, these results may help focus search efforts. We note here that our analysis encompasses all viable WP (i.e. those of multiplicity two or greater), and is thus distinct from and expands upon the work of \cite{Roig24_AMMinMod}.

\begin{table}[h!]
\caption{Point groups supporting altermagnetic phases, corresponding space groups as they appear on Bilbao Crystallographic Server~\cite{BilbaoGenPosWP}, and the irreducible representation $\Gamma_{\mathbf{N}}$ under which the N\'eel vector ${\bf N}$ transforms. Note: non-conjugate space groups arise with conjugate point groups $\mathbf{\overline{4}2m}$ and $\mathbf{\overline{4}m2},$ as well as $\mathbf{\overline{6}2m}$ and $\mathbf{\overline{6}m2}.$ If these point groups are treated as distinct, then in total, there are 54 altermagnetic point groups irreps. Otherwise, there are 48.
    }
\centering
\begin{tabularx}{\columnwidth}{ >{\centering\arraybackslash}X  >{\centering\arraybackslash}X  >{\centering\arraybackslash}X}
\hline
\hline
Point group & Space group & $\Gamma_{\bf N}$ irrep. of $\mathbf{N}$ \\
\hline
$\mathbf{2}$ & $3$--$5$ & $\{B\}$ \\
$\mathbf{m}$ & $6$--$9$ & $\{A''\}$ \\
$\mathbf{2/m}$ & $10$--$15$  & $\{B_{g}\}$ \\
$\mathbf{222}$ & $16$--$24$    & $\{B_{1}, B_{2}, B_{3}\}$ \\
$\mathbf{mm2}$ & $25$--$46$ &    $\{A_{2}, B_{1}, B_{2}\}$ \\
$\mathbf{mmm}$ & $47$--$74$ & $\{B_{1g}, B_{2g}, B_{3g}\}$ \\
$\mathbf{4}$ & $75$--$80$ & $\{B\}$ \\
$\mathbf{\overline{4}}$ & $81$, $82$ & $\{B\}$ \\ 
$\mathbf{4/m}$ & $83$--$88$ &  $\{B_{g}\}$ \\ 
$\mathbf{422}$ & $89$--$98$  & $\{ A_{2},B_{1}, B_{2}\}$ \\
$\mathbf{4mm}$ & $99$--$110$  & $\{A_{2}, B_{1}, B_{2}\}$ \\
$\mathbf{\overline{4}2m}$ & $111$--$114$  & $\{A_{2}, B_{1}, B_{2}\}$ \\
$\mathbf{\overline{4}m2}$ & $115$--$122$  & $\{A_{2}, B_{1}, B_{2}\}$ \\
$\mathbf{4/mmm}$ & $123$ --$142$ & $\{B_{1g}, A_{2g}, B_{2g}\}$ \\
$\mathbf{32}$ & $149$--$155$ & $\{A_{2}\}$ \\ 
$\mathbf{3m}$ & $156$--$161$  & $\{A_{2}\}$ \\ 
$\mathbf{\overline{3}m}$ & $162$-- $167$  & $\{A_{2g}\}$ \\ 
$\mathbf{6}$ & $168$--$173$  & $\{B\}$ \\ 
$\mathbf{\overline{6}}$ & $174$  & $\{A''\}$ \\ 
$\mathbf{6/m}$ & $175$, $176$  & $\{B_{g}\}$ \\ 
$\mathbf{622}$ & $177$--$182$  & $\{A_{2}, B_{1}, B_{2}\}$ \\
$\mathbf{6mm}$ & $183$--$186$  & $\{A_{2}, B_{1}, B_{2}\}$ \\
$\mathbf{\overline{6}m2}$ & $187$, $188$ & $\{A_{2}', A_{1}'', A_{2}''\}$ \\
$\mathbf{\overline{6}2m}$ & $189$, $190$  & $\{A_{2}', A_{1}'', A_{2}''\}$ \\
$\mathbf{6/mmm}$ & $191$--$194$  & $\{ B_{1g}, A_{2g}, B_{2g}\}$ \\
$\mathbf{432}$ & $207$--$214$  &  $\{A_{2}\}$ \\ 
$\mathbf{\overline{4}3m}$ & $215$--$220$  & $\{A_{2}\}$ \\ 
$\mathbf{m\overline{3}m}$ & $221$--$230$  & $\{A_{2g}\}$ \\ 
\hline
\hline
    \end{tabularx} 
    \label{tab:AM_PGs_and_irreps}
\end{table}

A few general results can narrow our search. First, we can immediately rule out crystals with point groups $1$, $\overline{1}$, $3$, $\overline{3}$, $23$, and $\frac{2}{m}\overline{3}$ because they contain no irreps satisfying the conditions for altermagnetism. This omission leaves $26$ of the $32$ point groups that may host altermagnetic order and $210$ of the possible $230$ space groups compatible with altermagnetism listed in Table~\ref{tab:AM_PGs_and_irreps}.

We find that each of the 210 possible space groups has at least one Wyckoff position that can support altermagnetism. Of the 1731 space group WPs, 1197 may host altermagnetic order. More specifically, if we were to count all sublattice orders generated by irreps on the Wyckoff positions, 1941 out of 6714 options satisfy the altermagnetic constraints. Altermagnetism, at least at the level of symmetries, is therefore quite common and one may expect to find many altermagnetic materials.

We note that in non-centrosymmetric groups, all nontrivial, real, one-dimensional irreps correspond to symmetry-compensated collinear magnetic order, coinciding with altermagnetism.\footnote{We do not address compensated ferrimagnetism, which is possible when relaxing the constraint that compensation is symmetry-enforced.} If the magnetic unit cell is not enlarged, \emph{any} collinear antiferromagnet in these space groups will necessarily be altermagnetic.

This analysis can be simplified by realizing that there are 54 real, one-dimensional, nontrivial, and inversion-even (where applicable) irreducible representations of the 26 viable point groups, providing only 54 distinct SO-free Landau theories. Studying these 54 cases, as opposed to studying each structure defined by the possible Wyckoff positions, allows us to develop a broader understanding of altermagnets, more clearly delineate their common properties, and identify what distinguishes different realizations. 

\section{\label{subsec:AMLTOverview}Altermagnetic Landau Theory at Zero SOC}

Landau theory is a general framework for understanding symmetry-broken states of matter in terms of their order parameters alone. Constrained only by the nature of the symmetry breaking, Landau theory allows for generic predictions of properties near the phase transition, as well as the dependence of symmetry-allowed response functions on the order parameter. We are interested in a second-order (or weakly first-order)transition passing from a high-symmetry phase to an ordered phase whose symmetries form a subgroup of those present in the original phase. 

For ideal altermagnets, the appropriate order parameter is the Néel vector $\mathbf{N}$, which describes a staggered magnetization. The high-symmetry paramagnetic phase is invariant under all possible global spin transformations (rotations and time inversion), as well as all crystal symmetries (i.e. the space group). Therefore, the general Landau theory for the thermodynamic potential ($\Phi$) takes the form
\begin{equation}
\Phi(\mathbf{N}) = a_2 (\mathbf{N}\cdot\mathbf{N}) + a_4 (\mathbf{N}\cdot\mathbf{N})^2 + \dots,
\end{equation}
where we have assumed $\Phi$ is an analytic function of the order parameter, $\mathbf{N}$. In the ordered phase, the order parameter acquires a nonzero value, $\mathbf{N} \neq 0$. Because ideal altermagnets lack SOC, the symmetries that leave $\mathbf{N}$ invariant do not correspond to a magnetic space group, where all transformations act simultaneously on spin and lattice degrees of freedom. 
Instead, they belong to a more general group of transformations: a \emph{spin-space group}~\cite{Brinkman1966,Brinkman1966b, spinPointLitvin,schiff2023}. 
A spin-space group consists of all operations on real space and spin space that leave the magnetic structure invariant, allowing for operations that transform the spins and the lattice differently. All terms in the free energy and all combinations of $\mathbf{N}$ with other quantities that  $\mathbf{N}$ can couple to must transform trivially under the spin group. 

In the next brief subsection, we first show
that these conditions may be recast so that the Landau theory can avoid using the language of spin groups, instead only requiring the more familiar point group symmetries.

\subsection{Spin groups to point groups}

In the ideal altermagnetic phase, $\mathbf{N}$ transforms as the trivial irrep of a spin group. We may alternatively view $\mathbf{N}$ as transforming under a nontrivial irrep of the SO-free paramagnetic group since this group is not the symmetry group of the ordered phase. Thus, to avoid using spin groups one can make a trade-off and construct the Landau theories using the nontrivial irreps of the SO-free paramagnetic group instead of the trivial irrep of the more complicated spin group. In this case, quantities allowed to couple linearly must have in common at least one irrep of the SO-free paramagnetic group.   

We note that the formal Landau theory in terms of spin groups can be recovered by restricting the SO-free paramagnetic group to elements of the appropriate spin group: with this restriction, $\mathbf{N}$ will transform trivially.  Appendix~\ref{app:circumventingSPGs} details how this restriction reproduces the Landau theory based on the spin group and provides an in-depth justification of bypassing spin groups in the Landau theory.

The power of recasting the Landau theory in terms of the SO-free paramagnetic group lies in the fact that this group is a direct product of spin-space operations and the space group. It turns out that for the cases relevant to $\bm{Q}=\bf{0}$ collinear altermagnetism, the irreps of this group are direct products of the irreps of its factors (see Appendix~\ref{app:spin&realspace}). This factorization of irreps enables us to separate the spatial and spin degrees of freedom. Recalling that $\mathbf{N}$ transforms trivially under translations, we can restrict our focus to spatial symmetries of the SO-free paramagnetic {\it point} groups. Here, $\mathbf{N}$ transforms as a time-reversal odd vector under spin-space transformations, and under any $\Gamma_{\mathbf{N}}$ of the crystal point group satisfying the constraints in Sec.~\ref{sec:overview}.

So far, the Landau theory does not set altermagnets apart from other SO-free antiferromagnets, except in the particular transformation properties of $\mathbf{N}$ discussed above. We now see that the essential features of ideal altermagnets follow from the Landau theory formulated in this setting.

\subsection{Secondary Multipolar Order Parameters for zero SOC}\label{subsubsec:multipolarSecondaryOP}

Secondary multipolar order parameters have significant implications for the spin-splitting structure of electronic bands, and they determine entire classes of observable quantities that couple to $\mathbf{N}$ in the presence of SOC~\cite{McClarty2024}. 
Momentum space multipoles have been utilized to classify spin-splitting~\cite{Radaelli2024TensorialApproach,Roig24_AMMinMod} and as order parameters \cite{Fernandes2024_Zeeman} in altermagnets and in the broader context of electronic band structures in magnetic materials~\cite{HayamiMultipole1, HayamiMultipole2, HayamiMultipole3,clusterMultipoles2017}. Our results differ from these in that they fully exhaust all possibilities for collinear $\bm{Q}=\bf{0}$ altermagnets, and are applicable beyond the analysis of electronic spin-splitting. 

The spirit of Landau theory is to identify all couplings allowed by the choice of primary order parameter which itself is defined through its symmetry properties. Here the primary order parameter is $\mathbf{N}$. For each of the 26 viable spin groups admitting altermagnetism, identified in Table \ref{tab:AM_PGs_and_irreps}, we may identify a multipolar order parameter that couples linearly to $\mathbf{N}$. We consider the time-reversal breaking, spin symmetric, (magnetoelectric) multipoles of the form 
\begin{equation}\label{eq:multipoleDef}
\int d^{3}r\, \, [r_{\mu_{1}}\dotsm r_{\mu_{n}}]  \, \mathbf{m}(\mathbf{r}),
\end{equation}
where $n$ is a positive integer, $r_{\mu}$ are spatial coordinates ($x, y,$ or $z$), and $\mathbf{m}$  is the local magnetization density. The square brackets indicate symmetrization under permutations of the spatial coordinates $x$, $y$, and $z$. We refer to this quantity as a magnetoelectric $(n\!+\!1)$-multipole, that is composed of a rank-1 time-reversal breaking spin-dipole and a rank-$n$ spatial multipole. For example,  $n\!=\!0$  corresponds to the magnetization ${\bf M}=\int d^{3}r\, \,\mathbf{m}(\mathbf{r})$, $n\!=\!1$ is an inversion-breaking magnetoelectric quadrupole that transforms as a rank-1 tensor in both spin and real space $\int d^{3}r\, \,r_{\mu} \, \mathbf{m}(\mathbf{r})$, and $n\!=\!2$ is an inversion-symmetric octupole with a rank-2 spatial component $\int d^{3}r\, \, [r_{\mu_{1}}r_{\mu_{2}}]  \, \mathbf{m}(\mathbf{r})$.  

\begin{figure}[t]
    \centering
    \includegraphics[width=.48\textwidth]{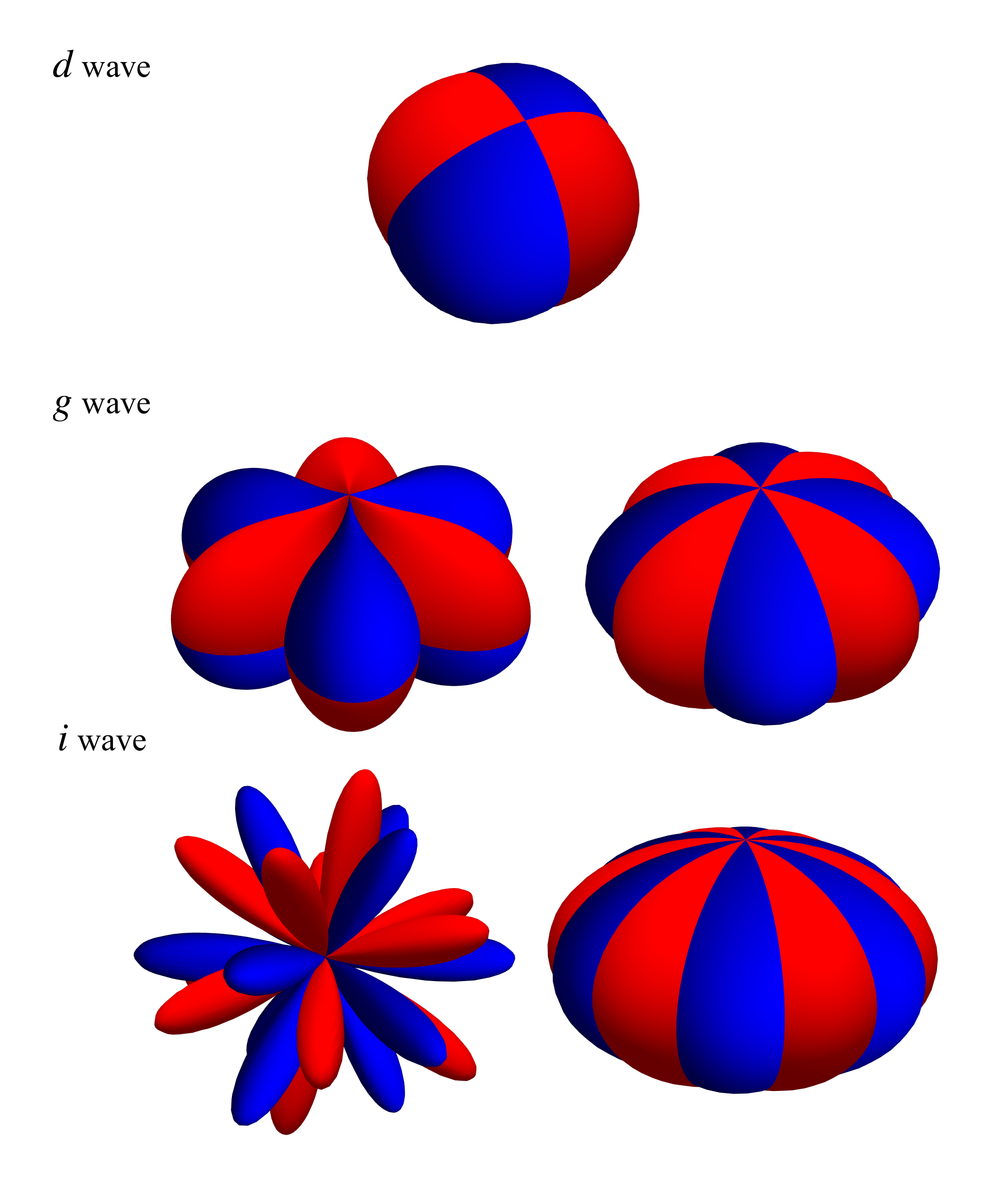}
    \caption{Illustrations of all possible altermagnetic spin-splitting anisotropies in momentum space allowed by symmetry. These correspond to the spatial anisotropies of the lowest order multipole that can couple to the aletermagnetic order parameter ${\bf N}$.}
    \label{fig:couplingReps}
\end{figure}

For ideal altermagnets, we can always find some multipole of the form Eq.~\ref{eq:multipoleDef} that couples linearly to $\mathbf{N}$~\cite{McClarty2024}. A linear coupling requires that the decompositions of the representations of $\mathbf{N}$ and the multipole into irreps of the SO-free paramagnetic group have at least one irrep in common. In spin space, the multipoles and $\mathbf{N}$ already transform identically: the local magnetization density $\mathbf{m}(\mathbf{r})$ and the Néel vector $\mathbf{N}$ transform as time-reversal odd vectors under spin-space rotations and time inversion. Now, we must only check for compatibility between $\mathbf{N}$ and the multipole under point group transformations, noting that without SOC the magnetization density $\mathbf{m}(\mathbf{r})$ transforms trivially under real space operations.

Because $\mathbf{N}$ transforms as $\Gamma_{\mathbf{N}}$ under point group symmetries, the condition for linear coupling to an $(n+1)$-multipole amounts to checking that 
$\Gamma_{\mathbf{N}}$ is contained in the representation under which  $[r_{\mu_{1}}\dotsm r_{\mu_{n}}]$ transforms\footnote{As $\mathbf{N}$ and the multipole component transform identically the latter is strictly not a secondary order parameter but a {\it pseudo-primary} order parameter. ``Secondary" typically denotes an order parameter that transforms under a different irrep than the primary order parameter~\cite{landautheorybook}.}. 

\begin{mdframed}
\begin{center}
{\bf  Jahn Notation} \\
\end{center}
In the following, we use the Jahn symbols~\cite{Jahn1949,mtensor} to denote the intrinsic symmetry properties of a tensor. In this notation, the symbol $a$ marks the time-reversal odd property, and $e$ specifies that the tensor is axial (i.e. inversion-even). The exponent of $V^n$ corresponds to the rank of the tensor. For example, the magnetization transforms as $aeV$ (in the typical, SOC case), corresponding to a time-reversal odd axial vector (rank-1 tensor), and the electric polarization would belong to $V$, a polar and time-reversal even vector. Additionally, symmetry (antisymmetry) of pairs of indices is denoted by square (curly) brackets.  In this notation,  $[r_{\mu_{1}}\dotsm r_{\mu_{n}}]$ transforms as $[V^{n}]$, a time-reversal even rank-$n$ polar tensor that is symmetric in all of its indices.
\end{mdframed}

We find all SO-free Landau theories by determining the $n$ for which the multipole in Eq.~\ref{eq:multipoleDef} couples to $\mathbf{N}$, for every possible altermagnetic structure. Each altermagnetic structure found in Sec.~\ref{sec:overview} is identified in Table \ref{tab:wpGN} by a Wyckoff position and an irrep $\Gamma_{\mathbf{N}}$ of the crystal point group. We check for all $n \leq 6$ whether $\Gamma_{\mathbf{N}}$ is contained in $[V^{n}]$, with the results listed in table \ref{tab:socfreeMultipoles}.

We then focus on the minimal multipole (i.e. with the smallest possible $n$) and find the specific multipole components (by specifying the $r_{\mu_{i}}$ appearing in Eq.~\ref{eq:multipoleDef}) that couple to $\mathbf{N}$. The technical details of this procedure are provided in Appendix~\ref{app:SAB}, and the multipole components are given for each $\Gamma_{\mathbf{N}}$ of every point group in the third column of Table~\ref{tab:CouplingComps}. These results fully determine the SO-free Landau theory.

\begin{figure}[t!]
    \centering
    \includegraphics[width=.48\textwidth]{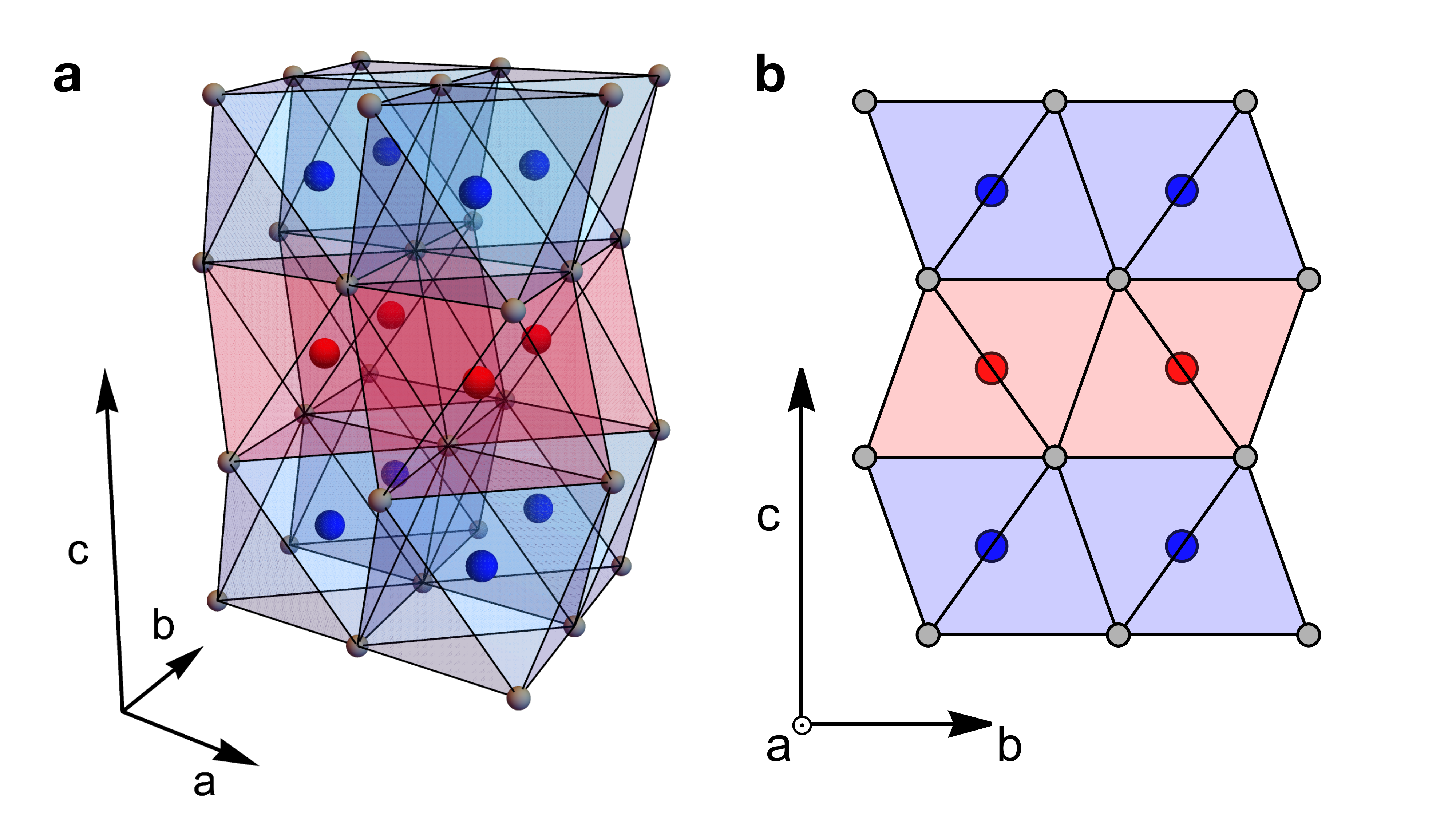}
    \caption{The crystal structure of MnTe with space group symmetry  $P6_{3}/mmc$.  Magnetic Mn ions (red and blue denote magnetic sublattices) reside on the $2a$ Wyckoff positions, at $\{0,0,0\}$ and $\{0, 0, \frac{1}{2}\}$ within the unit cell. The Te ions (gray) occupy the Wyckoff positios $4e$, at $\{\frac{1}{3},\frac{2}{3},\frac{1}{4}\}$, and $\{\frac{2}{3},\frac{1}{3},\frac{3}{4}\}$.}
    \label{fig:MnTe}
\end{figure}

As an example, let us consider the SO-free Landau theory for the semiconductor MnTe, whose crystal structure and magnetic sublattices are given in Fig.~\ref{fig:MnTe}. MnTe has a Néel temperature of about 307 K. The space group for this material is $P6_{3}/mmc$ (No. 194), corresponding to the point group $6/mmm$. The magnetic Mn atoms reside at the $2a$ Wyckoff position~\cite{MnTe_Aoyama2023_Piezomag, MnTe_Betancourt2023_AHE,MnTe_Kriegner2016_MagRes,MnTe_Greenwald1953_Structure,MnTe_Podgorny1983_El,MnTe_Przezdiecka2005_Structure,MnTe_Reichlova2017_MagAniso,MnTe_Komatsubara1963_MagProp,MnTe_Szuszkiewicz2005_NeutronMag}. By considering which point group elements swap magnetic sublattices, we find that the irrep of the altermagnetic N\'eel vector for this crystal structure is $\Gamma_{\mathbf{N}} = B_{2g}$ (in agreement with the tabulated result in Table~\ref{tab:wpGN}). We systematically check for which $n$ the $[V^{n}]$ representation contains the $B_{2g}$ irrep. The spatial part of the $n\!=\!1$ multipole transforms as $V$, the polar vector representation of $6/mmm$, whose decomposition $A_{2u} \oplus E_{1u}$ lacks $\Gamma_{\mathbf{N}}$. Neither $[V^{2}]$ nor $[V^{3}]$ contain $B_{2g}$ in their decompositions, excluding the $n\!=\!2$ and $\!n=\!3$ multipoles. The first multipole allowed to couple with $\mathbf{N}$ is the $n\!=\!4$ multipole. Here, $[V^{4}]$ decomposes as $3 A_{1g} \oplus B_{2g} \oplus B_{1g} \oplus 3 E_{2g} \oplus 2 E_{1g}$, containing $B_{2g}$. For each of the irreps appearing in this decomposition, there is an appropriately transforming (set of) fourth-order polynomials in the $r_{\mu}$. The polynomial transforming as $B_{2g}$ is $yz(y^{2}-3x^{2})$ as described in Appendix~\ref{app:SAB}. Thus, the precise SOC-free multipole coupling to $\mathbf{N}$ in MnTe is $\int d^{3}r\, yz(y^{2}-3x^{2}) \mathbf{m}(\mathbf{r}).$ The next allowed multipole has $n\!=\!6$, as shown in Table~\ref{tab:socfreeMultipoles}.

The spatial polynomials appearing in the secondary (or pseudo-primary) multipolar order parameter are related to the spin-splitting pattern of electronic bands. In centrosymmetric structures, we can identify the spin-splitting pattern with the $n-$order of the secondary multipole~\cite{McClarty2024}. This correspondence follows from the symmetry equivalence of real-space terms $r_{\mu_{1}}...r_{\mu_{A}} \mathbf{m}(\mathbf{r})$ and reciprocal space terms $k_{\mu_{1}}...k_{\mu_{A}}\mathbf{s},$ where $\mathbf{s}$ is the spin of a band~\cite{McClarty2024, HayamiMultipole1,Radaelli2024TensorialApproach}. In this context, the lowest order multipole being $n=4$ in the above example of MnTe is consistent with the observed $g-$wave spin-splitting pattern~\footnote{We briefly comment that the choice of axes differs between this work and that of Ref.~\cite{McClarty2024}, which results in a relabelling of the $B_{1g}$ and $B_{2g}$ irreps}~\cite{McClarty2024}. When inversion is a symmetry, $n$ is always even because a polar vector changes sign under inversion.  

The non-centrosymmetric altermagnets, allowing for both even and odd $n$ multipoles, deserve an additional remark in connection to the spin splitting. The spin splitting is always even in momentum regardless of whether the system is centrosymmetric or non-centrosymmetric, due to the  $\tau 2_{\perp\mathbf{n}}^{s}$ symmetry present for collinear spins. We find, however, that the lowest order multipole is often of odd $n$. In these cases, the spin splitting is \emph{not} dictated by the lowest order multipole, but by the dominant \emph{even} multipole. 

In addition to capturing the pattern of spin splitting in momentum space, the multipolar order parameter has a more direct interpretation as a local multipole in the magnetization density of altermagnetic materials, expected to be observable experimentally~\cite{Bhowal2024,McClarty2024}. 

\section{\label{sec:finiteSOC} Altermagnetic Landau Theory at Finite SOC}

So far we have focused on the zero spin-orbit coupled limit where altermagnetism is most clearly defined. In this limit, we have been able to determine all possible crystalline symmetries compatible with altermagnetism and we have found the finite number of Landau theories and multipolar order parameters corresponding to these cases. 

In real materials, spin-orbit coupling is finite. What this means for the magnetic properties at the microscopic level is somewhat involved. The specifics depend, among other things, on the precise orbital content, the nature of the spin-orbit coupling, and the crystal field. Here we side-step these details and focus on the consequences of symmetry alone.

We identify the lowest-order multipolar order at finite SOC, to see what intrinsic features of zero SOC altermagnets are inherited by real materials. Then, we concern ourselves with the physics of real materials by determining, on symmetry grounds, what responses are expected in altermagnets. For example, noteworthy features of certain metallic altermagnets are that they support spin currents or anomalous Hall conductivity among other exotic transport properties. 

We organize this section by first making some general remarks about the nature of Landau theories for altermagnets at finite SOC. Then we provide a group theoretic result that allows us to generalize the observations of the next section to the full class of altermagnets. Then we discuss the finite SOC analogs of the multipolar order at zero SOC thus connecting the ideal limit to realistic systems.  Finally, we give an overview of the observable quantities that might be of interest in the context of altermagnetism including their symmetry properties. In the following section, we apply all these ideas to specific materials candidates. 

\subsection{Altermagnetic Landau Theories at Finite SOC: General Remarks}

Previously we observed that Landau theories at zero SOC are completely determined from the transformation properties of the N\'{e}el vector $\mathbf{N}$, described by some irreducible representation, $\Gamma_{\mathbf{N}}$, of the crystal point group. Crucially, in the paramagnetic phase, this Landau theory is completely symmetric under spin rotations of $\mathbf{N}$. The full symmetry of the problem is the group of all rotations $\mathbf{SO}(3)$ in spin space, along with $\mathbf{G}+\tau \mathbf{G}$, where $\mathbf{G}$ is the space group of the crystal acting purely on the lattice, and $\tau$ denotes time reversal. Here, transformations of the system may differ in spin-space and real space. 

When SOC is finite, the symmetry group of the paramagnetic phase is lower because spatial transformations and spin transformations are locked: transformations on spins and the lattice are identical, with the caveat that the spins transform axially. Pure spin rotation symmetry is lost, meaning that the full symmetry of the SO-coupled paramagnetic phase is given by $\mathbf{G} + \tau \mathbf{G}$. Restricting our attention again to $\bm{Q}=\bm{0}$ orders, the moments transform under the time reversal odd axial vector representation of the point group of the lattice, denoted as $aeV$ in Jahn notation~\cite{Jahn1949,mtensor} (introduced briefly in Sec.~\ref{subsec:AMLTOverview}), and the N\'eel vector $\mathbf{N}$ transforms as 
\begin{equation} \label{eq:NRepSOC}
aeV \otimes \Gamma_{\mathbf{N}}\;.
\end{equation}

For example, in the previous section, we saw that for MnTe, the irrep $\Gamma_{\mathbf{N}}$ corresponds to $B_{2g}$ of $6/mmm.$ The axial vector representation for $6/mmm$ decomposes as $A_{2g}\oplus E_{1g},$ where the 2D irrep corresponds to axial $x$ and $y$ components. Thus, by taking the product of $\Gamma_{\mathbf{N}}$ and $aeV$ we find that the Néel vector components $\{N_{x}, N_{y}\}$ and $N_{z}$ transform as $E_{2g}$ and $B_{1g}$ respectively. If we restrict to an in-plane $\mathbf{N} = N_{x}\hat{\mathbf{x}} + N_{y}\hat{\mathbf{y}}$, as is experimentally observed~\cite{Kunitomi1964_MnTe}, then the Landau theory is given by
\begin{equation}
\Phi = a_2 (N_{x}^{2} + N_{y}^{2}) + a_4 (N_{x}^{4} + N_{y}^{4}) + \ldots
\end{equation}
In the following subsection, we determine which tensors can couple linearly to components of $\mathbf{N}.$

\subsection{Coupling to the N\'{e}el Vector at Finite SOC\label{sec:SOCLT_general}}

In this section, we give a simple criterion that allows one to assess whether some components of a tensor observable $\xi$ couple linearly to $\mathbf{N}$, based on knowledge of the multipolar order parameter of the SO-free theory. In other words, we tie together features of the spin-splitting at zero SOC and physical properties at finite SOC. 

To set the stage, let $\xi$ be a tensor that transforms under a representation $\Gamma_\xi$ of the spin-orbit coupled paramagnetic group. This tensor corresponds to some physical observable of the altermagnetic phase that we wish to probe, such as electrical conductivity, magnetoresistance, etc. We also suppose that the SO-free theory has a spin symmetric multipole with a spatial component transforming as $[V^{n}]$ where $n$ is the lowest rank that appears in the Landau theory. 

Linear coupling between $\mathbf{N}$ and $\xi$ is allowed if their representations share at least one common irrep in their decompositions. This criterion is equivalent to the trivial irrep appearing in the decomposition of $\Gamma_{\xi}\otimes(aeV \otimes \Gamma_{\mathbf{N}})$. Recall that $aeV$ is the time reversal odd axial vector representation. We may recast this condition into a more practical form: that $\Gamma_{\mathbf{N}}$ \emph{must appear in the decomposition of} $\Gamma_\xi\otimes aeV$. That this condition is equivalent may be seen by first invoking associativity of the direct product, so that we seek the trivial irrep in $(\Gamma_{\xi}\otimes aeV)\otimes \Gamma_{\mathbf{N}}.$ Then, it is clear that $(\Gamma_{\xi}\otimes aeV)$ must contain $\Gamma_{\mathbf{N}}$ in its decomposition for the trivial irrep to appear in this product.

In our analysis of the SO-free limit, we established that the lowest order altermagnetic multipolar order parameter coupling linearly to $\mathbf{N}$ has the smallest $n$ for which $\Gamma_{\mathbf{N}}$ is contained in $[V^{n}]$. Therefore, if $[V^{n}]$ is fully contained in $\Gamma_{\xi}\otimes aeV$ then $\Gamma_{\bf N}$ will also be contained in $\Gamma_{\xi}\otimes aeV$, meaning ${\bf N}$ will couple to $\xi$. This criterion
\begin{equation}\label{eq:SOCguaranteedCoupling}
[V^{n}] \subseteq \Gamma_{\xi}\otimes aeV
\end{equation}
connects the Landau theories with and without SOC, and allows us to identify quantities $\xi$ \emph{directly predicted by the SO-free analysis}. We shall additionally see that these $\xi$ can differentiate between altermagnetic and non-altermagnetic phases.

For a given $(n+1)$-multipole from the SO-free theory, we identify representations $\Gamma_{\xi}$ 
for which $[V^{n}]$ is contained in $\Gamma_{\xi}\otimes aeV.$ Viable $\Gamma_{\xi}$ meet this condition for \emph{all} point groups; the presence of the $(n+1)$-multipole without SOC then guarantees coupling between $\mathbf{N}$ and $\xi$ when SOC is included, and this feature is a universal property of the $(n+1)$-multipole. Observables $\xi$ obtained in this fashion are fundamental in altermagnets; they arise due to secondary multipolar order present in the ideal altermagnetic phase.  

In Table~\ref{tab:guaranteedsoc}, we list the representations $\Gamma_{\xi}$ of the tensors that can couple linearly to ${\bf N}$, based on the presence of an $(n+1)$-multipole. This table is somewhat spartan containing only Jahn symbols of coupled quantities at each multipolar rank. Later, we demonstrate the utility of this table and spell out examples of explicit components of particular physical quantities that are relevant to altermagnetism. A partial list of physical quantities of interest is given in Table~\ref{tab:properties} together with their transformation properties labeled by $\Gamma_{\xi}$. 

To give a flavor of how this table can be used, we return to the case of MnTe. Recall from Sec.~\ref{sec:overview} that the minimal SO-free multipole for this system has spatial rank $n\!=\!4.$ From Table~\ref{tab:guaranteedsoc}, any tensor transforming as $aeV[V^{2}]$ can couple to $\mathbf{N}.$ In words, these are spatially symmetric rank-$2$ tensors times an axial time-reversal odd vector. In Sec.~\ref{subsec:expsigs} we make explicit the coupling between components of $\xi$ and the components $N_{i}$ of the Néel vector, and to make concrete the physical quantities corresponding to $\xi$.

\begin{table}[t]
\caption{\label{tab:guaranteedsoc}%
The representation $\Gamma_\xi$ for quantities $\xi$ is guaranteed to couple linearly to $\mathbf{N}$ in the presence of SOC, based on the rank-$n$ multipole in the SOC-free limit. The representations are denoted by Jahn symbols, where $aV$ is a time-reversal odd ($a$) polar vector ($V$), $aeV$ is a time-reversal odd ($a$) axial vector ($eV$), $aeV^{2}$ is a time-reversal odd axial tensor of rank 2, and $aeV[V^{2}]$ is a time-reversal odd axial rank-3 axial tensor that is symmetric in two indices. The $n=5$ case is absent because this multipole is not minimal for any irrep of any point group (see Table~\ref{tab:socfreeMultipoles}).}
\centering
\begin{tabularx}{\columnwidth}{ >{\centering\arraybackslash}X  >{\centering\arraybackslash}X }
\hline
\hline
Multipole rank ($n$) &
Representation $\Gamma_{\xi}$\\
\hline
1 &  $aV$, $aeV^{2}$\\
2 & $aeV$\\
3 & $aeV^{2}$\\
4, 6 & $aeV[V^{2}]$\\
\hline
\hline
\end{tabularx}
\end{table}

We emphasize that the observables $\xi$, derived from the SO-free limit, are not the \emph{only} quantities that are allowed to couple to $\mathbf{N}$ in the spin-orbit coupled altermagnetic phase. There are other quantities $\mathbf{N}$ can couple to, but we would not view these quantities as being fundamentally related to the altermagnetism as they do not follow from the idealized limit. Up to tensors of rank three, the representations $\Gamma_{\xi}$ listed in Table~\ref{tab:guaranteedsoc} are the only types of tensors fundamentally implied by the SO-free theory.

\subsection{\label{subsec:distAFM&AM}Distinguishing between N\'eel AFMs and Altermagnets}

Both altermagnetic and non-altermagnetic AFMs are collinear, compensated magnetic structures. This makes them difficult to distinguish in experiment. Here, we underline that the $\xi$ found in the previous section are unique to altermagnets, in the sense that a non-altermagnetic N\'eel AFM would \emph{not} have a linear coupling between $\mathbf{N}$ and these quantities. We only need to distinguish between these cases in centrosymmetric crystals, since there is no distinction between altermagnetic and non-altermagnetic collinear AFM order in non-centrosymmetric crystals, as discussed in Sec.~\ref{sec:overview}. The distinction between these two types of orders is a consequence of their parity under inversion symmetry. When $\bm{Q}=\bf{0}$, altermagnets are even, while non-altermagnetic orders are odd.

In the end, the distinction is simple to state: any inversion-even tensor couples linearly exclusively to altermagnets, while an inversion-odd tensor will couple only to non-altermagnetic $\mathbf{N},$ provided $\bm{Q}=\bf{0}.$

\subsection{Multipolar Order in Altermagnets at Finite SOC}

We briefly question whether the secondary multipolar order parameter, crucial to the SO-free theory, plays a role in the finite SOC limit. Consider, again, multipoles of mixed polar and magnetic character as in Eq.~\ref{eq:multipoleDef}. Now, with SOC these multipoles transform as $aeV\otimes[V^{n}].$ The multipoles for $n$ equal to that of the SO-free case are still able to couple to $\mathbf{N}$ in the presence of SOC, as both share $aeV,$ and we know $\Gamma_{\mathbf{N}} \in [V^{n}]$. As such, even in the presence of SOC, the multipoles act as a secondary (or pseudoprimary) order parameter.

\begin{table}[b!]
    \caption{\label{tab:MnTeSOC_Multipole}
    Transformation properties of the order parameter ${\bf N}$ in MnTe and the part of the integrand of the $n\!=\!4$ spin-orbit coupled multipole in Eq.~\ref{eq:multipoleDef} to which the N\'{e}el vector component couples.
    }
    \centering
    \begin{tabularx}{\columnwidth}{>{\hsize=0.3\hsize\raggedright\arraybackslash}X >{\hsize=0.6\hsize\centering\arraybackslash}X >{\hsize=1\hsize\centering\arraybackslash}X}
    \hline
    \hline
      Irrep. & Néel component & Multipole component\\
      \hline
      \multirow{4}{*}{$B_{1g}$ }  & \multirow{4}{*}{$N_{z}$} & $(x^3-3xy^2)z m_z$ \\
       & & $(x^3-3xy^2)\left(\begin{smallmatrix}
           x\\
           y
       \end{smallmatrix}\right)^{\intercal}
       \left(\begin{smallmatrix}
           m_x\\
           m_y
       \end{smallmatrix}\right)$\\
       & & $(y^3-3yx^2)\left(\begin{smallmatrix}
           x\\
           y
       \end{smallmatrix}\right)^{\intercal}
       \left(\begin{smallmatrix}
           m_y\\
           -m_x
       \end{smallmatrix}\right)$\\
       & & $z^2\left(\begin{smallmatrix}
           x^2-y^2\\
           -2xy
       \end{smallmatrix}\right)^{\intercal}
       \left(\begin{smallmatrix}
           m_x\\
           m_y
       \end{smallmatrix}\right)$\\
      \hline
     \multirow{8}{*}{$E_{2g}$} & \multirow{8}{*}{$\begin{pmatrix}N_{x}\\N_{y}\end{pmatrix}$} &  $z (x^2y-\frac 1 3 y^3)\left(\begin{smallmatrix}
           m_x\\
           m_y
       \end{smallmatrix}\right)$ \\
        {} & {} & $z (xy^2-\frac 1 3 x^3) \left(\begin{smallmatrix}
           m_y \\
           -m_x
       \end{smallmatrix}\right)$\\
       {} & {} & $z^2 \left(\begin{smallmatrix}
           2 xy \\
           x^2-y^2
       \end{smallmatrix}\right) m_z$\\
       {} & {} & $\left(\begin{smallmatrix}
           x^3y - \frac 1 3 xy^3 \\
           x^2y^2 -\frac 1 3 y^4
       \end{smallmatrix}\right) m_z$\\
       {} & {} & $\left(\begin{smallmatrix}
           2 x^3y +2 xy^3 \\
           x^4-y^4
       \end{smallmatrix}\right)m_z$\\
      {} & {} & $z^3 \left(\begin{smallmatrix}
           x m_y +y m_x\\
           x m_x -y m_y
       \end{smallmatrix}\right)$  \\
        {} & {} & $z (x^2+y^2) \left(\begin{smallmatrix}
           x m_y +y m_x\\
           x m_x -y m_y
       \end{smallmatrix}\right)$  \\
      \hline
      \hline
    \end{tabularx}
\end{table}

To be concrete, we find the components of the multipole coupling to $\mathbf{N}$ for MnTe with SOC, which has $\Gamma_{\mathbf{N}} = B_{2g}$. One can show that from Eq.~\ref{eq:NRepSOC}, the $N_{x}$ and $N_{y}$ components of the Néel vector transform as $E_{2g}$, while $N_{z}$ transforms as $B_{1g}$. Since $n\!=\!4$ is the spatial order of the SO-free multipole, the SOC multipole transforms as the $aeV\otimes [V^{4}]$ representation of $6/mmm.$ The irrep decomposition for this representation is $2A_{1g} \oplus 5 A_{2g} \oplus 4 B_{2g} \oplus 
 4B_{1g} \oplus 7 E_{2g} \oplus 8 E_{1g}.$ Because this decomposition contains $B_{1g}$ and $E_{2g}$, the SOC multipole for MnTe can couple to all components of the N\'{e}el vector. From this calculation, it follows that there are four $B_{1g}$ multipoles and seven $E_{2g}$ multipoles that are relevant to the spin-orbit coupled case, a much richer selection than the spin-orbit free case.

Despite the complexity of the allowed multipoles in MnTe with SOC it is instructive to see how to compute the multipolar components that couple linearly to the N\'{e}el vector in at least one case. This can be accomplished using the procedure outlined in Appendix~\ref{app:SAB}. Because $B_{1g}$ squares to the trivial irrep, $N_{z}$ may couple to any of the four $B_{1g}$ multipole components listed in Table~\ref{tab:MnTeSOC_Multipole}. And, similarly, any of the seven symmetry-allowed multipolar components may couple to the $(N_x, N_y)$ components.  

The transformation properties of the N\'{e}el components and multipole components are shown in Table~\ref{tab:MnTeSOC_Multipole}, where we provide the integrand of the multipole from definition Eq.~\ref{eq:multipoleDef}. In each case, we have expressed the multipole components in a simple basis, such that the dot product with $(N_{x}, N_{y})$ yields the allowed coupling. As the moments in the ordered phase of MnTe lie in the triangular planes~\cite{Kunitomi1964_MnTe} the $E_{2g}$ multipoles are the ones that are experimentally relevant. As we should expect, when SOC is present, the multipole is tied to the direction of the local magnetization density. In common with the SOC-free case, the relevant multipoles are time-reversal odd with rank $n=4$ though the pattern of nodes is very different to the case of the SOC-free multipole.

In general, the condition of Eq.~(\ref{eq:SOCguaranteedCoupling}) is equivalent to the condition that the SOC multipole representation is contained within $\Gamma_{\xi}.$ This fortifies the notion that the multipolar order parameter plays a central role in dictating the behavior of altermagnets.

As the example of MnTe indicates, we should expect a considerable increase of complexity in the allowed multipoles in passing from the SO-free to the spin-orbit coupled case. While the SO-free analysis provides simple, direct information about spin-splittings of the bands of ideal altermagnets we do not expect detailed information about multipoles in materials to shed much light on the general phenomenon of altermagnetism. Therefore, we do not tabulate the SOC multipole couplings in general. However, because multipoles with SOC may be of interest in specific instances we emphasize that they may be obtained using the same technique as all other tabulated couplings that is described in Appendix~\ref{app:SAB}.

\subsection{Experimental Signatures of Altermagnetism\label{subsec:expsigs}}

The goal of this section is to spell out the framework that will allow us to make experimental predictions about the behavior of altermagnets based on symmetry alone. To this end, we now identify some concrete physical quantities corresponding to the tensor $\xi$ from Sec.~\ref{sec:SOCLT_general}. Further, we predict which components of $\xi$ are generically non-zero.

In Table~\ref{tab:properties} we provide a list of common equilibrium, transport, and optical material properties transforming under the representations $\Gamma_{\xi}$ identified to be relevant for altermagnets in Table~\ref{tab:guaranteedsoc} (sourced from MTENSOR~\cite{mtensor}). For each property, we list its name and defining equation. In some cases, the full tensor has one of the desired transformation properties. In other cases, it is only a \emph{part} of the tensor that transforms under a $\Gamma_{\xi};$ we specify this in the fourth column of Table~\ref{tab:properties}.

For some tensors, the (anti)symmetric part may be `repackaged' into a smaller object. A canonical example is the anomalous Hall conductivity (AHC), the antisymmetric part of the electrical conductivity tensor. In Jahn notation, the full conductivity tensor transforms as $[V^{2}]^{*}$, a rank-2 polar tensor, with the time-reversal property $\tau\sigma_{ij}=\sigma_{ji}$, denoted by the starred square bracket $[~~]^*$\footnote{In `generalized' Jahn notation~\cite{mtensor}, the star denotes that time-reversal relates a tensor element to some other tensor element, potentially of a different tensor (such as for the Seebeck and Peltier effect).}. The AHC tensor, $\sigma_{ij}^{A} = \frac{1}{2}(\sigma_{ij} - \sigma_{ji})$ transforms as an antisymmetric time-reversal odd rank-2 tensor $a\{V^{2}\}$, whose three independent components, $\sigma_{yz}$, $\sigma_{zx}$, and $\sigma_{xy}$, can be ``repackaged" into a magnetic axial vector, $\bm{\sigma} = \{\sigma_{yz}, \sigma_{zx},\sigma_{xy}\}$, transforming as $aeV$. For details about the repackaging of tensor components in Table~\ref{tab:properties} see Appendix~\ref{app:repackaging}.

Having fixed a set of observables, we compute the components of these quantities that couple linearly to components $N_{i}$ of the Néel vector. These results are provided for each point group in the final column of Table~\ref{tab:CouplingComps}.

\begin{table}[t!]
    \caption{\label{tab:MnTe_N_and_R}Transformation properties of the order parameter ${\bf N}$ in MnTe and the part of the magnetoresistance that is symmetric in the first two indices.}
    \centering
    \begin{tabularx}{\columnwidth}{>{\hsize=0.2\hsize\raggedright\arraybackslash}X >{\hsize=0.4\hsize\centering\arraybackslash}X >{\hsize=1\hsize\centering\arraybackslash}X}
    \hline
    \hline
      Irrep. & Order parameter & Magnetoresistance component\\
      \hline
      $B_{1g}$   & $N_{z}$ & $2R^{S}_{xyx}+R^{S}_{xxy}-R^{S}_{yyy}$ \\
      \hline
     \multirow{4}{*}{$E_{2g}$} & {} &  \multirow{2}{*}{\makecell{$\begin{pmatrix}
  2R_{xyz}^{S}\\
  R_{xxz}^{S}-R_{yyz}^{S}
\end{pmatrix}$}} \\
      {} & \multirow{2}{*}{ \makecell{$\begin{pmatrix}N_{x}\\N_{y}\end{pmatrix}$}} & {}\\
      {} & {} &  \makecell{$\begin{pmatrix}
  R_{yzx}^{S} + R_{xzy}^{S}\\
  R_{xzx}^{S}-R_{yzy}^{S}
\end{pmatrix}$}\\
      \hline
      \hline
    \end{tabularx}
\end{table}

To see how this information may be of use, we again consider MnTe. In Sec.~\ref{sec:SOCLT_general} we concluded that the MnTe order parameter, $\mathbf{N}$, couples to $aeV[V^{2}]$ tensors due to its $n\!=\!4$ SO-free multipolar order parameter. One may be interested, for example, in the non-zero components of the magnetoresistance, $R_{ijk},$ for spintronics applications. We focus on the part that is symmetric in the first two indices, $R^{S}_{ijk}$, as this part transforms as $aeV[V^{2}]$. We have seen already that $\{N_{x},N_{y}\}$ transform under the $E_{2g}$ irrep of $6/mmm,$ while $N_{z}$ transforms as $B_{1g}$. Our task now is to find the components $R_{ijk}^{S}$ that may couple to $N_{i}$, i.e. components of these observables that transform under the same irrep. The transformation properties of $R_{ijk}^{S}$ and $N_{i}$
are listed in Table~\ref{tab:MnTe_N_and_R}.

\clearpage

\onecolumngrid
\begin{center}
\begin{table}[t]
\caption{\label{tab:properties}%
Tensors transforming under the representations in Table~\ref{tab:guaranteedsoc}.  In the last column, we denote the tensor part transforming under $\Gamma_\xi$ (details can be found in Bilbao's MTENSOR package~\cite{mtensor} and in Appendix~\ref{app:repackaging}). 
Superscripts $A$ and $S$ indicate the symmetric and antisymmetric parts of a tensor, respectively.  Furthermore, $\varepsilon_{\alpha ij}$ is the Levi-Civita symbol, $J_{i}$ denotes an electric current density, $E_{i}$ an electric field, $q_{i}$ a thermal current, $H_{i}$ a magnetic field, $T$ temperature, $\Sigma_{ij}$ the stress, $\varepsilon_{ij}$ the dielectric tensor, and $\rho_{ij}$ the resistivity tensor. Most notation coincides with that of Ref~\cite{mtensor}. For $aeV[V^{2}]$ our Jahn symbol does \emph{not} indicate which indices are symmetrized. All inverse effects have the same transformation properties and are omitted from the table for brevity. All of these observables may appear in non-centrosymmetric altermagnets, while in centrosymmetric altermagnets only those corresponding to even $n$ multipoles may appear.
}
\footnotesize
\begin{tabularx}{\textwidth}{m{1.5cm} m{1cm} m{6.5cm} m{4cm} m{3cm}}
\hline
\hline
$\Gamma_\xi$ & $n$ & Quantity ($\xi$) & Defining equation & Tensor part   \\
\hline
\multirow{2}{1.5cm}{$aV$} & \multirow{2}{1cm}{1} & Polar Toroidal Moment $T_{i}$ & - & Full  \\
& & Pyrotoroidic tensor $r_{i}$ & $T_{i} = r_{i} \Delta T$ & Full  \\
\hline
\multirow{7}{1.5cm}{$aeV$} & \multirow{7}{1cm}{2} & Magnetization $M_{i}$ & -  & Full  \\
& & Electric conductivity $\sigma_{ij}$ & $J_{i} = \sigma_{ij}E_{j}$ & $ \sigma_{\alpha}^{A} = \frac{1}{2}\varepsilon_{\alpha ij}\sigma_{ij}^{A}$  \\
& & Soret thermodiffusion tensor $s_{ij}$ & $J_{i} = s_{ij}(\nabla T)_{j}$ & $s_{\alpha}^{A} = \frac{1}{2}\varepsilon_{\alpha ij} s_{ij}^{A}$  \\
& & Thermal conductivity $\kappa_{ij}$ & $q_{i} = \kappa_{ij}(\nabla T)_{j}$ & $\kappa^{A}_{\alpha} = \frac{1}{2} \varepsilon_{\alpha ij} \kappa^{A}_{ij}$  \\
& & Peltier tensor $\pi_{ij}$  & $q_{i} = \pi_{ij}J_{j}$  & \multirow{2}{4cm}{$\tilde{S}_{\alpha} = \frac{1}{2}\varepsilon_{\alpha ij}(\pi_{ij}^{A} + \beta_{ij}^{A})$} \\
& & Seebeck tensor $\beta_{ij}$ &  $E_{i}=\beta_{ij}(\nabla T)_{j}$ &  \\
& & Spontaneous Faraday effect $F_{ij}$ & - & $F_{\alpha} = \frac{1}{2} \varepsilon_{\alpha ij}F_{ij}$ \\
\hline
 $aeV^{2}$ & 1 \& 3 & Magnetoelectric tensor $\alpha_{ij}$ & $M_{i}=\alpha_{ij}E_{j}$ & Full \\
 \hline
\multirow{13}{1.5cm}{$aeV[V^{2}]$} & \multirow{13}{1cm}{4 \& 6} & Piezomagnetic tensor $\Lambda_{ijk}$ & $M_{i}=\Lambda_{ijk}\Sigma_{jk}$ & Full \\
& & Second order magnetoelectric tensor $\alpha_{ijk}$ & $M_{i} = \alpha_{ijk}E_{j}E_{k}$ & Full  \\
& & Magneto-optic Kerr effect $z_{ijk}^{S}$ & $\varepsilon_{ij} = iz_{ijl}^{S} H_{l}$ & Full  \\
& & Quadratic magneto-optic Kerr effect $iC^{A}_{ijkl}$ & $\varepsilon_{ij} = C_{ijkl}^{A}H_{k}H_{l}$ & $C_{\alpha kl} = \frac{1}{2}\varepsilon_{\alpha ij} C_{ijkl}^{A}$   \\
& & Magnetoresistance $R_{ijk}$ & $E_{i} = R_{ijk}J_{j}H_{k}$ & $R_{ijk}^{S} = \frac{1}{2}(R_{ijk}+R_{jik})$  \\
& & Righi-Leduc magnetorhermal tensor $Q_{ijk}$ & $q_{i} = Q_{ijk}(\nabla T)_{j}H_{k}$ & $Q_{ijk}^{S}  = \frac{1}{2}(Q_{ijk}+Q_{jik})$  \\
& & Ettinghausen tensor $M_{ijk}$  & $q_{i} = M_{ijk}J_{j}H_{k}$ & \multirow{2}{4cm}{$S_{ijk} = \frac{1}{2}(M_{ijk}^{S} + N_{ijk}^{S})$ } \\
& &  Nernst tensor $N_{ijk}$  & $E_{i}=N_{ijk}(\nabla T)_{j} H_{k}$ &   \\
& & Magnetic resistance tensor $T_{ijkl}$ & $E_{i}=T_{ijkl}J_{j}H_{k}H_{l}$ & $T_{\alpha kl}^{A} = \frac{1}{2}\varepsilon_{\alpha ij} T_{ijkl}^{A}$  \\
& & Magneto-heat-conductivity tensor $\mathcal{S}_{ijkl}$ & $q_{i}=\mathcal{S}_{ijkl}(\nabla T)_{j}H_{k}H_{l}$ & $\mathcal{S}^{A}_{\alpha kl} = \frac{1}{2}\varepsilon_{\alpha ij} \mathcal{S}^{A}_{ijkl}$  \\
& & Piezoresistivity tensor $\Pi_{ijkl}$ & $\Delta \rho_{ij} = \Pi_{ijkl}\Sigma_{kl}$ & $\Pi_{\alpha kl}^{A} = \frac{1}{2}\varepsilon_{\alpha ij}\Pi_{ijkl}^{A}$   \\ 
& & Magneto--Seebek tensor $\alpha_{ijkl}$   & $E_{i}=\alpha_{ijkl}(\nabla T)_{j}H_{k}H_{l}$   & \multirow{2}{4cm}{$\tilde{\mathcal{A}} = \frac{1}{4}\varepsilon_{\alpha ij}(\alpha_{ijkl}^{A} - P_{ijkl}^{A})$}  \\
& &  Magneto-Peltier tensor $P_{ijkl}$ & $q_{i}=P_{ijkl}H_{k}H_{l}H_{j}$ &   \\
\hline
\hline
\end{tabularx}
\end{table}
\vspace{2em}
\end{center}
\twocolumngrid

The direct product $B_{1g}\otimes B_{1g}$ is $A_{1g}$, providing the invariant term $N_{z}\left( 2R^{S}_{xyx} + R^{S}_{xxy}-R^{S}_{yyy} \right)$. The products of the $E_{2g}$ irreps decompose as $A_{1g}\oplus A_{2g} \oplus E_{2g}$, and so we expect one invariant coupling for each of the two $E_{2g}$ irreps. The pairs of components in Table~\ref{tab:MnTe_N_and_R} are expressed such that their dot product with the in-plane Néel vector gives rise to the allowed couplings.

The couplings in Table~\ref{tab:MnTe_N_and_R} indicate that $R^{S}_{xyx},$ $R^{S}_{xxy},$ $R^{S}_{yyy}$, $R^{S}_{xyz}$, $R^{S}_{xxz}$, $R^{S}_{yyz}$, $R^{S}_{yzx}$, $R^{S}_{xzy},$ $R^{S}_{xzx}$ and $R^{S}_{yzy}$ may all generically be non-zero in MnTe. However, with $\mathbf{N}$ restricted to an in-plane $\mathbf{N} = N_{x}\hat{\mathbf{x}} + N_{y}\hat{\mathbf{y}}$ in accordance with experimental data~\cite{Kunitomi1964_MnTe}, we expect $R^{S}_{xyx}$, $R^{S}_{xxy}$ and $R^{S}_{yyy}$ to be zero.

Similar couplings between the order parameter and any of the possible tensors can be found using the procedure outlined in Appendix~\ref{app:SAB}. In Table~\ref{tab:CouplingComps}, we explicitly list the couplings between $N_{\alpha}$ and tensor components of $\xi$ transforming under each possible representation $\Gamma_{\xi}.$ In this table, polar vector ($V$) components are expressed as $x$, $y$ and $z$, while axial vector ($eV$) components are denoted by $R_{x}$, $R_{y}$ and $R_{z}$. For example, the first coupling in Table~\ref{tab:MnTe_N_and_R} appears in Table~\ref{tab:CouplingComps} in a more general form, applicable to \emph{any} $aeV[V^{2}]$ tensor, as
\begin{equation}\label{eq:MnTeSOCFromTable}
    N_z \left(2 x y R_x+\left(x^2-y^2\right)R_y \right).
\end{equation}
In constructing the couplings in Table~\ref{tab:MnTe_N_and_R}, then, $N_{i}$ components couple to $R^{S}_{ijk}$ where $i,j$ are given by the polar components, and $k$ is given by the axial components appearing in Eq.~\ref{eq:MnTeSOCFromTable}. For example, the first term in Eq.~\ref{eq:MnTeSOCFromTable} corresponds to the $N_{z}R_{xyx}^{S}$ term in the $B_{1g}$ coupling from Table~\ref{tab:MnTe_N_and_R}.
Tables~\ref{tab:guaranteedsoc} and~\ref{tab:properties} can guide the experimental diagnosis of altermagnetic phases once the rank $n$ of the minimal SO-free multipole is determined. Taking the representations $\Gamma_\xi$ guaranteed by the rank-$n$ multipole from Table~\ref{tab:guaranteedsoc}, one finds the corresponding measurable quantities in Table~\ref{tab:properties}.

We have reduced the analysis of altermagnets from hundreds of Wyckoff positions to 54 SO-free Landau theories, and to four cases of measurable responses we may expect, as shown in Table~\ref{tab:guaranteedsoc}. For example, a minimal multipole with $n\!=\!2$ in the SO-free theory guarantees coupling to \emph{any} magnetic axial vector, such as the anomalous Hall conductivity (AHC), magnetization, pyromagnetic tensor, etc., as listed in Table~\ref{tab:properties}. As we have just seen, coupling with an $n\!=\!4$ SO-free multipole, as in MnTe, guarantees coupling to any $aeV[V^{2}]$ quantity, including magnetoresistance, piezomagnetism, the magneto-optic Kerr or Nernst effects, among others listed in Table~\ref{tab:properties}. By combining the physical properties in Table~\ref{tab:properties} with the explicit tensor components in Table~\ref{tab:CouplingComps}, we have laid foundations for the prediction of an abundance of experimentally accessible features of collinear altermagnets.

For example, one can consider the generation of spin-currents by
the application of electric fields -- an important potential
application of altermagnetic materials. Here the spin conductivity $\bm{\sigma}^s_{\mu\nu}$ has three indices: two spatial indices $\mu$,$\nu$, and an index in spin space (here written as a vector). If we consider the symmetric components of this tensor, they are axial and odd under time-reversal and thus transform as $ae[V^2]$ spatially and as an axial vector
in spin space. Since $\mathbf{N}$ is time odd, axial, transforms as a vector in spin space and as $\Gamma_{\mathbf{N}}$ spatially, a linear coupling of $\bm{\sigma}^s_{\mu\nu}$ and $\mathbf{N}$ thus requires that 
$$
\Gamma_{\mathbf{N}} \subseteq [V^2] 
$$
This linear coupling then requires that the lowest-rank multipole is
$n=2$ (a quadrupole). 

This result can be made considerably stronger: altermagnets whose lowest rank multipoles have $n>2$ have vanishing spin conductivity. To see this, note that the only axial, time-odd quantities that transform as a vector spin space that can be created using $\mathbf{N}$ are of the form $f(|\mathbf{N}|^2)\mathbf{N}$ where $f$ is an arbitrary function. As $|\mathbf{N}|^2$ transforms trivially, this again transforms as $\Gamma_{\mathbf{N}}$ and so can only appear in $\bm{\sigma}^s_{\mu\nu}$ if
$\Gamma_{\mathbf{N}} \subseteq [V^2]$. Thus altermagnets with $n=4$ or $n=6$ (i.e. with $g$-wave or $i$-wave spin splitting) do not have spin currents generated by electric fields in the SO-free limit.

We comment on the comparison between our framework and standard techniques at finite SOC using the black \& white groups. One may just as well analyze all possible couplings at finite SOC; the result would be a potentially longer list of quantities than those listed in Table~\ref{tab:properties}. However, this approach would not distinguish between properties arising from altermagnetism, and those simply arising from finite SOC. Herein lies the primary benefit of our framework: whereas the properties listed in Table~\ref{tab:properties} highlight \emph{only} those quantities originating from ideal altermagnetism in the SO-free limit, our framework offers a bridge between the SO-free and SO-coupled theories. In addition, this perspective allows one to predict altermagnetic couplings solely on the basis of the lowest order pseudoprimary multipole.

There is a growing body of literature examining the connection between the Néel vector and various physical properties in altermagnets. For example, the anomalous Hall conductivity or magnetization in the presence of SOC has been explored in works including Ref.~\cite{Roig2024_QSymmFMinAM,Xiao2025_AHNeelAM,MnTe_Reichlova2017_MagAniso,Han2024_ElSwitchingNwAHE,Smejkal2020,Attias2024_AMIntrinsicAHE, McClarty2024}, among others. Here, we expand upon the previous literature by making a series of predictions applicable to \emph{all} symmetry classes of tensors that may be seen as arising from altermagnetism, and by providing a coherent framework for understanding how these couplings arise.

\section{\label{sec:examples}Examples of Materials}

In the previous sections, we derived a Landau theory of $\bm{Q}=\bf{0}$ collinear altermagnets, capitalizing on the philosophy that the SO-free behavior dictates features in a real material with weak SOC. These Landau theories serve as a guide for experiments to identify altermagnetic phases.

We have already shown how to put Landau theory to use through the example of MnTe. In general, based on the WP of the magnetic ions, we can identify the irrep $\Gamma_{\mathbf{N}}$ under which $\mathbf{N}$ transforms in the SO-free theory. This data can be found in Table~\ref{tab:wpGN}. This irrep dictates the order $n$ of the SO-free $(n+1)$-multipolar secondary order parameter as defined in Eq.~\ref{eq:multipoleDef}. Knowledge of this order $n$ is sufficient to identify physical quantities $\xi$ that may couple linearly to $\mathbf{N}$ when SOC is included. The viable representation $\Gamma_{\xi}$ of $\xi$ is listed in Table~\ref{tab:guaranteedsoc} for each $n$, and is linked to physical properties in Table~\ref{tab:properties}. Finally, predictions of specific non-zero tensor components, as well as the explicit form of the coupling are found in the final column of Table~\ref{tab:CouplingComps}.

In the following, we illustrate how to apply these results to further examples of candidate altermagnetic materials and, in the process, make measurable predictions. We focus on materials appearing in the altermagnetic literature, such as those appearing in Ref.~\cite{Smejkal2022b}, many of which also appear in Refs.~\cite{HayamiMultipole2} and~\cite{Guo23_AMCandidateMaterials}. Additionally, we emphasize that such results, as well as our previous conclusions for MnTe, rely only on the magnetic symmetries of the material and are therefore independent of the microscopic details of any particular material. In the case of MnTe, for example, our results apply to apply to any other $6/mmm$ material with $\Gamma_{\mathbf{N}} = B_{2g}$.

\subsection{Point group $\mathbf{2/m}$}

\begin{figure}[b!]
    \centering
    \includegraphics[width=.48\textwidth]{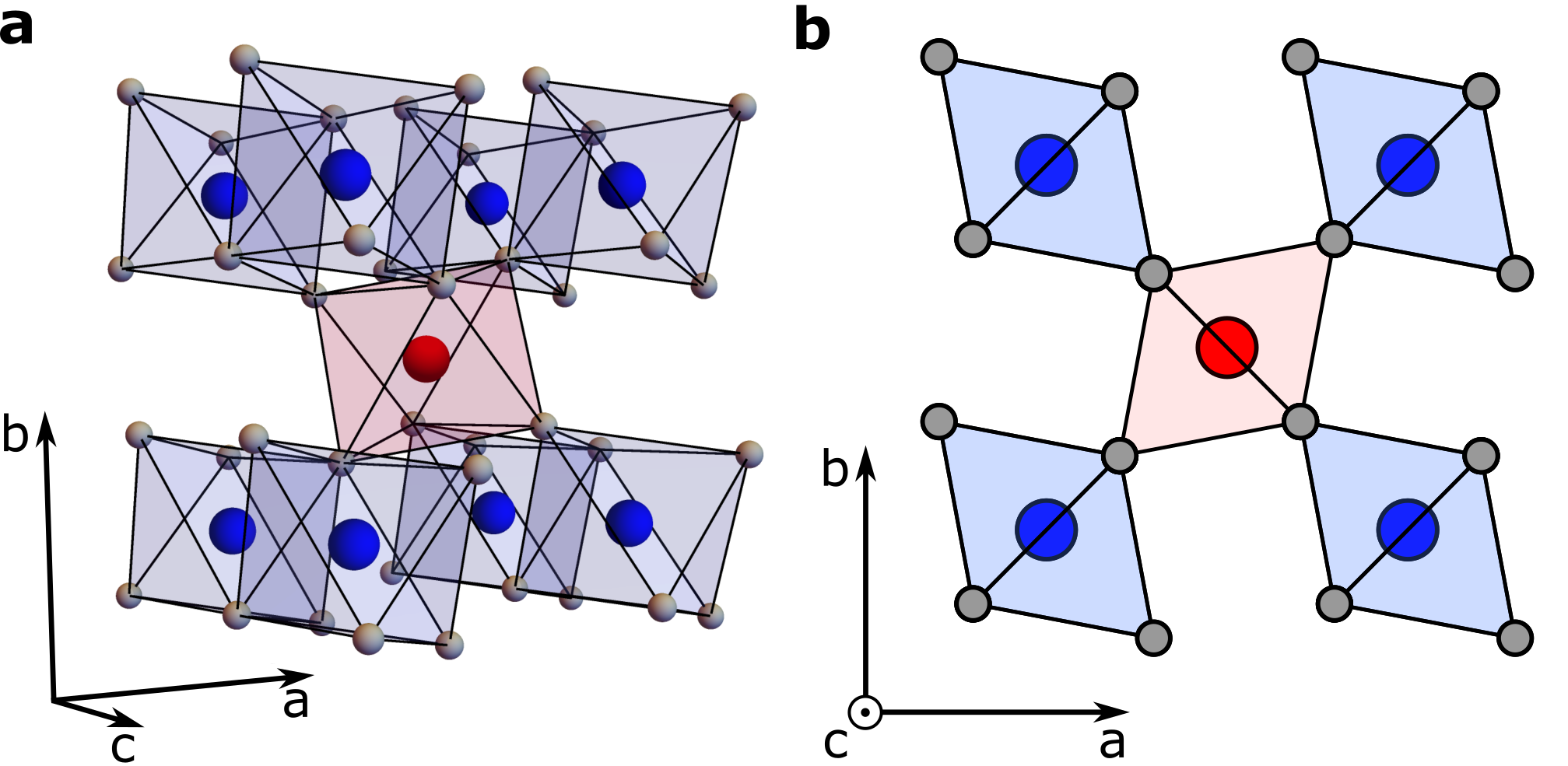}
    \caption{The crystal structure of CrF$_{2}$ with space group symmetry  $P2_{1}/c$. We use the setting $P1\,2_{1}/n\,1,$ related to the original setting by $\{\mathbf{a},\mathbf{b},\mathbf{c}\}\rightarrow\{-\mathbf{a}-\mathbf{c},\mathbf{b},\mathbf{a}\}.$  Magnetic Cr ions (red and blue denote magnetic sublattices) reside on the $2b$ Wyckoff positions, at $\{0,0,0\}$ and $\{\frac 1 2, \frac 1 2, \frac 1 2\}$ within the unit cell. The F ions (gray) occupy the Wyckoff positions $4e$, at $\pm \{x,y,z\}$, and $\pm \{x+\frac{1}{2},\frac{1}{2}-y,z+\frac{1}{2}\}$, forming a distorted  octahedral environment tilting out of the $bc$ plane.}
    \label{fig:CuF2}
\end{figure}

Among the transition metal fluorides $X$F$_{2}$ ($X=$ Cr, Cu, Mn, F, Co, Ni, V) most are rutiles but two cases (those with $X=$ Cr and Cu) have a distorted rutile structure~\cite{CrF2_Chatterji2011_JTE} such that the crystal has monoclinic space group  $P2_{1}/c$ (No. 14), with point group $2/m$. The magnetic order is different in these two materials. We focus on insulating CrF$_{2}$ in this section, as is it a $\mathbf{Q}=\bm{0}$ altermagnetic candidate. CrF$_{2}$ has a Néel temperature of roughly $T_{N} = 53K$~\cite{CrF2_CableWilkinson1960_Neutron}. The crystal and magnetic sublattice structure is depicted in Fig.~\ref{fig:CuF2}.  For more details on the material properties see Refs.~\cite{CrF2_CableWilkinson1960_Neutron,CrF2_Chatterji2011_JTE}.

Because the two-fold rotation $\{2_{010}|\frac{1}{2}\,\frac{1}{2}\,\frac{1}{2}\}$ and mirror elements $\{m_{010}|\frac{1}{2}\,\frac{1}{2}\,\frac{1}{2}\}$ swap sublattices, these elements are represented by $-1$ in the irrep $\Gamma_{\mathbf{N}}$ describing the N\'eel vector's spin-orbit free sublattice properties. Further, inversion leaves the sublattice structure invariant, so this order is inversion-even. This corresponds to the $B_{g}$ irrep of $2/m$, so $\Gamma_{\mathbf{N}} = B_{g}$ for CrF$_{2}$. This is consistent with the entry in Table \ref{tab:wpGN} corresponding to the $2b$ WP of space group 14.

Our next step is to determine the SO-free multipole. The minimal multipole coupling to $\mathbf{N}$ in absence of SOC has $n\!=\!2$ according to Table~\ref{tab:socfreeMultipoles}, meaning that the multipole's generic form is $\int d^{3}r\, [r_{\mu}r_{\nu}]\,\mathbf{m}(\mathbf{r}).$ To determine the polynomial $[r_{\mu}r_{\nu}]$, one must find the order two polynomial in $x,\,y,$ and $z$ that transforms as the $B_{g}$ irrep of $2/m.$ Either by explicit checking or by using the procedure outlined in Appendix \ref{app:SAB}, one finds that $xy$ and $yz$ transform as $B_{g}$ (matching the entry in Table~\ref{tab:CouplingComps}). These SO-free multipoles are consistent with a $d-$wave spin-splitting pattern in the band structure, matching predictions in Refs~\cite{Smejkal2022a,Smejkal2022b}.

We are now prepared to find experimentally measurable responses of CrF$_{2}$ due to altermagnetism when SOC is included. The $n\!=\!2$ SO-free multipole tells us that in the presence of SOC, the Néel vector may couple to any $aeV$ tensor (according to Table~\ref{tab:guaranteedsoc}). Many responses, listed in Table~\ref{tab:properties}, abide by this symmetry. We use the thermal Hall conductivity (THC), $\bm{\kappa}^{A}$ as a representative example.

Non-zero components $\kappa^{A}_{i} = \frac{1}{2}\varepsilon_{ijk}\kappa^{A}_{jk}$ of the THC couple to components $N_{i}$ of the N\'{e}el vector. Our task is to determine which $\kappa^{A}_{i}$ are non-zero, and to which $N_{i}$ components they couple in the Landau theory. Table~\ref{tab:CuF2Coupling} lists the irreps under which these components transform.

\begin{table}[t!]
\caption{\label{tab:CuF2Coupling}%
Irreps of $2/m$ describing the transformation properties of the Néel vector components $N_{i}$ and THC components $\kappa^{A}_{i}$ in CuF$_{2}$. Recall that $\mathbf{N}$ transforms under $ aeV\otimes\Gamma_{\mathbf{N}}$ with $\Gamma_{\mathbf{N}} = B_{g}$, while $\bm{\kappa}^{A}$ transforms under $aeV.$}
\centering
\begin{tabularx}{\columnwidth}{>{\centering\arraybackslash}X  >{\centering\arraybackslash}X  >{\centering\arraybackslash}X  >{\centering\arraybackslash}X}
\hline
\hline
Component $i$ & $x$ &
$y$ & $z$\\
\hline
$N_{i}$ irrep & $A_{g}$ &  $B_{g}$ & $A_{g}$\\
$\sigma^{A}_{i}$ irrep & $B_{g}$ & $A_{g}$ & $B_{g}$\\
\hline
\hline
\end{tabularx}
\end{table}

To this end, we look for components $\kappa^{A}_{j}$ and $N_i$ that transform in the same way. Alternatively, we can ask for the product of thermal Hall and N\'{e}el vector components that transform trivially (as $A_{g}$) under the point group $2/m$. With the knowledge that $B_{g}$ squares to the trivial irrep, we find the following allowed couplings:
\begin{equation}
    \kappa^{A}_{y} \sim N_{x},\,\, \kappa^{A}_{y} \sim N_{z}, \,\, \kappa^{A}_{x}\sim N_{y},\,\,\kappa^{A}_{z} \sim N_{y}.
\end{equation}

The neutron diffraction study of Ref.~\cite{CrF2_CableWilkinson1960_Neutron} reports a nearly collinear antiferromagnetic structure with zero propagation vector. A symmetry analysis reveals that a single primary order parameter would have either moments in the $x, z$ plane or in the $y$ plane.
In the study of Ref.~\cite{CrF2_CableWilkinson1960_Neutron}, it is noted that the best fit for their data indicates order in the $\bf{ac}-$plane, at an angle of $32^{\circ}$ from the $\bf{c}$-axis; consistent with ordering in the $A_{g}$ irrep, and in this case, a THC signal and weak magnetization would be expected along the $\pm y$ direction. 

We note that the same neutron study additionally reports a possible magnetic structure with moments aligned and anti-aligned along one of the long Cr--F bonds~\cite{CrF2_Chatterji2011_JTE}. It may therefore be interesting to revisit the problem of the precise magnetic order in this material. In any case, one expects a thermal Hall effect in this material either with components $\kappa^{A}_{x}$, $\kappa^{A}_{z}$ for ordering in the $B_g$ irrep or, as seems more likely, a $\kappa^{A}_{y}$ component coming from order in the $A_g$ irrep. In both cases, weak ferromagnetism is anticipated. 

We further note that the case of CuF$_2$ which has the same parent (paramagnetic) space group as CrF$_2$ has magnetic order with propagation vector $\mathbf{Q}=(1/2,0,0)$~\cite{CuF2Fischer1974} which requires a separate analysis that we leave for future study.

\subsection{Point group $\mathbf{mmm}$}

CaCrO$_{3}$, LaMnO$_{3}$, and La$_{2}$CuO$_{4}$ were proposed as candidate altermagnetic materials with point group symmetry $mmm$ in Ref.~\cite{Smejkal2022b}, and magneto-optical effect in La{\it M}O$_3$ ({\it M}= Cr, Mn, and Fe) has been reported as early as Ref.~\cite{Solovyev1997}. CaCrO$_{3}$ and LaMnO$_{3}$ have space group symmetry $Pnma$ (No. 62), while La$_{2}$CuO$_{4}$ belongs to the space group symmetry $Cmce$ (or $Bmab$). For concreteness, we consider La$_{2}$CuO$_{4}$ though our predictions based on symmetry are equally applicable to LaMnO$_{3}$ and CaCrO$_{3}$.

\begin{figure}[t!]
    \centering
    \includegraphics[width=.48\textwidth]{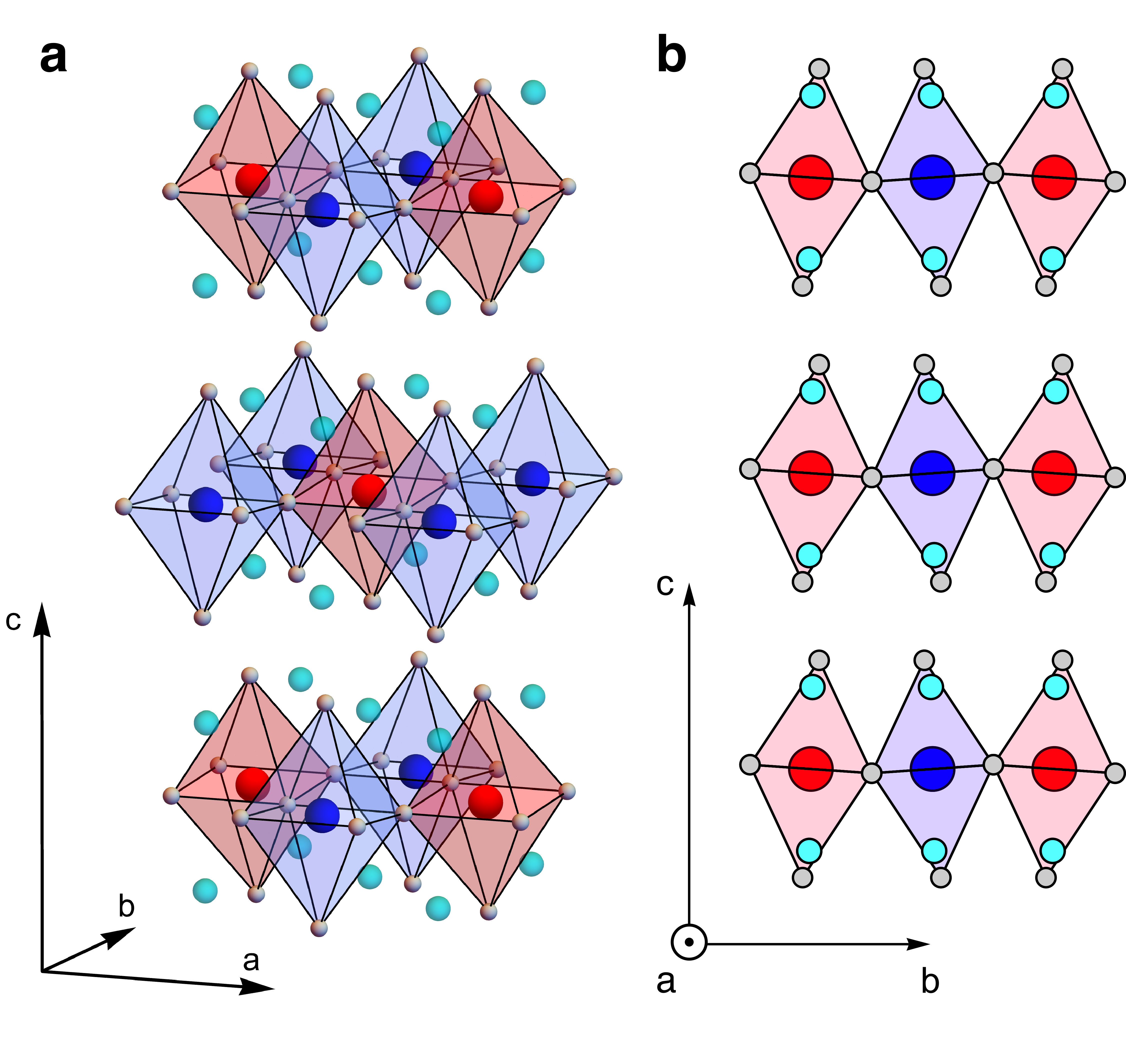}
    \caption{La$_{2}$CuO$_{4}$ structure and magnetic sublattices. The space group is $\mathbf{G} = Bmab$ (No. 64). This setting is related to $Cmce$ by  $\mathbf{c} \leftrightarrow -\mathbf{b},$ and has a pure half-translation $\{\frac{1}{2},0,\frac{1}{2}\}$. Magnetic Cu atoms (red and blue) occupy the $4a$ WP $\{0,0,0\}$ and $\{\frac{1}{2},\frac{1}{2}, 0\}.$ La atoms (cyan) reside on the $8f$ WP $\pm \{x,y,0\}$, $\pm \{x + \frac{1}{2}, -y+\frac{1}{2},0\}$. O atoms (grey) occupy two WP, $8f$ and $8e$, at $\{x,\frac{1}{4},\frac{1}{4}\}$, $\{x+\frac{1}{2}, \frac{1}{4},\frac{1}{4}\}$, $\{-x,\frac{3}{4},\frac{3}{4}\}$, $\{-x+\frac{1}{2}, \frac{3}{4}, \frac{3}{4}\}$).}
\label{fig:La2CuO4AFM}
\end{figure}

\begin{table}[b!]
\caption{\label{tab:La2CuO4_SOC}%
Irreps of $mmm$ describing the transformation properties of the Néel vector components $N_{i}$ and magnetization components $M_{i}$ in La$_{2}$CuO$_{4}$. Recall that $\mathbf{N}$ transforms under $ aeV\otimes\Gamma_{\mathbf{N}}$, while $\bm{M}$ transforms under $aeV.$}
\centering
\begin{tabularx}{\columnwidth}{>{\centering\arraybackslash}X  >{\centering\arraybackslash}X  >{\centering\arraybackslash}X  >{\centering\arraybackslash}X}
\hline
\hline
Component $i$ & $x$ &
$y$ & $z$\\
\hline
$N_{i}$ irrep & $A_{g}$ & $B_{1g}$ & $B_{2g}$ \\
$M_{i}$ irrep & $B_{3g}$ & $B_{2g}$ & $B_{1g}$ \\
\hline
\hline
\end{tabularx}
\end{table}

The compensated magnetic order in insulating La$_{2}$CuO$_{4}$ has a N\'eel temperature of $T_{\mathbf{N}} = 325K$, and is discussed in Ref.~\cite{La2CuO4_Yamada1987,La2CuO4_Vaknin1987,La2CuO4_Jorgensen1988, La2CuO4_KastnerReview1998,La2CuO4_Reehuis2006,La2CuO4_Lane2018,La2CuO4_Thio1988,La2CuO4_Thio1994}. The crystal and sublattice structure is
as shown in Fig.~\ref{fig:La2CuO4AFM}. We have shown the crystal structure in the $Bmab$ setting, whereas the irreps and WP in Table~\ref{tab:wpGN} are derived in the standard setting ($Cmce$ in this case). Changes between settings can be achieved using the tools available in the Bilbao crystallographic server~\cite{BilbaoGenPosWP}.

Group elements $\{2_{100}|000\},$ $\{\overline{1}|000\}$ and $\{m_{100}|000\}$ preserve the sublattice structure, while $\{2_{001}|\frac{1}{2}\,\frac{1}{2}\,0\}$, $\{2_{010}|\frac{1}{2}\,\frac{1}{2}\,0\}$, $\{m_{001}|\frac{1}{2}\,\frac{1}{2}\,0\}$ and $\{m_{010}|\frac{1}{2}\,\frac{1}{2}\,0\}$ swap the sublattices. To find the irrep $\Gamma_{\mathbf{N}}$ describing the sublattice-swapping properties of $\mathbf{N}$, we assign $-1$ to each of the sublattice-swapping elements. We find that $\Gamma_{\mathbf{N}} = B_{3g}$ in $mmm$, consistent with our findings in Table~\ref{tab:wpGN} for space group 64 and the copper WP ($4a$).

Having found $\Gamma_{\mathbf{N}},$ our next step is to determine the order $n$ of the SO-free multipole. From Table~\ref{tab:socfreeMultipoles}, we find that the minimal multipole has $n=2$, and from Table~\ref{tab:CouplingComps} we can see that this multipole is of the form $\int d^{3}r\, yz\, \mathbf{m}(\mathbf{r}).$ This is consistent with a $d$-wave spin-splitting pattern, aligning with the {\it ab initio} prediction in Ref.~\cite{Smejkal2022b}.

We are now prepared to determine the spin-orbit coupled theory. For $n\!=\!2$ multipoles, any $aeV$ tensor has components that couple linearly to $\mathbf{N}$ (see Table~\ref{tab:guaranteedsoc}). Physical properties of this type include weak ferromagnetism $\mathbf{M}$, among others listed in Table~\ref{tab:properties}. We use the magnetization as a representative example. The components of $\mathbf{N}$ and of $\mathbf{M}$ transform according to the irreps of $mmm$ listed in Table~\ref{tab:La2CuO4_SOC}.

By virtue of the 1D irreps squaring to the trivial irrep, $M_{y}$ can couple to $N_{z}$, and $M_{z}$ can couple to $N_{y}$. As a consequence, the $y$ and $z$ components of the magnetization, and any relevant $aeV$ tensor, may generically be nonzero in La$_{2}$CuO$_{4}$.

Experimentally, in Refs.~\cite{La2CuO4_Reehuis2006} and~\cite{La2CuO4_Vaknin1987} it was found that the moments align along the crystallographic $b$-axis, corresponding with our Cartesian $y$-axis. For this reason, we also expect a weak ferromagnetic component $M_{z}$ along the $c$-axis, consistent with the predictions and measurements of Refs.~\cite{La2CuO4_Thio1988,La2CuO4_Thio1994}. Theoretical and experimental aspects of La$_{2}$CuO$_{4}$ are reviewed in~\cite{La2CuO4_KastnerReview1998}.

\subsection{Point group $\mathbf{4/mmm}$}

\begin{figure}[b!]
    \centering
    \includegraphics[width=.48\textwidth]{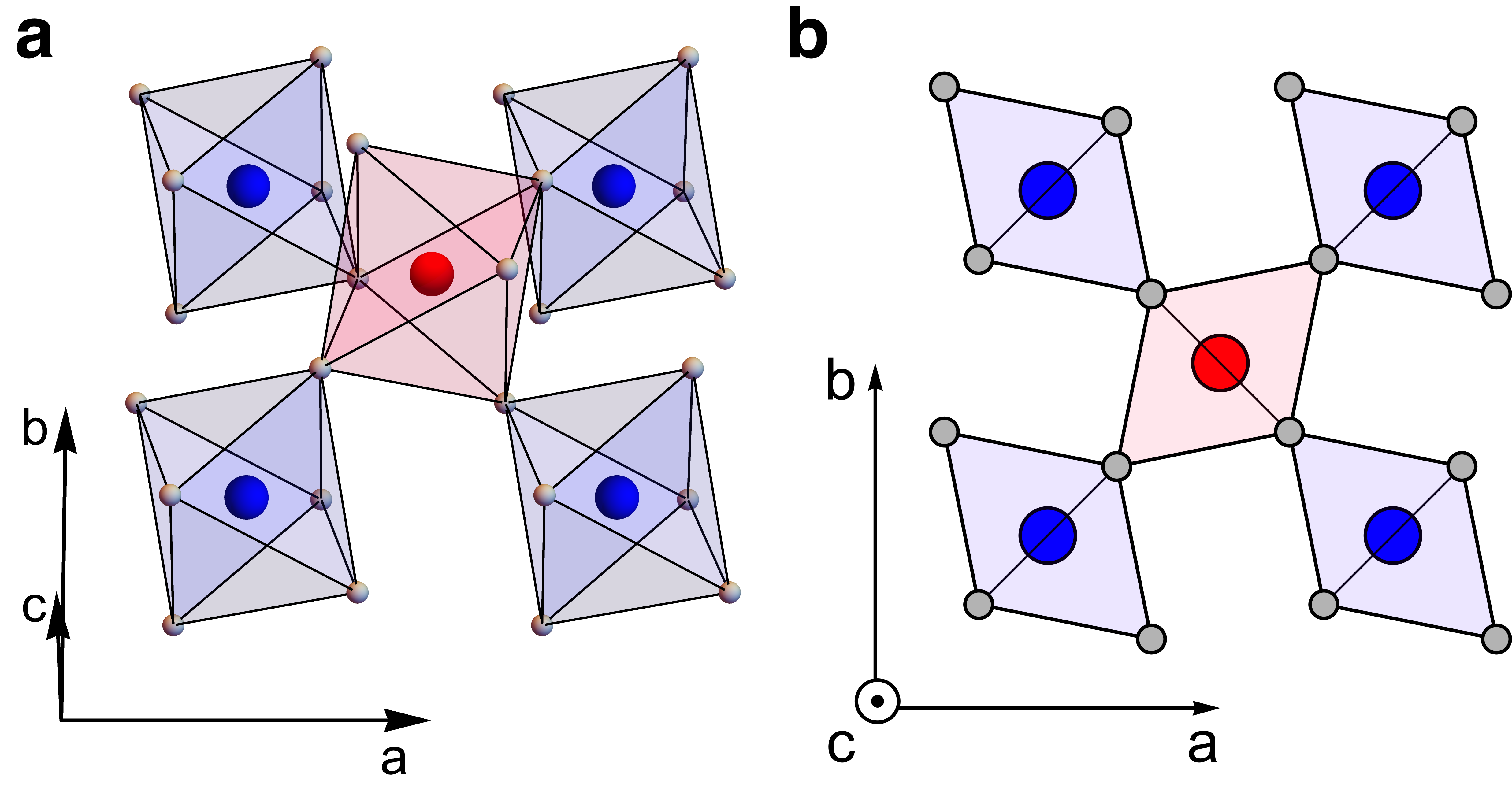}
    \caption{The crystal and magnetic sublattice structure of MnF$_{2}$, with space group $P4_{2}/mnm$ (No. 136). Mn atoms (red and blue denote magnetic sublattices) reside on the $2a$ WP $\{0,0,0\}$ and $\{\frac{1}{2}, \frac{1}{2}, \frac{1}{2}\}$, while F atoms (grey) occupy the $4f$ WP with positions $\pm\{x,x,0\}$ and $\pm\{-x+\frac{1}{2},x+\frac{1}{2},\frac{1}{2}\}$.}
    \label{fig:MnF2}
\end{figure}

Three candidate altermagnetic materials with point group symmetry $4/mmm$ are suggested in Ref.~\cite{Smejkal2022b}: MnF$_{2}$, MnO$_{2}$, RuO$_{2}$.  We concentrate on the insulator MnF$_{2}$, whose crystal structure~\cite{MnF2_Baur1958_Crystal,MnF2_Baur1971_ComparingRutiles} is shown in Fig.~\ref{fig:MnF2}, to illustrate this class of examples.

The onset of antiferromagnetic ordering in MnF$_{2}$ occurs at roughly $T_{\mathbf{N}}=67\ K$~\cite{MnF2_Stout1942_TN}. We begin our analysis by determining the sublattice preserving and sublattice swapping elements of the space group $P4_{2}/mnm$ (No. 136)~\cite{MnF2_Baur1958_Crystal, MnF2_Baur1971_ComparingRutiles, MnF2_Yuan2020}. The non-symmorphic elements $\{4_{001}|\frac{1}{2}\frac{1}{2}\frac{1}{2}\},$ $\{2_{100}|\frac{1}{2}\frac{1}{2}\frac{1}{2}\}$, $\{2_{010}|\frac{1}{2}\frac{1}{2}\frac{1}{2}\}$ swap up- and down-spin sublattices, while the symmorphic $\{I|000\}$, $\{2_{110}|000\}$ and $\{2_{1\overline{1}0}|000\}$ preserve the sublattice structure. By ascribing the non-symmorphic elements with the representation $-1$, we can identify the irrep $\Gamma_{\mathbf{N}}$ for MnF$_{2}$ as $B_{2g}$. This matches the finding for magnetic ions at WP $2a$ in Table~\ref{tab:wpGN}.

Next, we develop the SO-free Landau theory by identifying the lowest order $n$ multipole coupling to the Néel vector. From Table~\ref{tab:socfreeMultipoles} we see that $n\!=\!2$, and using Table~\ref{tab:CouplingComps} we find that the multipole is of the form $\int d^{3}\,xy\,\mathbf{m}(\mathbf{r})$. The $xy$ integrand indicates a $d$-wave spin-splitting pattern, consistent with the predictions in Refs.~\cite{Smejkal2022b} and~\cite{MnF2_Yuan2020}. Ab initio studies on MnF$_{2}$ may be found in Refs.~\cite{MnF2_Moreira2000_AbInitio} and~\cite{MnF2_Moreno2016}.

\begin{table}[b!]
\caption{\label{tab:MnF2_SOC}%
Irreps of $4/mmm$ describing the transformation properties of the Néel vector components $N_{i}$ and magnetization components $M_{i}$ in MnF$_{2}$. Recall that $\mathbf{N}$ transforms under $ aeV\otimes\Gamma_{\mathbf{N}}$, while $\bm{M}$ transforms under $aeV.$}
\centering
\begin{tabularx}{\columnwidth}{>{\centering\arraybackslash}X  >{\centering\arraybackslash}X  >{\centering\arraybackslash}X  >{\centering\arraybackslash}X}
\hline
\hline
Component $i$ & $x$ &
$y$ & $z$\\
\hline
$N_{i}$ irrep & $E_{g}$ & $E_{g}$ & $B_{1g}$ \\
$M_{i}$ irrep & $E_{g}$ & $E_{g}$ & $A_{2g}$ \\
\hline
\hline
\end{tabularx}
\end{table}

As in our previous examples, the presence of an $n\!=\!2$ multipole in the SO-free theory dictates that when SOC is included, components of $\mathbf{N}$ may couple to an $aeV$ tensor (see Table~\ref{tab:guaranteedsoc}). We will use the magnetization $\mathbf{M}$ as an example, though other quantities may be found in Table~\ref{tab:properties}. The irreps under which components of $\mathbf{N}$ and $\mathbf{M}$ transform are provided in Table~\ref{tab:MnF2_SOC}. No linear coupling is allowed with $N_{z}$ and $M_{z}$, while we may use the procedure outlined in Appendix \ref{app:SAB} to determine that the $x$- and $y$-components may couple as 
\vspace{-0.1cm}
\begin{equation}
N_{x}M_{x}-N_{y}M_{y},
\end{equation}
\vspace{-0.05cm}where $N_{i}$ and $M_{i}$ components correspond to the choice of crystallographic axes depicted in Fig.~\ref{fig:MnF2}. If the crystallographic axes are chosen to point in the directions $\mathbf{a'} = \mathbf{a} + \mathbf{b}$, $\mathbf{b'} = \mathbf{a}-\mathbf{b}$ and $\mathbf{c'}=\mathbf{c}$ (which corresponds to the setting in Bilbao~\cite{BilbaoIrreps}), then the coupling is of the form 
\vspace{-0.2cm}
\begin{equation}
    N'_{x}M'_{y}+N'_{y}M'_{x},
\end{equation}
which matches the entry for the $B_{2g}$ irrep of $4/mmm$ in Table~\ref{tab:CouplingComps}, as well as the reported coupling in~\cite{McClarty2024}. As a consequence of this coupling, a weak ferromagnetic moment may develop in the crystallographic $ab$-plane.

Recalling that the thermal Hall conductivity $\bm{\kappa}^{A}$ transforms identically to the magnetization $\mathbf{M}$, we see that this result also implies generically non-zero allowed values of $\kappa^{A}_{x}$ and $\kappa^{A}_{y},$ consistent with the theoretical results of Ref.~\cite{MnF2_Hoyer2024_THE} examining thermal transport at zero field via magnons in insulating rutile systems. Indeed, they find that when $\mathbf{N}$ is aligned with the crystallographic $\mathbf{c}$-axis, $\kappa_{x}^{A}$ and $\kappa_{y}^{A}$ are zero while any canting gives rise to a non-zero value of these thermal conductivities.

It has been experimentally determined in Ref.~\cite{MnF2_Erickson1953_NeutronScattering} that antiferromagnetic order in MnF$_{2}$ is aligned along the crystallographic $c$-axis, corresponding to a dominant $N_{z}^{2}$ term in the free energy. This is likely due to the magnetostatic dipolar coupling. This coupling, while significantly smaller than the exchange scale, pins the moments along $c$ and gaps out the magnon spectrum~\cite{Nikotin1969}.

We note additionally that altermagnetic band structure in rutiles has been studied using spin-groups in Ref.~\cite{schiff2023} where distinctive degeneracies of the band structure at zero SOC are discussed. The spin splitting and momentum-space spin texture have been studied using DFT in Ref.~\cite{MnF2_Yuan2020}.

\subsection{Point group $\mathbf{\overline{3}m}$}

In the point group $\overline{3}m,$ it has been suggested that the insulating collinear antiferromagnetic state of hematite Fe$_{2}$O$_{3}$ below the Morin temperature $T_{M}=265K$~\cite{Fe2O3_Morin1950_MorinTransition} is altermagnetic~\cite{Smejkal2022b}. Magnetism in hematite has been a longstanding and ongoing topic of research~\cite{Fe2O3_Dzyaloshinsky1958_ThermoOfWeakM,Fe2O3_Moriya_AnisoSuperex_WeakFM,Fe2O3_Hill2008_Neutrons,Fe2O3_Dannegger2023_MagAbInitio,Fe2O3_Lebrun2020_SpinTransport,Fe2O3_Santoshi2024_StrucMagExp}. Proposed altermagnetic features of hematite have been investigated in Ref.~\cite{Fe2O3_Galindez2024_AMHoppingRegime}, and recently, chiral splitting of magnons in hematite has been investigated \cite{Fe2O3_Hoyer2025_magnonSplitting}. Here, we develop the SO-free and SOC Landau theories for hematite and compare them with known material properties.

\begin{figure}[t!]
    \centering
    \includegraphics[width=.48\textwidth]{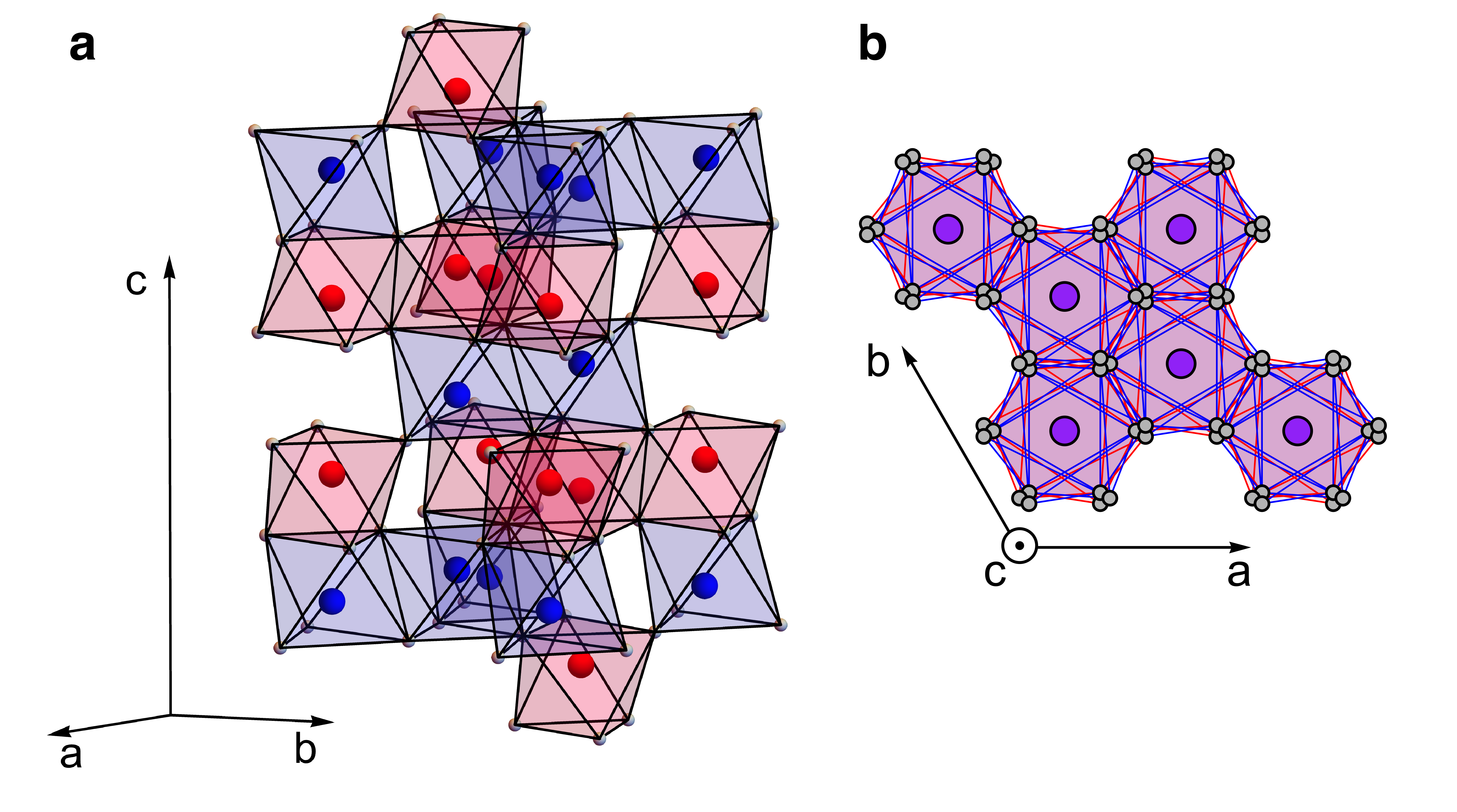}
    \caption{The crystal and magnetic sublattice structure of Fe$_2$O$_3$, with space group $R\overline{3}c$ (No. 167) in the hexagonal setting. Fe atoms (red and blue denote magnetic sublattices) reside on the $12c$ WP $\{0, 0, z\}, \{0, 0, \frac{1}{2} - z\}, \{0, 0, -z\}$, and  $\{0, 0, \frac{1}{2} + z\}$, while O atoms (grey) occupy the $18e$ WP with positions $\{x, 0, \frac{1}{4}\}, \{0, x, \frac{1}{4}\}, \{-x, -x, \frac{1}{4}\}, \{-x, 0, \frac{3}{4}\}, \{0, -x, \frac{3}{4}\}$, and $\{x, x, \frac{3}{4}\}$. Note that the hexagonal setting has pure lattice translations $\{\frac{2}{3},\frac{1}{3},\frac{1}{3}\}$ and $\{\frac{1}{3},\frac{2}{3},\frac{2}{3}\}$.}
    \label{fig:Fe2O3}
\end{figure}

To begin, we determine the irrep $\Gamma_{\mathbf{N}}$ under which the Néel vector transforms in the SO-free limit. The crystal and magnetic sublattice structure for hematite is shown in Fig.~\ref{fig:Fe2O3}. This structure has the symmetry of space group $R\overline{3}c$ (No. 167). The threefold element $\{3_{001}|000\}$ and inversion $\{I|000\}$ preserve the sublattice structure, while all three non-symmorphic two-fold axes $\{2_{100}|00\frac{1}{2}\}$, $\{2_{010}|00\frac{1}{2}\}$ and $\{2_{110}|00\frac{1}{2}\}$ (and corresponding mirrors) swap the sublattices. Assigning $-1$ to the sublattice swapping elements, we may deduce that the Néel vector transforms under $\Gamma_{\mathbf{N}} = A_{2g},$ matching the entry for magnetic ions at the $12c$ WP of space group 167 in Table~\ref{tab:wpGN}.

We now seek the secondary multipolar order parameter in the SO-free limit. From Table~\ref{tab:socfreeMultipoles} we see that the minimal multipole in $\overline{3}m$ with $\Gamma_{\mathbf{N}} = A_{2g}$ has order $n\!=\!4,$ and in Table~\ref{tab:CouplingComps} we see that this multipole is of the form $\int d^{3}r\, y(y^{2}-3x^{2})z\,\mathbf{m}(\mathbf{r}).$ An SO-free multipole with $n\!=\!4$ corresponds to a $g-$wave spin-splitting pattern, matching the pattern predicted in Refs.~\cite{Smejkal2022b} and~\cite{Smejkal2022a}.

When SOC is included, we would expect an altermagnetic N\'{e}el vector in hematite to couple with tensors transforming as $aeV[V^{2}]$, on the basis of the order $n\!=\!4$ of the SO-free multipole and Table~\ref{tab:guaranteedsoc}. Many physical properties, listed in Table~\ref{tab:properties} obey this transformation law; here, we will use the piezomagnetic tensor $\Lambda_{ijk}$ as an illustrative example, where indices $j$ and $k$ are polar and symmetrized, corresponding to components of the strain tensor, while the $i$ index denotes the magnetic axial component. The transformation properties of the N\'{e}el components $N_{i}$ and of $\Lambda_{ijk}$ are shown in Table~\ref{tab:Fe2O3SOC_Coupling}. Both couplings in the $A_{1g}$ irrep from Table~\ref{tab:Fe2O3SOC_Coupling} are allowed. For each of the six $E_{g}$ irreps, one specific coupling between $(N_{x},N_{y})$ and the $\Lambda_{ijk}$ is allowed. We have expressed the twelve basis linear combinations such that their dot product with the in-plane Néel components gives rise to the allowed coupling.

\begin{table}[t!]
    \caption{\label{tab:Fe2O3SOC_Coupling} Transformation properties of the piezomagnetic tensor $\Lambda_{ijk}$ and Néel vector $N_{i}$ components in $\overline{3}m$, for Fe$_{2}$O$_{3}$.}
    \centering
    \begin{tabularx}{\columnwidth}{>{\hsize=0.3\hsize\raggedright\arraybackslash}X >{\hsize=0.6\hsize\centering\arraybackslash}X >{\hsize=1\hsize\centering\arraybackslash}X}
    \hline
    \hline
      Irrep. & Néel component & Piezomagnetic tensor component\\
      \hline
      \multirow{2}{*}{$A_{1g}$ }  & \multirow{2}{*}{$N_{z}$} & $2\Lambda_{xxy}+\Lambda_{yxx}-\Lambda_{yyy}$ \\
       & & $\Lambda_{yxz}-\Lambda_{xyz}$\\
      \hline
      \multirow{6}{*}{$E_{g}$} & \multirow{6}{*}{$\begin{pmatrix}N_{x}\\N_{y}\end{pmatrix}$}& $\left(\begin{smallmatrix}
     -\Lambda_{yzz}\\
\Lambda_{xzz}
\end{smallmatrix}\right)$   \\
& & $\left(\begin{smallmatrix}
\Lambda_{zzy}\\
-\Lambda_{zzx}
\end{smallmatrix}\right)$  \\
& & $\left(\begin{smallmatrix}
      \Lambda_{xyz}+\Lambda_{yxz}\\
 \Lambda_{xxz}-\Lambda_{yyz}
\end{smallmatrix}\right)$  \\
     & & $\left(\begin{smallmatrix}
4\Lambda_{zxy}\\
\Lambda_{zxx}-\Lambda_{zyy}
\end{smallmatrix}\right)$   \\
       & & $\left(\begin{smallmatrix}
-2\Lambda_{xxy}+\Lambda_{yxx}\\ 2\Lambda_{yyx}-\Lambda_{xyy}
\end{smallmatrix}\right)$  \\
& & $\left(\begin{smallmatrix}
\Lambda_{xxy}+\Lambda_{yyy}\\
-\Lambda_{xxx}-\Lambda_{yyx}
\end{smallmatrix}\right)$  \\
      \hline
      \hline
    \end{tabularx}
\end{table}

These results may be derived using the method outlined in Appendix~\ref{app:SAB}, and are consistent with the listing in Table~\ref{tab:CouplingComps} for $\overline{3}m$ and irrep $\Gamma_{\mathbf{N}}=A_{2g}$. As an example of the correspondence with Table~\ref{tab:CouplingComps} we consider the last SOC coupling for $\overline{3}m,$
$$z^{2}(N_{y}R_{x}-N_{x}R_{y}) = N_{y}R_{x}z^{2}-N_{x}R_{y}z^{2},$$ which corresponds to a coupling of the form $-N_{x}\Lambda_{yzz}+N_{y}\Lambda_{xzz}.$ This is precisely the coupling we find from the last $E_{g}$ pair in Table~\ref{tab:Fe2O3SOC_Coupling}. 

Below the Morin transition, the magnetic order in hematite has been measured to be collinear and compensated, with the Néel vector pointing along the crystallographic $c-$axis~\cite{Fe2O3_Morin1950_MorinTransition,Fe2O3_Hill2008_Neutrons,Fe2O3_Galindez2024_AMHoppingRegime}. This implies that only the components of the piezomagnetic tensor appearing in the $A_{1g}$ irrep are non-zero, corresponding to the strain applied in the $xy$-plane.

The absence of magnetization below the Morin temperature~\cite{Fe2O3_Morin1950_MorinTransition,Fe2O3_Hill2008_Neutrons,Fe2O3_Galindez2024_AMHoppingRegime} may also be understood on the basis of our results. 
We begin by noticing that $aeV$ tensor coupling is not guaranteed by the SO-free Landau theory with $n\!=\!4$ multipolar order, as can be seen from Table~\ref{tab:properties}. Nevertheless, this does not reflect that coupling to $aeV$ quantities forbidden. The symmetry allowed form of the coupling between the Néel vector and the magnetization $\mathbf{M}$ is $N_{x}M_{y}-N_{y}M_{x}$. Since $N_{x}$ and $N_{y}$ are zero below the Morin temperature, no linear coupling to the magnetization exists. Thus, we can conclude that $M_{x}$ and $M_{y}$ vanish. The $z$-component of magnetization transforms as $A_{2g}$, and so it cannot couple to any $N_{i}$, implying that it also vanishes.

\section{\label{sec:discussion}Discussion}

Landau theory has rightly been central to condensed matter physics since its inception; it supplies a unifying framework for all symmetry-broken states of matter and, as we have seen, it can be adapted to provide insights on altermagnets as well. One distinctive feature of altermagnets is that they are most cleanly defined in the limit of zero spin-orbit coupling. Nevertheless, materials tend to have finite SOC and therefore one is interested in those properties of altermagnets that are inherited from the ideal limit. For these reasons, in this paper, we have taken the dual approach of analyzing Landau theories at both zero and finite SOC. 

We began by specifying a simple criterion for determining altermagnetism in the ideal limit, in terms of the transformation properties of the N\'eel vector.  This rule allows one to determine all magnetocrystalline symmetries compatible with altermagnetism, and to tabulate all altermagnets from their space group, Wyckoff position, and magnetic structure in the case where the magnetic order does not enlarge the magnetic unit cell (which covers almost all cases considered to date). 

Although the set of possible altermagnetic structures is large, the Landau theories depend only on the (1D) irrep of the crystal point group. This leads to a much more manageable set of $54$ possible Landau theories. For these theories, we have determined the leading multipole that couples to the N\'{e}el vector. This directly reveals the pattern of spin splittings in the band structures in the zero SOC limit. This work therefore supplies a classification of altermagnets based on symmetry alone and the resulting Landau theories are tied to various observable properties even in the ideal limit.

Turning to the realistic finite SOC limit, we have established a further criterion that ties the appearance of the minimal allowed multipole in the zero SOC to linear couplings between the primary antiferromagnetic order parameter and a given response function. In other words, we have made precise the notion that certain features of altermagnets at finite SOC are inherited from the ideal limit and tabulated these features across all possible $\bm{Q}=\bm{0}$ altermagnetic orders.

To illustrate all of these ideas we have shown how to identify altermagnetism given a magnetic structure in a crystal and then establish its basic properties both including and stemming from the spin splitting in momentum space. Spin splitting on its own is directly measureable using (spin-polarized) ARPES. However, the value of the symmetry analysis is that one can directly compute symmetry-allowed components of electronic and spintronic responses coupling spin, charge, and heat. We have exemplified how to make experimentally relevant predictions based on the symmetry analysis presented for a number of different altermagnetic candidate materials.

Having determined the Landau theories describing altermagnets whose crystal and magnetic unit cells coincide, some questions for future investigation remain. A natural extension of this work would consider the $\bm{Q}\neq \bm{0}$ ``supercell" altermagnets introduced in Ref.~\cite{supercellAltermagnets2024}. In this case, altermagnets may arise even in structures whose point group is one of the six forbidden $\bm{Q}=\bm{0}$ point groups. 

Further, the nature of non-centrosymmetric altermagnets has received limited attention \cite{Yu2025_OddParityMag}. Due to the emergent inversion-symmetry of the band structure, there is a discrepancy between the lowest order allowed SO-free multipolar order parameter and the spin-splitting pattern in reciprocal space. It may be worth exploring different properties that would inherit the lowest-order multipolar symmetry. 

Finally, the list of tensors corresponding to physical properties used in this work is far from exhaustive. Future studies may seek to expand the present symmetry analysis to other experimentally relevant features of altermagnetic systems.

\begin{acknowledgments}
H.S. and J.R. were supported by the NSF through grant DMR-2142554. Work at
the University of Windsor (J.G.R.) was funded by the Natural Sciences and Engineering Research Council of Canada (NSERC)
(Funding Reference No. RGPIN-2020-04970). P.M. acknowledges funding from the CNRS.
\end{acknowledgments}

\appendix 
\section{\label{app:circumventingSPGs}Circumventing Spin Groups}

In the Landau theory, conventionally, one uses the symmetry group of the ordered phase, in which the (primary and secondary) order parameters transform as the fully symmetrical (trivial) irreducible representation. The symmetry groups of ideal altermagnetic phases correspond to spin groups~\cite{Smejkal2022a,Smejkal2022b}. Spin point groups describing SO-free cases are subgroups of $\mathbf{O}^{s}(3)\times \mathbf{O}(3)$, where $\mathbf{O}^{s}(3)$ and $\mathbf{O}(3)$ contain the proper and improper rotations in spin space ($R_s$) and real space ($R_l$), respectively~\cite{spinPointLitvin, Liu}. The improper rotations in spin space contain the time-reversal operator (spin-inversion), $\tau$, and the improper rotations in real space include the inversion element, $I$. Group elements of $\mathbf{O}^{s}(3)\times \mathbf{O}(3)$ are usually written as $[R_s\|R_l]$, where the first element acts on spin and the second on lattice degrees of freedom~\cite{spinPointLitvin, Liu,schiff2023}.

When SOC is zero, the spin point group can be written as $\mathbf{b}\times\mathbf{S}$, where the spin-only group, $\mathbf{b}$, describes symmetries dictated by mutual spin orientations (collinear, coplanar, and non-collinear non-coplanar), and $\mathbf{S}$ is one of the 598 non-trivial spin point groups~\cite{spinPointLitvin, Liu,schiff2023}.

There are 58 spin point groups describing collinear antiferromagnetic spin arrangements, corresponding to $\mathbf{b}_{\infty}\times\mathbf{S}$, with
\begin{align}
    &\mathbf{b}_{\infty} = [\mathbf{SO}(2)\|E] \rtimes \{ \,[E\|E], [\tau 2_{\perp\mathbf{n}}\|E]\,\}\;\;\text{and}\\
    &\mathbf{S} = [E\|\mathbf{H}]+ [\tau\|a][E\|\mathbf{H}]\;,\label{eq:nontrivi}
\end{align} 
where $[\mathbf{SO}(2)\|E]$ is the group of spin-only rotations about the shared spin axis $\mathbf{n}$, and $2_{\perp\mathbf{n}}$ is a $\pi-$rotation about an axis perpendicular to the spin axis. 
$\mathbf{H}$ is a standard crystallographic point group and the group $\mathbf{H} + a\mathbf{H}$ is isomorphic to $\mathbf{F}$, the point group of the underlying crystal structure ~\cite{spinPointLitvin, Liu,schiff2023}.

The elements in the coset $\mathbf{H}$ preserve the sublattices, while the elements in $a\mathbf{H}$ swap them. Thus the group element $a$ must be paired with time-reversal $\tau$ in the spin point group $\mathbf{b}_{\infty}\times\mathbf{S}$ (see Eq.~(\ref{eq:nontrivi})), so that the coset $[\tau\|a][E\|\mathbf{H}]$ leaves the antiferromagnetic arrangement invariant.

The N\'eel vector describing an altermagnetic order must be inversion-even. This constraint means that $\mathbf{S}$ cannot contain the group element $[\tau\|I]$, disqualifying $21$ of the $58$ possible spin groups corresponding to collinear antiferromagnetism. These include any spin group based on $\mathbf{F} = \overline{1},\overline{3},$ and $\frac{2}{m}\overline{3}$. These symmetry considerations result in 37 spin point groups that are compatible with altermagnetism.

It is possible to avoid complications associated with the spin groups for collinear altermagnets as the Landau theory is based on long-range order developing out of the paramagnetic phase.  Using the representation theory of the SO-free paramagnetic group\footnote{When referring to groups containing antiunitary time-reversal, the correct terminology is a ``co-representation." In these appendices, we use use representation and co-representation interchangeably, as antiunitarity is apparent from the group.}, the altermagnetic order parameter, $\mathbf{N}$, does not belong to the fully symmetrical trivial irrep but instead transforms as a nontrivial irrep.  This nontrivial representation of the paramagnetic group becomes the trivial one if we restrict the group elements of the paramagnetic point group to those of the spin point group corresponding to the order. 

The advantage of the SO-free paramagnetic group is that it can be written as a direct product of spin-only and lattice-only transformations.  The spin-only group is $\mathbf{O}^{s}(3)$, containing the proper and improper spin rotations and the lattice-only transformations encompass the space group of the crystal, with point group $\mathbf{F}$. Because of the constraint that $\mathbf{N}$ transforms trivially under translations, it is sufficient to consider the properties of $\mathbf{N}$ under the spin point group $\mathbf{O}^{s}(3)\times \mathbf{F}$, describing the SO-free paramagnetic phase.

The N\'eel vector ${\bf N}$ transforms as a nontrivial irrep of $\mathbf{O}^{s}(3)\times\mathbf{F}$, which can be expressed as a direct product of the irreps of $\mathbf{O}^{s}(3)$ and $\mathbf{F}$. This is a non-trivial fact; the co-irreps of direct product groups containing time-reversal (or any antiunitary element) are generally not tensor products of the groups that are multiplied. In Appendix~\ref{app:spin&realspace} we give a detailed argument as to why the irreps can be written in such tensor-product form here.

Similar to the irreps of $\mathbf{SO}(3)$, the irreps of $\mathbf{O}^{s}(3)$ are labelled by angular momentum integers $l\in\mathbb{N}^{+}$.  Because $\mathbf{N}$ is the three-component staggered magnetization, in spin-space $\mathbf{N}$ transforms like a vector $(l=1)$ that is odd under time-reversal symmetry. 
Furthermore, following the main text notation, $\mathbf{N}$ transforms as the $\Gamma_{\mathbf{N}}$ irreducible representation of the point group ${\bf F}$.
Altogether, the N\'eel vector belongs to the $\Gamma_{l=1}\otimes \Gamma_{\mathbf{N}}$ irrep of the SO-free paramagnetic group $\mathbf{O}^{s}(3) \times \mathbf{F}$. 

We will now show that there is a one-to-one correspondence between the spin point groups and the non-trivial irreducible representations of crystallographic point groups, with $\Gamma_{l=1}^{s} \otimes \Gamma_{\mathbf{N}}$ irrep reducing to the trivial irrep of the true spin group of the ordered phase. This correspondence allows us to derive the Landau theory of altermagnets starting from the paramagnetic phase, using the irreps of $\mathbf{O}^{s}(3) \times \mathbf{F}$, and avoid using spin groups altogether. This approach provides a conceptual simplification in the study of altermagnetism.

To encode a bipartite sublattice structure (necessary for collinear antiferromagnetism), $\mathbf{F}$ must have a one-dimensional real irreducible representation where the elements of $\mathbf{H}$ are represented by $1$ and the elements of $a\mathbf{H}$ are represented by $-1$. Three point groups, $1$, $3$, and $23$, are immediately eliminated because they do not have any nontrivial real one-dimensional irreducible representations. Consequently, there are no collinear antiferromagnetic spin point groups based on any of these three point groups.

To encode the inversion-even criterion of altermagnetism, when $\mathbf{F}$ contains the inversion element $I$, i.e.  $\mathbf{F}$ is centrosymmetric, there must be at least one nontrivial one-dimensional real irreducible representation that is also inversion even~\cite{McClarty2024}. This condition disqualifies three additional point groups: $\overline{1}$, $\overline{3}$, and $\frac{2}{m}\overline{3}$, as these do not have any non-trivial one-dimensional real irreps that are even under inversion.

Altogether, we have 26 remaining point groups $\mathbf{F}$ that are compatible with altermagnetism.
The question is whether there is a correspondence between these point groups 
and the 37 collinear spin point groups that can describe altermagnetism. The answer is affirmative: the non-trivial inversion-even one-dimensional real irreps of the viable 26 point groups $\mathbf{F}$ are -- up to relabelling coordinate axes -- in a one-to-one correspondence with the remaining nontrivial spin point groups $\mathbf{S}$. We show this correspondence in Table~\ref{tab:F_irreps_vs_S_irreps}.

\begin{table}
    \centering
        \caption{Point groups ${\bf F}$ that are compatible with altermagnetism and the nontrivial one-dimensional real inversion-even irreps of ${\bf N}$ in them. The irreps inside the curly brackets are identical up to axes relabelling. The last column contains the nontrivial spin group that corresponds to the altermagnetic order described by the $\Gamma_{\bf N}$ irrep of the paramagnetic point group.}
    \begin{tabularx}{\columnwidth}{>{\hsize=0.3\hsize\raggedright\arraybackslash}X >{\hsize=0.6\hsize\centering\arraybackslash}X >{\hsize=0.9\hsize\centering\arraybackslash}X}
    \hline
    \hline
       ${\bf F}$  & $\Gamma_{\bf N}$ & corresponding ${\bf S}$\\
       \hline
       ${\bf 2}$ & $B$  & $^{1}1 + [\tau\|2]\;^{1}1$\\
       ${\bf m}$ & $A''$  & $^{1}1 + [\tau\|m]\;^{1}1$\\
       ${\bf 2/m}$ & $B_g$  & $^{1}\overline{1} + [\tau\|2]\;^{1}\overline{1}$\\
       ${\bf 222}$ & $\{B_1, B_2, B_3\}$ & $^{1}2 + [\tau\|2]\;^{1}2$\\
       ${\bf mmm}$ & $\{B_{1g}, B_{2g}, B_{3g}\}$  & $^{1}2_z/^{1}m_z + [\tau\|2_x]\;^{1}2_z/^{1}m_z$\\
       ${\bf 4}$ &  $B$ & $^{1}2 + [\tau\|4]\;^{1}2$\\
       ${\bf \overline{4}}$ & $B$ & $^{1}2 + [\tau\|\overline{4}]\;^{1}2$\\
       ${\bf 4/m}$ &  $B_g$ & $^{1}2/^{1}m + [\tau\|4]\;^{1}2/^{1}m$ \\
       ${\bf 32}$ &  $A_2$ & $^{1}3+ [\tau\|2]\;^{1}3$\\
       ${\bf 3m}$ &   $A_2$ & $^{1}3+ [\tau\|m]\;^{1}3$\\
       ${\bf \overline{3}m}$ &   $A_{2g}$ & $^{1}\overline{3}+ [\tau\|m]\;^{1}\overline{3}$\\
       ${\bf 6}$ &  $B$ & $^{1}3+ [\tau\|6]\;^{1}3$\\
       ${\bf \overline{6}}$ &  $A''$ & $^{1}3+ [\tau\|\overline{6}]\;^{1}3$\\
       ${\bf 6/m}$ &  $B_g$ & $^{1}\overline{3}+ [\tau\|6]\;^{1}\overline{3}$\\
       ${\bf m\overline{3}m}$ & $A_{2g}$ & $^{1}2/^{1}m\;\overline{3}+ [\tau\|{4}]\;^{1}2/^{1}m\;\overline{3}$\\
       ${\bf 432}$ &  $A_{2}$ & $^{1}2^{1}3+ [\tau\|{4}]\;^{1}2^{1}3$\\
       ${\bf \overline{4}3m}$ &  $A_{2}$ & $^{1}2^{1}3+ [\tau\|\overline{4}]\;^{1}2^{1}3$\\
       \hline
       \multirow{2}{2cm}{${\bf mm2}$} &  $A_2$  & $^{1}2 + [\tau\|m]\;^{1}2$\\
       {} &  $\{B_1$, $B_2\}$  & $^{1}m + [\tau\|2]\;^{1}m$\\
       \multirow{2}{2cm}{${\bf 422}$} & $A_2$ & $^{1}4 + [\tau\|2_x]\;^{1}4$\\
       {} & $\{B_1, B_2\}$  & $^{1}2^{1}2^{1}2 + [\tau\|4]\;^{1}2^{1}2^{1}2$\\
       \multirow{2}{2cm}{${\bf 4mm}$} &  $A_2$ & $^{1}4 + [\tau\|m_{x}]\;^{1}4$ \\
       {} & $\{B_1$, $B_2\}$  & $^{1}m^{1}m^{1}2 + [\tau\|4]\;^{1}m^{1}m^{1}2 $\\
       \multirow{2}{2cm}{${\bf 4/mmm}$} &  $A_{2g}$ & $^{1}4/^{1}m + [\tau\|m_x]\;^{1}4/^{1}m$\\
       {} & $\{B_{1g}, B_{2g}\}$ & $^{1}m^{1}m^{1}m + [\tau\|4]\;^{1}m^{1}m^{1}m $\\
       \multirow{2}{2cm}{${\bf 622}$} &  $A_2$  & $^{1}6+ [\tau\|2_x]\;^{1}6$\\
       {} & $\{B_1, B_2\}$  & $^{1}3^{1}2+ [\tau\|6]\;^{1}3^{1}2$\\
        \multirow{2}{2cm}{${\bf 6mm}$} &  $A_2$ & $^{1}6+ [\tau\|m_x]\;^{1}6$\\
        {} & $\{B_1, B_2\}$  & $^{1}3^{1}m+ [\tau\|6]\;^{1}3^{1}m$\\
       \multirow{2}{2cm}{${\bf 6/mmm}$} &  $A_{2g}$ & $^{1}6/^{1}m+ [\tau\|m_x]\;^{1}6/^{1}m$\\
       {} & $\{B_{1g}, B_{2g}\}$ & $^{1}\overline{3}{}^{1}m+ [\tau\|6]\;^{1}\overline{3}{}^{1}m$\\
       \hline
       \multirow{3}{2cm}{${\bf \overline{4}2m}$ (and~${\bf \overline{4}m2}$)} & $A_2$ & $^{1}\overline{4} + [\tau\|2_x]\;^{1}\overline{4}$\\
       {} & $B_1$ & $^{1}2^{1}2^{1}2 + [\tau\|\overline{4}]\;^{1}2^{1}2^{1}2$\\
       {} & $B_2$ & $^{1}m^{1}m^{1}2 + [\tau\|\overline{4}]\;^{1}m^{1}m^{1}2$\\
       \multirow{3}{2cm}{${\bf \overline{6}m2}$ (and~${\bf \overline{6}2m}$)} &  $A'_2$ & $^{1}\overline{6}+ [\tau\|m_x]\;^{1}\overline{6}$\\
       {} & $A''_1$ & $^{1}3^{1}2+ [\tau\|\overline{6}]\;^{1}3^{1}2$\\
       {} & $A''_2$ & $^{1}3^{1}m+ [\tau\|\overline{6}]\;^{1}3^{1}m$\\
       \hline
       \hline
    \end{tabularx}
    \label{tab:F_irreps_vs_S_irreps}
\end{table}

We demonstrate the correspondence between the $\Gamma_{\bf N}$ irreps and the spin point groups on the example of point group $4mm$. There are two collinear spin groups corresponding to antiferromagnetic arrangements that can be derived from $4mm$: $^{1}4^{\overline{1}}m^{\overline{1}}m = \,^{1}4 + [\tau\|m_{x}]^{1}4$, and $^{\overline{1}}4^{1}m^{\overline{1}}m =\, ^{1}m^{1}m^{1}2 + [\tau\|4]^{1}m^{1}m^{1}2 $.  In this notation $^{g}f$ indicates that the point group generator $f$ appears with spin-space element $g$, i.e. $[g\|f]$ is one of the generators of the spin point group~\cite{spinPointLitvin}. The $\overline{1}$ superscript indicates that the spin-space element is the time-reversal operator, $\tau$. The point group $4mm$ has three non-trivial, one-dimensional irreps (inversion is not present in this group): $A_{2}$, $B_{1}$, and $B_{2}$. 

The irrep $A_{2}$ of $4mm$ assigns $1$ to the $\frac{\pi}{2}$ and $\pi$ rotations about the $z-$axis, and $-1$ to the four reflections. This irrep is in direct correspondence with the spin point group $^{1}4^{\overline{1}}m^{\overline{1}}m$, where the mirrors are paired with time-reversal $\tau$. 

The $B_{1}$ and $B_{2}$ irreps of $4mm$ assign $1$ to the $\pi$ rotation about the $z-$ axis as well as two of the four mirrors, while the four-fold rotations and remaining two mirrors are assigned $-1$. To establish a connection to a spin point group, the four elements represented by $-1$ in the point group need to be composed with $\tau$ in the spin point group. The two spin point groups obtained in this way are conjugate to each other in $\mathbf{O}_{3}^{s} \times \mathbf{O}_{3}$ and so they correspond to the same (class of) spin point groups~\cite{spinPointLitvin,schiff2023}, $^{\overline{1}}4^{1}m^{\overline{1}}m$. The equivalence of these groups effectively amounts to a relabelling of the $x-$axis to the axis at an angle of $45^{\circ}$ between the $x-$ and the $y-$axes. Any collinear antiferromagnet whose N\'eel vector transforms under the $A_{2}$ irrep of $4mm$ will have spin group symmetry given by $^{1}4^{\overline{1}}m^{\overline{1}}m,$ whereas if $\mathbf{N}$ transforms under $B_{2}$ or $B_{3}$ of $4mm$ it will have spin point group symmetry given by $^{\overline{1}}4^{1}m^{\overline{1}}m,$ with appropriately chosen axes.

Another class of examples that clarifies this correspondence are the non-centrosymmetric point groups with only one associated (collinear antiferromagnetic) spin point group. These are $2$, $m$, $222$, $4$, $\overline{4},$ $32,$ $3m$, $\overline{6},$ $6$, $\overline{4}3m$, and $432$. Aside from $222$, each of these point groups only has one non-trivial real one-dimensional irrep. This is precisely why they only have one corresponding (collinear antiferromagnetic) spin point group. In the case of $222$, there are three valid irreducible representations, but they give rise to spin point groups that are conjugates in $\mathbf{O}_{3}^{s} \times \mathbf{O}_{3}$.

The one-to-one correspondence between the $\Gamma_{\bf N}$ irreps of the paramagnetic point group ${\bf F}$ and the possible altermagnetic spin point groups enables the derivation of Landau theory using ${\bf F}$. Based on symmetry arguments, we ruled out six point groups that cannot support altermagnetic phases. The six nonviable point groups belong to 20 space groups, therefore, we expect 210 out of 230 space groups to have at least one Wyckoff position that can support altermagnetism. This is consistent with our results shown in Appendix~\ref{app:wpGNtable}.

We note that while we may avoid the use of spin groups in the Landau theory, the representation theory of spin groups becomes essential when discussing certain symmetry properties in the ordered phase $-$ for example band degeneracies $-$ and phase transitions from the altermagnetic phase. 

\section{\label{app:spin&realspace}Direct product representations of the SOC-free paramagnetic group}

In this section, we clarify the argument that for quantities we are interested in, representations of the SO-free paramagnetic group $\mathbf{O}^{s}(3) \times \mathbf{F}$ can be expressed as the direct product of representations of $\mathbf{O}^{s}(3)$ and representations of $\mathbf{F}$, where $\mathbf{O}^{s}(3) \cong \mathbf{SO}(3) + \tau \mathbf{SO}(3)$ and $\mathbf{F}$ is a crystallographic point group. 

The crucial point is that we are only interested in quantities whose real-space transformation properties are described by real representations of $\mathbf{F}$, denoted by $\Gamma^{(\nu)}$.

The SO-free paramagnetic group can be expressed in a coset decomposition of its unitary halving subgroup: 
\begin{align}
\mathbf{O}^{s}(3)\times\mathbf{F} = (\mathbf{SO}(3) \times \mathbf{F}) + \tau\; (\mathbf{SO}(3) \times \mathbf{F})\;.
\end{align}
The co-irreps of $\mathbf{O}^{s}(3)\times\mathbf{F}$ will be induced from the irreps of $\mathbf{SO}(3)\times\mathbf{F}$, $\Delta^{(l)} \otimes \Gamma^{(\nu)}$, where $l \in \mathbb{N}_{+}$ labels the irreps of $\mathbf{SO}(3)$ and $\nu$ labels the irreps of $\mathbf{F}$. The induction scheme for each irrep depends on its reality because the coset representative is simply $\tau$, and so Dimmock's test reduces to the Frobenius-Schur indicator~\cite{wigner2012group,DimmockTest,DamnjanovicSymm,BradleyCracknell, schiff2023}. 

Since $\Delta^{(l)}$ are all real, the induction scheme depends only on the reality of the point group irrep $\Gamma^{(\nu)}$. When the irrep $\Gamma^{(\nu)}$ is real,  an element $[a R\|f]$ of this group (where $R\in\mathbf{SO}(3)$, $f\in\mathbf{F}$, and $a$ is either the identity element or time-reversal $\tau$) can be chosen to be represented in the co-irrep by $(-1)^{\pi(a)} \Delta^{(l)}(R) \times \Gamma^{(\nu)}(f)$, where $\pi(E) = 0$, and $\pi(\tau) = 1$. This choice of $\pi(a)$ corresponds to time-reversal inverting spins. Notice that $(-1)^{\pi(a)}\Delta^{(l)}(R)$ corresponds to the ``polar" $l$ co-irrep of $\mathbf{O}^{s}(3)$, where $\tau$ corresponds to inversion element and is represented by a scalar matrix $-1$ of appropriate dimension. These are the $\Gamma_{l}^{(s)}$ irreps referred to in Ref.~\cite{McClarty2024}. 

We have shown here that for real point group irreps, the co-irrep of the SO-free paramagnetic group is simply expressed as the direct product of the $\Gamma_{l}^{(s)}$ co-irrep of $\mathbf{O}^{s}(3)$ and the point group irrep $\Gamma^{(\nu)}.$

We emphasize that without SOC, the irreducible representation of $\mathbf{F}$ describing real-space transformation properties of the N\'eel vector must be real, and so the co-irrep of the SOC-free paramagnetic point group will be of the direct-product form above. 

We are also interested in the representations under which the multipoles transform. We will now demonstrate that the representations describing SO-free multipoles can also be expressed in direct-product form.

A multipole's real-space transformation properties under $\mathbf{F}$ are given by a generically reducible representation $D = \bigoplus_{\nu} a_{\nu}\Gamma^{(\nu)}$, where the irreps with non-zero multiplicity $a_{\nu} \neq 0$ are real irreps of $\mathbf{F}$. In fact, this may be chosen by using only the ``physically irreducible" representations of the point groups~\cite{PhysicallyIrrStokesHatchKim87}, which are the irreps allowed over $\mathbb{R}$ as opposed to $\mathbb{C}$, and are appropriate for a tensor constructed out of real-space coordinates. A multipole's spin-space transformation properties will be given by a reducible (real) representation $\Delta = \bigoplus_{l} b_{l} \Gamma_{l}^{(s)}$ of $\mathbf{O}^{s}(3)$. 

The direct product representation of $\mathbf{O}^{s}(3)\times\mathbf{F}$ given by $\Delta \otimes D$ can then be expressed as
\begin{align}
\Delta \otimes D &= \left(\bigoplus_{l} b_{l}\Gamma_{l}^{(s)}\right) \otimes \left(\bigoplus_{\nu}a_{\nu} \Gamma^{(\nu)}\right)\nonumber\\
&= \bigoplus_{l,\nu} b_{l}a_{\nu} \Gamma_{l}^{(s)} \otimes \Gamma^{(\nu)}.
\end{align}

Due to the reality of $\Gamma_{l}^{s}$ and $\Gamma^{(\nu)}$, $\Gamma_{l}^{(s)}\otimes\Gamma^{(\nu)}$ are co-irreps of $\mathbf{O}^{s}(3)\times\mathbf{F}$, and we have found the co-irrep decomposition of $\Delta \otimes D$. 

Formally, our claim that we can use direct product representations of $\mathbf{O}^{s}(3)\times \mathbf{F}$ for quantities we are interested in reduces to the fact that we only need co-irreps falling into case (a) of Wigner's co-irrep classification scheme~\cite{wigner2012group, BradleyCracknell, DamnjanovicSymm, DimmockTest, schiff2023}, as these are the co-irreps appearing in the decompositions of any multipole's representation. These case (a) co-irreps can be expressed as a direct product of $\mathbf{O}^{s}(3)$ co-irreps and $\mathbf{F}$ irreps.

\section{\label{app:WPalg}Altermagnetic Structures Algorithm: Technical Details}

In this section, we outline an algorithm for identifying all crystal structures capable of supporting ($\bm{Q}=0$) altermagnetism. This means that we can identify the Wyckoff positions in each space group $\mathbf{G}$ whose sublattices satisfy the symmetry constraints outlined in Sec.~\ref{sec:overview}: the spin sublattices, and consequently the Néel vector $\mathbf{N}$ (both in absence of spin-orbit coupling) transform under a 1D, real irrep of the crystal point group $\mathbf{F}$, that is inversion-even in centrosymmetric cases. These Wyckoff positions are candidates for positions of magnetic ions in an altermagnet. The results of this algorithm are summarized in Tables~\ref{tab:AM_PGs_and_irreps} and~\ref{tab:wpGN}. The Wyckoff positions and space group elements used in our algorithm were obtained from the Bilbao Crystallographic Server~\cite{BilbaoGenPosWP}. 

By selecting a Wyckoff position $\vec{w}$ and acting on it with all transformations of the space group, a lattice is generated; in one unit cell, there will be $n_{\vec{w}}$ atoms. For these $n_{\vec{w}}$ atoms to be compatible with altermagnetism, it must be possible to place `up' and `down' spins on each site, implying that the multiplicity $n_{\vec{w}}$ must be even. The symmetry constraints of altermagnetism described in Sec.~\ref{sec:overview} dictate that these two sublattices must not be mapped into one another by pure spatial translation or inversion. 

When we ask how the sublattices are mapped into one another under lattice transformations, we are examining the \emph{permutation} action of the space group on the atoms. Then naturally we are concerned with the permutation representation of the space group on the lattice.

The group elements of $\mathbf{G}$ can be expressed in Wigner-Seitz notation as $[f|\vec{t}]$, where $f\in\mathbf{F}$ is some $\mathbf{O}(3)$ matrix, and $\vec{t}$ is a three-dimensional translation vector\footnote{the single vertical bar distinguishes space group elements from the more general spin group notation}~\cite{BradleyCracknell, DamnjanovicSymm}. This group element transforms the atomic position $\vec{r}$ to $f\vec{r} + \vec{t}$. 

There is a great deal of redundancy in the action of $\mathbf{G}$ on our lattice. Without any loss of information, we may restrict our attention to the action of $\mathbf{G}$ on the $n_{\vec{w}}$ atoms within a single unit cell, by treating every element of $\mathbf{G}$ \emph{modulo translations}. This means that we identify as equivalent all elements $[f|\vec{t}]$ with $\vec{t}$ vectors of the form $\vec{t} = \vec{q} + n_{1}\vec{a}_{1} + n_{2}\vec{a}_{2} + n_{3}\vec{a}_{3}$ for $\{\vec{a}_{i}|i\in\{1,2,3\}\}$ representing the primitive lattice vectors, $n_{i} \in \mathbb{Z},$ $i\in\{1,2,3\}$ and $|\vec{q}| < |\vec{a}_{i}|$. The group composition is also treated modulo this equivalence relation. This has the effect of reducing the space group $\mathbf{G}$ to the quotient group $\tilde{\mathbf{F}} = \mathbf{G}/\mathbf{T}^{(3)}$ where $\mathbf{T}^{(3)}$ is the Abelian group of translations of the lattice. This quotient group $\tilde{\mathbf{F}}$ is isomorphic to the point group $\mathbf{F}$ of the lattice, and it is this group $\tilde{\mathbf{F}}$ for which we would like to construct a permutation representation.

Each element $[f|\vec{q}] \in \tilde{\mathbf{F}}$ will send an atom $\vec{w}_{i}$ within the unit cell to another atom $\vec{w}_{j}$ within the unit cell. The permutation representation $\Delta(f)$ of this element will be given by $\Delta(f) \vec{w}_{i} = \vec{w}_{j},$ resulting in a $n_{\vec{w}} \times n_{\vec{w}}$ matrix whose $i-$th row contains exactly one 1 in the $j-$th column.

Let $\Gamma_{\mathbf{N},\alpha}$ denote irreps of $\mathbf{F}\cong\tilde{\mathbf{F}}$ that satisfy the altermagnetic constraints. There may be several such irreps in $\mathbf{F}$ and we index these by $\alpha$. The Wyckoff position $\vec{w}$ is compatible with altermagnetism if and only if the permutation representation $\Delta(\mathbf{\tilde{F}})$ contains any of the irreps $\Gamma_{\mathbf{N},\alpha}.$ This condition is easily checked by taking the inner product of the characters $\chi(\Delta) = \{\, \mathrm{Tr}(\Delta(f)) \, |\,\, [f|\vec{q}]\in\mathbf{\tilde{F}}\}$ of the permutation representation with the characters $\Gamma_{\mathbf{N},\alpha}(f)$ of $\Gamma_{\mathbf{N},\alpha}$ \footnote{Because $\Gamma_{\mathbf{N},\alpha}$ is one-dimensional, the representation is equal to its characters.}~\cite{BradleyCracknell, DamnjanovicSymm, GTPackBook}: $$a_{\Gamma_{\mathbf{N},\alpha}} = \left(\chi(\Delta),\Gamma_{\mathbf{N},\alpha}\right) = \frac{1}{|\mathbf{\tilde{F}}|}\sum_{f\in\mathbf{\tilde{F}}} \Gamma_{\mathbf{N},\alpha}(f) \chi(\Delta(f)).$$ If the natural number $a_{\Gamma_{\mathbf{N},\alpha}} \neq 0,$ then this Wyckoff position $\vec{w}$ can support an altermagnetic order with sublattice transformation properties dictated by the irrep $\Gamma_{\mathbf{N},\alpha}$. The result of applying this algorithm to all Wyckoff positions in all 230 space groups are summarized in table \ref{tab:wpGN}.

This technique can be adapted to study structures supporting any magnetic order of interest, so long as translational symmetry is preserved (i.e. translations act trivially on the level of permutations within the unit cell). The extension of this technique to structures with an enlarged magnetic unit cell is relatively straightforward, but irrelevant to collinear altermagnets: the procedure is modified only by the choice of translational group with which $\mathbf{G}$ is quotiented.

\section{Consistency of SOC Landau Theory with Magnetic Symmetry Analysis}\label{app:noMagGroups}

In Appendix \ref{app:circumventingSPGs} we demonstrated that the SO-free Landau theory derived in Sec.~\ref{subsec:AMLTOverview} is justified; all conclusions based on our analysis with ordinary point groups are consistent with a Landau theory using a spin point group in the ordered phase. Here, we provide the sibling argument for the spin-orbit coupled Landau theory. This scenario is more involved from the perspective of symmetries, than the SO-free case.

In Sec.~\ref{sec:finiteSOC}, we formulate Landau theories for altermagnets when SOC is included. By turning on spin-orbit coupling, we implicitly \emph{lock} the spins to the lattice, making it impossible to transform lattice and spin degrees of freedom separately. This reduces the symmetry of the paramagnetic phase to a so-called \emph{grey group}. When assuming translations act trivially, the spin-orbit coupled paramagnetic group is $\mathbf{F} + \tau \mathbf{F}$, with $\mathbf{F}$ being the crystallographic point group and $\tau$ being time-reversal. With SOC, in passing from the high symmetry paramagnetic phase to the collinear altermagnetic phase, the symmetry is reduced to a \emph{black \& white} magnetic group~\cite{BradleyCracknell,BWPG,Litvin_MagGps}. 

In the presence of SOC, each component of the Néel vector may, in principle, transform under different irreps of the paramagnetic grey group. Recall that the spin-orbit coupled Néel vector $\mathbf{N}$ transforms as $aeV\otimes\Gamma_{\mathbf{N}}$. For each point group relevant to altermagnets, this representation decomposes into three one dimensional irreps, one 1D and one 2D irrep, or a singular 3D irrep. Having multiple order parameters, and having order parameters whose irreps are larger than 1D makes the SOC Landau theory slightly more subtle than in the SO-free case. 
A one-to-one correspondence between the ordered symmetry group and the paramagnetic co-irrep is not guaranteed when SOC is included, due to the more complicated nature of the order parameters.

Without this one-to-one correspondence, it may be useful to remind the reader that there are two \emph{equivalent} ways of formulating Landau theories. The (direct) Landau problem is concerned with determining the possible symmetry groups of the ordered phase, given the high symmetry phase's group and the irrep under which the order parameter transforms. The inverse Landau problem starts with known high and low symmetry groups and asks which order parameters are possible. We have seen that in the SO-free case, both problems are exactly identical, not just equivalent~\cite{DamnjanovicSymm}. In the spin-orbit coupled case they are not identical, and this fact has been the root cause for decades of debate between the ``representation analysis" approach and the ``magnetic space group" approach to understanding magnetic structures~\cite{RAMSG_Petricek2010}.

This being said, we take the approach of the direct Landau problem. Each component $N_{i}$ of the Néel vector transforms under an irrep of the paramagnetic group. Necessarily, there will be at least one element that leaves each component invariant. The intersection of these elements for all three components gives the black \& white point group corresponding to all components $N_{i}$ being non-zero. 

Whether or not all three $N_{i}$ are non-zero in a given material, however, is not a question of symmetry: it is a question of the microscopic theory governing the magnetic interactions. With any $N_{i}$ being zero, the resulting black \& white symmetry groups of the possible orders may be larger. In this way, we can see that several black-and-white point groups may be identified with one paramagnetic (generically reducible) co-representation describing the ordered phase.

With this in mind, we may now proceed in justifying our use of ordinary point groups to determine the spin-orbit coupled Landau theory. To do so, we must first establish the co-irrep theory for the grey paramagnetic groups, and demonstrate that the co-irreps under which $\mathbf{N}$ transforms are completely determined by the decomposition of $aeV\otimes\Gamma_{\mathbf{N}}.$

The co-irreps for grey point groups $\mathbf{F} + \tau \mathbf{F}$ are generated (induced) from each irrep $\Gamma^{(\nu)}$ of $\mathbf{F}$. The induction algorithm~\cite{wigner2012group, BradleyCracknell, DimmockTest} depends on the reality of the $\Gamma^{(\nu)}$. Following the classification in Ref.~\cite{BradleyCracknell}, all irreps of the crystallographic point groups are of the first kind (real), \emph{except} those with complex characters, which are of the third kind (complex). The co-irreps arising from real $\Gamma^{(\nu)}$ are simple: we may choose that $\tau$ is represented by $-\mathbb{I}_{\mathrm{dim}\Gamma^{(\nu)}}$ (where $\mathbb{I}_{\mathrm{dim}\Gamma^{(\nu)}}$ is the identity matrix of dimension equal to that of $\Gamma^{(\nu)}$)\footnote{Formally, this choice corresponds to the single-valued co-irreps, which are appropriate for integer angular momentum. A full theory for half-integer angular momentum would use the double-valued co-irreps.}, physically corresponding to time-reversal inverting magnetic moments. This choice completely determines the irrep of the paramagnetic group, and no information is lost in derivations relying solely on the knowledge of $\Gamma^{(\nu)}$.

For the complex irreps $\Gamma^{(\nu)}$ of $\mathbf{F}$, the corresponding co-irrep of $\mathbf{F}+\tau\mathbf{F}$ is doubled. The elements of $\mathbf{F}$ are represented by matrices 
\begin{equation}\label{eq:coirrC_unitary}
\begin{bmatrix} \Gamma^{(\nu)}(f) & 0 \\ 0 & \Gamma^{(\nu)*(f)} \end{bmatrix},
\end{equation}
while the time-reversal element $\tau$, which satisfies $\tau^{2}=E,$ may be represented by 
\begin{equation}\label{eq:coirrC_anti}
\begin{bmatrix} 0 & -\mathbb{I}_{\mathrm{dim}\Gamma^{(\nu)}} \\ -\mathbb{I}_{\mathrm{dim}\Gamma^{(\nu)}} & 0 \end{bmatrix}.
\end{equation}

The equivalence of two co-representations of a magnetic group is determined entirely by the representation of the unitary coset (those elements without time-reversal, i.e. $\mathbf{F}$). If under the point group action $\mathbf{N}$ transforms as $aeV\otimes\Gamma_{\mathbf{N}}$, we then have a clear picture of the corresponding co-irrep of the paramagnetic group. When $aeV$ contains only real irreps in its decomposition, the true paramagnetic representation is generated by retaining $aeV(f)$ for elements of $\mathbf{F}$ while ascribing to the elements $\tau f$ the representation $-aeV(f).$ It's decomposition into paramagnetic co-irreps is given directly by the decomposition of $aeV$ into irreps $\Gamma^{(\nu)}$ of $\mathbf{F}.$ 

When $aeV$ contains a complex irrep\footnote{This is the case for the complex 1D irreps in point groups $4$, $\overline{4}$, $4/m$, $6$, $\overline{6}$, and $6/m$. The 2D irreps in centrosymmetric groups, as well as $422$, $4mm$, $\overline{4}2m$, $3m$, $\overline{3}m$, $622$, $6mm$, and $\overline{6}2m$ are all real. In $432$, $-43m$ and $m\overline{3}m,$ $aeV$ transforms as a real $3D$ irrep.} describing the transformation of a Néel component $N_{i}$, the paramagnetic co-irrep corresponding to $N_{i}$ will assign to the elements $f\in\mathbf{F}$ a matrix of the form Eq.~\ref{eq:coirrC_unitary}, and to the elments $\tau f$ the matrix given by composing Eq.~\ref{eq:coirrC_anti} with that of Eq.~\ref{eq:coirrC_unitary}. The decomposition into paramagnetic co-irreps is again determined entirely by the decomposition of $aeV$ in $\mathbf{F}$, though the co-irreps have greater dimensions. 

In both cases, whether $\Gamma^{(\nu)}$ is real or complex, the product of $aeV$ with $\Gamma_{\mathbf{N}}$ is no different than in the unitary case, owing to the reality of $\Gamma_{\mathbf{N}}.$ The true co-irrep in the spin-orbit coupled paramagnetic phase is uniquely determined by the decomposition of $aeV\otimes\Gamma_{\mathbf{N}}$ in $\mathbf{F}.$

Because we are concerned with the direct Landau problem, in principle we may then make predictions about the possible black-and-white point groups describing the low symmetry phase. We provide a simple example, using the CrF$_{2}$ example in Sec.~\ref{sec:examples}. The crystallographic point group is $2/m$, with elements $\{E, I, 2_{y}, m_{y}\}$. The SO-free irrep $\Gamma_{\mathbf{N}}$ is $B_{g}$, and $aeV$ decomposes as $A_{g} \oplus 2B_{2g}$, implying that $aeV\otimes \Gamma_{\mathbf{N}}$ decomposes as $B_{g} \oplus 2A_{g}$, with $N_{x}$ and $N_{z}$ belonging to $A_{g}$ and $N_{y}$ belonging to $B_{g}$. In Table~\ref{tab:CuF2_paramag_irrs} we show full co-irreps corresponding to $A_{g}$ and $B_{g}$ in $2/m + \tau 2/m$.

If all three components $N_{i}$ are non-zero, the only possible group that may describe the magnetic order is $\overline{1} = \{E,I\}$, as this is the intersection of trivially represented elements in $A_{g}$ and $B_{g}$. If only $N_{x}$ and $N_{z}$ are non-zero, then the trivially represented elements in $A_{g}$ define the ordered phase symmetry group, $2/m = \{E, I, 2_{y}, m_{y}\}.$ If, on the other hand, only $N_{y}$ is non-zero then $2'/m' = \{E, I, \tau 2_{y}, \tau m_{y}\}$ defines the symmetry of the ordered phase. All three of these cases are encapsulated by our SOC Landau theory, so it is not in conflict with an approach centered on magnetic groups.

\begin{table}[t!]
\caption{\label{tab:CuF2_paramag_irrs}%
Full co-irreps of $2/m + \tau 2/m$ corresponding to the point group irreps $A_{g}$ and $B_{g}$.}
\centering
\begin{tabularx}{\columnwidth}{>{\centering\arraybackslash}X  >{\centering\arraybackslash}X  >{\centering\arraybackslash}X  >{\centering\arraybackslash}X >{\centering\arraybackslash}X >{\centering\arraybackslash}X >{\centering\arraybackslash}X >{\centering\arraybackslash}X >{\centering\arraybackslash}X}
\hline
\hline
 & $E$ & $I$ &
$2_y$ & $m_y$ & $\tau$ & $\tau I$ & $\tau 2_{y}$ & $\tau m_{y}$\\
\hline
$A_{g}$  & 1 & 1 & 1 & 1 & -1 & -1 & -1 & -1 \\
$B_{g}$ & 1 & 1 & -1 & -1 & -1 & -1 & 1 & 1 \\
\hline
\hline
\end{tabularx}
\end{table}

\section{\label{app:symmTensorPowers}Symmetrization of tensor powers}

We summarize a well-known procedure for the symmetrization of tensor powers of any representation. 
Here, let $\mathbf{G}$ be any discrete group and let $D$ be a representation of $\mathbf{G}$ in some vector space $V$ with a basis $\{ \,|i\rangle\,|\,i\in\{1,...,\mathrm{dim}(V)\}\}.$ The $n$-th tensor power of $D$, denoted $D^{n}$, is a representation of $\mathbf{G}$ in the $n$-th Cartesian product of $V$, $V^{\times n} \equiv V\times ...\times V$. The basis in $V^{n}$ is $\{|i_{1}\rangle\otimes|i_{2}\rangle\otimes...\otimes|i_{n}\rangle\,|\, i_{1},i_{2},...,i_{n} \in \{1,...,\mathrm{dim}(V)\}\}.$ Symmetrizing means `equally representing' vectors that differ only by permutations of the components in different Cartesian factors of $V$. By this we mean that that the vectors $|i_{1}\rangle\otimes|i_{2}\rangle\otimes...\otimes|i_{n}\rangle$ and $|i_{\pi^{-1}(1)}\rangle\otimes|i_{\pi^{-1}(2)}\rangle\otimes...\otimes|i_{\pi^{-1}(n)}\rangle$ are treated as equivalent, where $\pi$ is an element of the permutation group on $n$ elements, $\mathcal{S}_{n},$ that is $$\pi = \begin{pmatrix}
1 & 2 & ... & N\\
\pi(1) & \pi(2) & ... & \pi(n)
\end{pmatrix} \in \mathcal{S}_{n}.
$$ This equivalence is achieved by projecting into the subspace of $V^{n}$ spanned by vectors transforming under the trivial irrep of $\mathcal{S}_{n}$. This projector is given~\cite{DamnjanovicSymm} by $P^{(+)}_{n} = \frac{1}{n!} \sum_{\pi\in\mathcal{S}_{n}} 1\cdot d^{(n)}(\pi)$, where $d^{(n)}(\pi)$ represents $\pi$ in $V^{n}$ by 
\begin{align*}
    d^{(n)}(\pi) |i_{1}\rangle&\otimes|i_{2}\rangle\otimes...\otimes|i_{N}\rangle\\
    &= |i_{\pi^{-1}(1)}\rangle\otimes|i_{\pi^{-1}(2)}\rangle\otimes...\otimes|i_{\pi^{-1}(N)}\rangle.
\end{align*}
Then, the symmetrized tensor power $[D^{n}]$ of $D$ is given by $$[D^{n}] = \left\{ P^{(+)}_{n} D^{n}(g) | g\in\mathbf{G}\right\}.$$ This is the technique used to calculate the symmetrized $n$-th tensor power of the polar vector representation when studying the multipoles.

The characters of symmetrized $n$-th tensor power representations can be easily computed using the ``bird-tracks" method~\cite{BirdTracks}. Up to $n=6,$ the character of an element $g\in\mathbf{G}$ in the $n$-th symmetrized tensor power, $\chi\left([D^{n}](g)\right)$, is given by 

\begin{align}\label{eq:chars1}
\chi\left([D^{ 2}](g)\right) &=  \frac{1}{2!}\left( (\chi(g))^{2} + \chi(g^{2}) \right) \nonumber \\
\chi\left([D^{ 3}](g)\right) &= \frac{1}{3!}\left( (\chi(g))^{3} + 3 \chi(g)\chi(g^{2}) + 2 \chi(g^{3}) \right) \\
\chi\left([D^{ 4}](g)\right) &= \frac{1}{4!}\left( (\chi(g))^{4} + 6(\chi(g))^{2}\chi(g^{2})  \right. \nonumber\\
&\quad \left.+\, 8 \chi(g)\chi(g^{3}) + 3(\chi(g^{2}))^{2} + 6\chi(g^{4}) \right)\nonumber\\
\chi\left([D^{ 5}](g)\right) &= \frac{1}{5!}\left( (\chi(g))^{5} + 10 (\chi(g))^{3}\chi(g^{2}) \right.\nonumber\\
&\quad \left.+ \,15 \chi(g) (\chi(g^{2}))^{2} + 20 (\chi(g))^{2}\chi(g^{3})+\right. \nonumber\\
&\quad \left. 20 \chi(g^{2})\chi(g^{3}) + 30 \chi(g)\chi(g^{4}) + 24 \chi(g^{5})\right) \nonumber 
\end{align}

\begin{align}
\chi\left([D^{ 6}](g)\right) &= \frac{1}{6!}\left( (\chi(g))^{6} + 15 (\chi(g))^{4}\chi(g^{2}) +\right. \nonumber\\
&\quad\left.45 (\chi(g))^{2}(\chi(g^{2}))^{2} + 15 (\chi(g^{2}))^{3} + \right. \nonumber \\
&\quad\left.40 (\chi(g))^{3}(\chi(g^{3}))+ 120\chi(g)\chi(g^{2})\chi(g^{3})\right. \nonumber\\
&\quad\left.+\, 40 (\chi(g^{3}))^{2} + 90 (\chi(g))^{2} \chi(g^{4})\right. \nonumber\\
&\quad\left. +\,90 \chi(g^{2})\chi(g^{4}) + 144 \chi(g)\chi(g^{5})\right. \nonumber\\
&\quad\left. + \,120 \chi(g^{6})\right) 
\label{eq:chars2}
\end{align}
in terms of the characters $\chi(D(g)) \equiv \chi(g)$ of the original representation $D$. With these character relations, one can demonstrate that the symmetrized tensor powers of $\Gamma_{l=1}^{(s)}$ have the following decompositions: 
\begin{align*}
[\Gamma_{l=1}^{(s)\,\otimes 2}] &= \Gamma_{l=0}^{(s)} \oplus \Gamma_{l=2}^{(s)}\\
[\Gamma_{l=1}^{(s)\,\otimes 3}] &= \Gamma_{l=1}^{(s)} \oplus \Gamma_{l=3}^{(s)} \\
[\Gamma_{l=1}^{(s)\,\otimes 4} ] &= \Gamma_{l=0}^{(s)} \oplus \Gamma_{l=2}^{(s)} \oplus \Gamma_{l=4}^{(s)}\\
[\Gamma_{l=1}^{(s)\,\otimes 5} ] &= \Gamma_{l=1}^{(s)} \oplus \Gamma_{l=3}^{(s)} \oplus \Gamma_{l=5}^{(s)}\\
[\Gamma_{l=1}^{(s)\,\otimes 6} ] &= \Gamma_{l=0}^{(s)} \oplus \Gamma_{l=2}^{(s)} \oplus \Gamma_{l=4}^{(s)} \oplus \Gamma_{l=6}^{(s)}.
\end{align*}
Only odd symmetrized tensor powers contain the $\Gamma_{l=1}^{(s)}$ representation, and so only these could couple to $\mathbf{N}.$ As we are looking for the minimal such multipole in the SOC-free limit, we can focus exclusively on the $\Gamma_{l=1}^{(s)}$ multipole, corresponding to $\mathbf{m}(\mathbf{r})$ in the integrand of Eq.~\ref{eq:multipoleDef} in \ref{subsec:AMLTOverview}. 

The character relations Eqs.~\ref{eq:chars1} and \ref{eq:chars2} also allow us to quickly decompose the characters of the symmetrized polar vector powers $[V^{n}]$ describing the spatial transformation properties of the SO-free multipoles of Sec.~\ref{subsec:AMLTOverview}.

\section{\label{app:SAB}Tensor \& multipole components coupling to $\mathbf{N}$}

Here we provide a brief overview of the well-known group projector techniques~\cite{DamnjanovicSymm, MDamnjanovicModGProj,BradleyCracknell,GTPackBook,KosterGroupProj, QGMGpThr} used to find the symmetry-adapted basis (SAB) any representation. We have used this technique to identify the multipole components coupling to $\mathbf{N}$ in the absence of SOC in Sec.~\ref{subsec:AMLTOverview}, as well as the components of tensors coupling to $\mathbf{N}$ when SOC is included as discussed in Sec.~\ref{sec:finiteSOC}. These couplings are all summarized in Table~\ref{tab:CouplingComps}.

A group projector $P_{11}^{\Gamma}(D)$ for a representation $D$ onto an irrep $\Gamma$ of the group $\mathbf{G}$ is given by 
\begin{equation}\label{eq:gpProj}
P_{11}^{\Gamma}(D) = \frac{|\Gamma|}{|\mathbf{G}|} \sum_{g\in\mathbf{G}} \Gamma_{11}^{*}(g) D(g),
\end{equation}
and it is non-zero provided $\Gamma$ is present in the decomposition of $D$. In Eq.~\ref{eq:gpProj}, $\Gamma_{11}$ is the matrix element of $\Gamma$ in the first row and first column. If the dimension $|\Gamma|$ of $\Gamma$ is one, then the SAB for the irrep $\Gamma$ of $D$ is given by a basis in the image of $P_{11}^{\Gamma}(D).$ If $\Gamma$ has dimension $|\Gamma| > 1$, then the SAB for $\Gamma$ will be given by a basis the image of $P_{11}^{\Gamma}(D),$ as well as those vectors obtained by acting on the previous vectors with each of the group operators $$P_{m1}^{\Gamma}(D) = \frac{|\Gamma|}{|\mathbf{G}|} \sum_{g\in\mathbf{G}} \Gamma_{m1}^{*}(g) D(g),$$ where $m \in \{2,.., |\Gamma|\}.$ To apply this procedure to $[V^{n}]$, for example, we first express these operators in matrix form in a vector space where each standard basis vector corresponds to one unique combination of $x,$ $y$, and $z$ of order $n$ (i.e. for $N\!=\!4$, $x^{2}yz$ is one basis vector, as opposed to distinct vectors for $xxyz$, $xyxz$, $xyzx$, $yxzx$, and $yzxx$). This step can be achieved for any power $n$ using the symmetrization procedure outlined in Appendix \ref{app:symmTensorPowers} on the space $\mathbb{R}^{3n}$, with basis elements given by ordered strings with characters $x$, $y$ or $z$. Then, the SAB vectors for $[V^{n}]$ will represent symmetrized polynomials of order $n$ that transform under the irrep $\Gamma$ of the point group. 

\section{\label{app:repackaging}``Repackaging" Tensor Components}

\newcommand{\fakesec}[1]{\begin{center} #1 \end{center}}

To produce Table~\ref{tab:properties}, we utilize the MTENSOR~\cite{mtensor} tables on the Bilbao Crystallographic Server. For each tensor type, we verify whether it is possible to ``repackage" the components of a tensor into a quantity transforming as one of the $d$ listed in Table~\ref{tab:guaranteedsoc}. Here we outline the various types of ``repacking" we can do, demonstrating specific examples. The definitions and transformation properties of the full tensors are discussed in Ref.~\cite{mtensor}.

\fakesec{
\textbf{Case 1:} $[V^{2}]^{*}\rightarrow a\{V^{2}\} \rightarrow aeV$
}

The classic example is repackaging the antisymmetric part of a $[V^{2}]^{*}$ tensor, which transforms as $a\{V^{2}\}$ into a magnetic axial vector $aeV.$ Such an example is that of the electrical conductivity, with defining equation $J_{i} = \sigma_{ij}E_{j}$. Using Onsager's reciprocity, under time-reversal symmetry $\tau$ the components of $\sigma_{ij}$ are related by $\tau\sigma_{ij} = \sigma_{ji}$, which gives it the $[ \,\,\cdot\,\,]^{*}$ unconventional Jahn symbol~\cite{mtensor}. The antisymmetric part of this tensor $\sigma_{ij}^{A} = \frac{1}{2}(\sigma_{ij} - \sigma_{ji})$ is a $a\{V^{2}\}$ tensor (where $\{ \,\,\cdot\,\,\}$ denotes antisymmetrization), responsible for the anomalous Hall conductivity. There are three independent tensor components $\sigma_{yz},$ $\sigma_{zx}$ and $\sigma_{xy}$ which we may arrange into a vector $\{\sigma_{yz},\sigma_{zx},\sigma_{xy}\}$ to form a magnetic axial vector $aeV$. To relate the rank two and rank one objects, we use the identity $\sigma^{A}_{\alpha} = \frac{1}{2}\varepsilon_{\alpha ij}\sigma^{A}_{ij}.$ We will use analogs of this identity for larger tensor quantities.

\fakesec{
\textbf{Case 2:} $(V^{2})^{*} \rightarrow a\{V^{2}\} \rightarrow aeV$}

This case is similar to Case 1, except that the initial is not related to itself under time-reversal but rather to another tensor quantity. A classic example of this case is that of the Peltier $\pi_{ij}$ and Seebeck $\beta_{ij}$ tensors, where by Onsager's reciprocity $\pi_{ij}$ is related to $\beta_{ij}$ under $\tau$ by $\tau \pi_{ij} = \beta_{ji}$ and vice versa. In this case, we first take the antisymmetric parts of each of these tensors, $\pi_{ij}^{A} = \frac{1}{2}(\pi_{ij} - \pi_{ji})$ and $\beta_{ij}^{A} = \frac{1}{2}(\beta_{ij} - \beta_{ji}).$ Next, we define a \emph{symmetric} combination of these two tensors: $\tilde{S}_{ij} =\frac{1}{2}(\pi_{ij}^{A} + \beta_{ij}^{A}).$ This tensor transforms as an $a\{V^{2}\}$ object, so by Case 1 we can repackage this into a magnetic axial vector $aeV$ by $\tilde{S}_{\alpha} = \frac{1}{2}\varepsilon_{\alpha ij}\tilde{S}_{ij}.$

\fakesec{
\textbf{Case 3:} $a\{V^{2}\}[V^{2}] \rightarrow aeV[V^{2}]$}

This case is a direct consequence of Case 1. An example is that of the Quadratic magneto-optic Kerr tensor $C_{ijkl}^{A}$. This tensor is defined as the antisymmetric part of the Cotton-Moutton tensor~\cite{mtensor}. By Onsager's relation, under $\tau$ the components are related by $\tau C_{ijkl}^{A} = -C_{jikl}^{A}.$ Using the Levi-Civita identity from Case 1, we obtain an $aeV[V^{2}]$ tensor via $C_{\alpha kl}^{A} = \frac{1}{2}\varepsilon_{\alpha ij} C_{ijkl}^{A}.$

\fakesec{
\textbf{Case 4:} $e\{V^{2}\}^{*}V \rightarrow ae[V^{2}]V$}

For tensors of type $e\{V^{2}\}^{*}V$ such as the magnetoresistance tensor $R_{ijk}$, the tensor symmetrized under exchange of the first two indices $R_{ijk}^{S} = \frac{1}{2}(R_{ijk} + R_{jik})$ transforms as a $ae[V^{2}]V$ tensor.

\fakesec{
\textbf{Case 5:} $(eV^{3})^{*}\rightarrow ae[V^{2}]^{*}V\rightarrow ae[V^{2}]V$}

Tensors transforming as an $(eV^{3})^{*}$ object such as the Ettinghausen $M_{ijk}$ and Nernst $N_{ijk}$ tensors are related by Onsagers relation under time-reversal symmetry: $\tau M_{ijk} = -N_{jik}$ and vice versa. We first extract the components of these tensors symmetrized under the first two indices, $M_{ijk}^{S} = \frac{1}{2}(M_{ijk} + M_{jik})$ and $N_{ijk} = \frac{1}{2}(N_{ijk} + N_{jik}),$ which both transform as $ae[V^{2}]^{*}V$ tensors. Then, we define a symmetric combination of these components: $\mathcal{S}_{ijk} = \frac{1}{2}(M_{ijk}^{S} + N_{ijk}^{S})$, which will now transform as an $ae[V^{2}]V$ tensor.

\fakesec{
\textbf{Case 6:} $[V^{2}]^{*}V^{2}\rightarrow a\{V^{2}\}V^{2}\rightarrow aeV$}

For tensors of the form $[V^{2}]^{*}V^{2}$ such as the magnetic resistance tensor $T_{ijkl}$ we apply the argument from Case 1. We first extract the component antisymmetric under exchange of the first two indices $T_{ijkl}^{A} = \frac{1}{2}(T_{ijkl} - T_{jikl})$ that transforms as a $a\{V^{2}\}[V^{2}]$ tensor. Then we use the Levi-Civita identity from Case 1 to obtain $T_{\alpha kl}^{A} = \frac{1}{2} \varepsilon_{\alpha ij} T_{ijkl}^{A}.$ This tensor transforms as $aeV[V^{2}]$.

\fakesec{
\textbf{Case 7:} $(V^{2}[V^{2}])^{*} \rightarrow ([V^{2}][V^{2}])^{*}\rightarrow aeV[V^{2}]$}

Quantities such as the magneto-Peltier $P_{ijkl}$ and magneto-Seebeck $\alpha_{ijkl}$ tensors are related to each other under time-reversal symmetry by Onsager's relations, $\tau \alpha_{ijkl} = P_{jikl}.$ We first extract the components of these tensors that are antisymmetric under exchange of the first two indices, $\alpha^{A}_{ijkl} = \frac{1}{2}(\alpha_{ijkl} - \alpha_{jikl})$ and $P_{ijkl}^{A} = \frac{1}{2}(P_{ijkl} - P_{jikl}),$ which transform as $([V^{2}][V^{2}])^{*}$ quantities. Then, we define an antisymmetric combination of these two tensors: $\tilde{\mathcal{A}} = \frac{1}{2}(\alpha_{ijkl}^{A} - P_{ijkl}^{A}),$ transforming as $a\{V^{2}\}[V^{2}].$ Finally, we use the identity from Case 1 to express this tensor as an $aeV[V^{2}]$ object, $\tilde{\mathcal{A}}_{\alpha kl} = \frac{1}{2}\varepsilon_{\alpha ij}\tilde{\mathcal{A}}_{ijkl}.$

\clearpage

\onecolumngrid
\section{\label{app:wpGNtable} Table of space groups and Wyckoff positions supporting altermagnetic order}

\begin{table}[h!]
 \caption{\label{tab:wpGN} Space group Wyckoff positions supporting altermagnetism, and the irreps $\Gamma_{\mathbf{N}}$ under which $\mathbf{N}$ transforms.}
\begin{minipage}{.45\columnwidth}
\centering
\begin{tabular}{|c|ccc|}
\hline
\hline
\rowcolor{dgray} PG & SG & WP & $\Gamma_{\mathbf{N}}$ \\
\hline
\hline
 $\mathbf{2}$ & $3$ & $\{2e\}$ & $\{B\}$ \\
 & $4$ & $\{2a\}$ & $\{B\}$ \\
 & $5$ & $\{4c, 2b, 2a\}$ & $\{B\}$ \\
 \hline
\rowcolor{mygray}$\mathbf{m}$ & $6$ & $\{2c\}$ & $\{A''\}$ \\
\rowcolor{mygray}& $7$ & $\{2a\}$ & $\{A''\}$ \\
\rowcolor{mygray}& $8$ & $\{4b\}$ & $\{A''\}$ \\
\rowcolor{mygray}& $9$ & $\{4a\}$ & $\{A''\}$ \\
\hline
$\mathbf{2/m}$ & $10$ & $\{4o\}$ & $\{B_{g}\}$ \\
 & $11$ & $\{4f, 2d, 2c, 2b, 2a\}$ & $\{B_{g}\}$ \\
& $12$ &  $\{8j, 4h, 4g, 4f, 4e, 2d, 2b\}$ & $\{B_{g}\}$ \\
& $13$ & $\{4g, 2d, 2c, 2b, 2a\}$ & $\{B_{g}\}$ \\
& $14$ & $\{4e, 2d, 2c, 2b, 2a\}$ & $\{B_{g}\}$ \\
& $15$ & $\{8f, 4e, 4d, 4c, 4b, 4a\}$ & $\{B_{g}\}$ \\
\hline
\rowcolor{mygray}$\mathbf{222}$ & $16$ &         $\{4u\}$             & $\{B_{1}, B_{3}, B_{2}\}$ \\
\rowcolor{mygray}                          &         &  $\{2t, 2s, 2r, 2q\}$     & $\{B_{1}\}$\\
\rowcolor{mygray}                          &         & $\{2p, 2o, 2n, 2m\}$ & $\{B_{2}\}$\\
\rowcolor{mygray}                          &         &  $\{2l, 2k, 2j, 2i\}$      & $\{B_{3}\}$ \\ 
\rowcolor{mygray} & $17$ &     $\{4e\}$    & $\{B_{1}, B_{3}, B_{2}\}$ \\
\rowcolor{mygray}  &         & $\{2d, 2c\}$  & $\{B_{2}\}$ \\
\rowcolor{mygray}  &         & $\{2b, 2a\}$  & $\{B_{3}\}$ \\
\rowcolor{mygray}  & $18$ &     $\{4c\}$   & $\{B_{1}, B_{3}, B_{2}\}$ \\
\rowcolor{mygray}  &          & $\{2b, 2a\}$ & $\{B_{1}\}$\\
\rowcolor{mygray} & $19$ & $\{4a\}$ & $\{B_{1}, B_{3}, B_{2}\}$ \\ 
\rowcolor{mygray} & $20$ & $\{8c, 4b\}$ & $\{B_{1}, B_{3}, B_{2}\}$ \\
\rowcolor{mygray}  &          & $\{4a\}$ & $\{B_{3}\}$ \\
\rowcolor{mygray} & $21$ & $\{8l, 4k, 4h, 4g\}$ & $\{B_{1}, B_{3}, B_{2}\}$ \\
\rowcolor{mygray}  &          & $\{4j, 4i, 2b\}$ & $\{B_{1}\}$\\
\rowcolor{mygray}  &          & $\{4f, 4e\}$     & $\{B_{3}\}$ \\
\rowcolor{mygray} & $22$ & $\{16k, 8j, 8i, 8h, 8g, 8f, 8e\}$ & $\{B_{1}, B_{3}, B_{2}\}$ \\
\rowcolor{mygray}  &          &                     $\{4d, 4b\}$        &       $\{B_{1}\}$ \\
\rowcolor{mygray} & $23$ & $\{8k, 4j, 4i, 4f, 2c\}$ & $\{B_{1}, B_{3}, B_{2}\}$ \\
\rowcolor{mygray}  &          &$\{4h, 4g\}$ & $\{B_{2}\}$\\
\rowcolor{mygray}  &          &$\{4e\}$       & $\{B_{3}\}$ \\
\rowcolor{mygray} & $24$  & $\{8d, 4c, 4a\}$ & $\{B_{1}, B_{3}, B_{2}\}$ \\
\rowcolor{mygray}  &          &        $\{4b\}$     &   $\{B_{2}\}$ \\
 \hline
$\mathbf{mm2}$ & $25$ &      $\{4i\}$    & $\{A_{2}, B_{2}, B_{1}\}$ \\
                           &          &  $\{2f, 2e\}$  & $\{B_{1}\}$\\
                           &          &  $\{2h, 2g\}$ & $\{B_{2}\}$ \\
                           & $26$ &    $\{4c\}$     & $\{A_{2}, B_{2}, B_{1}\}$ \\
                           &          & $\{2b, 2a\}$  & $\{B_{2}\}$ \\
                           & $27$ &              $\{4e\}$      & $\{A_{2}, B_{2}, B_{1}\}$ \\
                           &          & $\{2d, 2c, 2b, 2a\}$ & $\{ A_{2} \}$ \\          
\hline
                \end{tabular}
\end{minipage}\hfill 
\begin{minipage}{.45\columnwidth}
                \centering
                \begin{tabular}{|c|ccc|}
                \hline
                \hline
                \rowcolor{dgray} PG & SG & WP & $\Gamma_{\mathbf{N}}$ \\
                \hline
                \hline 
                  $\mathbf{mm2}$  & $28$ & $\{4d\}$       & $\{A_{2}, B_{2}, B_{1}\}$ \\
                           &          & $\{2b, 2a\}$ & $\{A_{2}\}$\\
                           &          &  $\{2c\}$      & $\{B_{2}\}$ \\  
                  & $29$ &   $\{4a\}$     & $\{A_{2}, B_{2}, B_{1}\}$ \\  
                    & $30$ &  $\{4c\}$      & $\{A_{2}, B_{2}, B_{1}\}$ \\
                           &          & $\{2b, 2a\}$ & $\{A_{2}\}$ \\
                           & $31$ & $\{4b\}$ & $\{A_{2}, B_{2}, B_{1}\}$ \\
                           &          & $\{2a\}$ & $\{B_{2}\}$ \\
                               & $32$ &      $\{4c\}$   & $\{A_{2}, B_{2}, B_{1}\}$ \\
                           &          & $\{2b, 2a\}$  & $\{A_{2}\}$ \\  
                            & $33$ & $\{4a\}$ & $\{A_{2}, B_{2}, B_{1}\}$ \\  
                           & $34$ & $\{4c\}$       & $\{A_{2}, B_{2}, B_{1}\}$ \\
                           &          & $\{2b, 2a\}$ & $\{A_{2}\}$ \\
                           & $35$ & $\{8f, 4e, 4c\}$ & $\{A_{2}, B_{2}, B_{1}\}$ \\
                           &          & $\{2b\}$            & $\{A_{2}\}$\\
                           &          & $\{4d\}$            & $\{B_{1}\}$ \\
                    & $36$ & $\{8b, 4a\}$  & $\{A_{2}, B_{2}, B_{1}\}$ \\  
                            & $37$ &  $\{8d, 4c\}$ & $\{A_{2}, B_{2}, B_{1}\}$ \\
                           &          &  $\{4b, 4a\}$ & $\{A_{2}\}$ \\
                           & $38$ & $\{8f\}$         & $\{A_{2}, B_{2}, B_{1}\}$ \\
                           &          &  $\{4c\}$       & $\{B_{1}\}$\\
                           &          &  $\{4e, 4d\}$ & $\{B_{2}\}$ \\
                         & $39$ & $\{8d, 4c\}$  & $\{A_{2}, B_{2}, B_{1}\}$ \\
                           &          &  $\{4b, 4a\}$ & $\{A_{2}\}$ \\
                        & $40$ & $\{8c\}$  & $\{A_{2}, B_{2}, B_{1}\}$ \\
                          &          &  $\{4a\}$ & $\{A_{2}\}$\\
                          &          &  $\{4b\}$ & $\{B_{2}\}$\\
                          & $41$ & $\{8b\}$  & $\{A_{2}, B_{2}, B_{1}\}$ \\
                          &          &  $\{4a\}$ & $\{A_{2}\}$ \\
                           & $42$ & $\{16e, 8d, 8c, 8b, 4a\}$ & $\{A_{2}, B_{2}, B_{1}\}$ \\ 
                          & $43$ & $\{16b, 8a\}$ & $\{A_{2}, B_{2}, B_{1}\}$ \\ 
                          & $44$ & $\{8e, 4d, 4c, 2b, 2a\}$ & $\{A_{2}, B_{2}, B_{1}\}$ \\ 
                           & $45$ &  $\{8c, 4b, 4a\}$ & $\{A_{2}, B_{2}, B_{1}\}$ \\ 
                           & $46$ &  $\{8c, 4b, 4a\}$ & $\{A_{2}, B_{2}, B_{1}\}$ \\ 
 \hline
\rowcolor{mygray}  $\mathbf{mmm}$ & $47$ & $\{4z, 4y\}$  & $\{B_{1g}\}$ \\
\rowcolor{mygray}                             &           & $\{8a\}$       & $\{B_{1g}, B_{3g}, B_{2g}\}$\\
\rowcolor{mygray}                             &           & $\{4x, 4w\}$ & $\{B_{2g}\}$ \\
\rowcolor{mygray}                             &           & $\{4v, 4u\}$  & $\{B_{3g}\}$ \\
\rowcolor{mygray} 	   	    	   & $48$ & $\{8m, 4f, 4e\}$ & $\{B_{1g}, B_{3g}, B_{2g}\}$ \\
\rowcolor{mygray}                             &           & $\{4l, 4k\}$   & $\{B_{1g}\}$ \\
\rowcolor{mygray}                             &           & $\{4j, 4i\}$    & $\{B_{2g}\}$ \\
\rowcolor{mygray}                             &           & $\{4h, 4g\}$  & $\{B_{3g}\}$ \\
\hline
 \end{tabular}
 \end{minipage}
 \end{table}

 \begin{table*}
 \begin{minipage}{.45\columnwidth}
\centering
\begin{tabular}{|c|ccc|}
\hline
\hline
\rowcolor{dgray}PG & SG & WP & $\Gamma_{\mathbf{N}}$ \\
\hline
\hline
\rowcolor{mygray} $\mathbf{mmm}$   & $49$ & $\{8r\}$ & $\{B_{1g}, B_{3g}, B_{2g}\}$ \\
\rowcolor{mygray}                             &           & $\{4q, 4p, 4o, 4n, 4m, 2d, 2c, 2b, 2a\}$ & $\{B_{1g}\}$\\
\rowcolor{mygray}                             &           & $\{4l, 4k\}$  & $\{B_{2g}\}$ \\
 \rowcolor{mygray}                            &           & $\{4j, 4i\}$   & $\{B_{3g}\}$ \\
\rowcolor{mygray} 	   & $50$ & $\{8m, 4f, 4e\}$ & $\{B_{1g}, B_{3g}, B_{2g}\}$ \\
\rowcolor{mygray}                             &           & $\{4l, 4k\}$  & $\{B_{1g}\}$ \\
 \rowcolor{mygray}                            &           & $\{4j, 4i\}$   & $\{B_{2g}\}$ \\
 \rowcolor{mygray}                            &           & $\{4h, 4g\}$ & $\{B_{3g}\}$ \\
 \rowcolor{mygray}                            & $51$  & $\{8l\}$ & $\{B_{1g}, B_{3g}, B_{2g}\}$ \\
  \rowcolor{mygray}                           &           & $\{4j, 4i, 4h, 4g, 2d, 2c, 2b, 2a\}$ & $\{B_{2g}\}$ \\
  \rowcolor{mygray}                           &           & $\{4k\}$     &    $\{B_{3g}\}$ \\
\rowcolor{mygray} 			   & $52$ & $\{8e, 4b, 4a\}$ & $\{B_{1g}, B_{3g}, B_{2g}\}$ \\
  \rowcolor{mygray}                           &           & $\{4c\}$ &   $\{B_{1g}\}$ \\
  \rowcolor{mygray}                           &           & $\{4d\}$ & $\{B_{3g}\}$ \\
\rowcolor{mygray} 			   & $53$ & $\{8i\}$   & $\{B_{1g}, B_{3g}, B_{2g}\}$ \\
 \rowcolor{mygray}                            &           & $\{4g\}$ & $\{B_{2g}\}$ \\
 \rowcolor{mygray}                            &           & $\{4h, 4f, 4e, 2d, 2c, 2b, 2a\}$ & $\{B_{3g}\}$ \\
\rowcolor{mygray} 			   &$54$ &  $\{8f, 4b, 4a\}$ & $\{B_{1g}, B_{3g}, B_{2g}\}$ \\
 \rowcolor{mygray}                            &           & $\{4e, 4d\}$     & $\{B_{1g}\}$ \\
 \rowcolor{mygray}                            &           & $\{4c\}$  	& $\{B_{2g}\}$ \\
 \rowcolor{mygray} 			   & $55$ & $\{8i\}$ & $\{B_{1g}, B_{3g}, B_{2g}\}$ \\
 \rowcolor{mygray}                            &           & $\{4h, 4g, 4f, 4e, 2d, 2c, 2b, 2a\}$ & $\{B_{1g}\}$ \\
 \rowcolor{mygray} 			   & $56$ & $\{8e, 4b, 4a\}$ & $\{B_{1g}, B_{3g}, B_{2g}\}$ \\
 \rowcolor{mygray}                            &           & $\{4d, 4c\}$       & $\{B_{1g}\}$ \\
\rowcolor{mygray} 			   & $57$ & $\{8e, 4b, 4a\}$ & $\{B_{1g}, B_{3g}, B_{2g}\}$ \\
\rowcolor{mygray}                             &           &  $\{4d\}$   	  & $\{B_{1g}\}$  \\
\rowcolor{mygray}                             &           & $\{4c\}$ 	 	  & $\{B_{3g}\}$ \\
\rowcolor{mygray} 	   & $58$ & $\{8h\}$ & $\{B_{1g}, B_{3g}, B_{2g}\}$ \\
\rowcolor{mygray}                             &           &  $\{4g, 4f, 4e, 2d, 2c, 2b, 2a\}$ & $\{B_{1g}\}$ \\
\rowcolor{mygray}  			   & $59$ & $\{8g, 4d, 4c\}$ & $\{B_{1g}, B_{3g}, B_{2g}\}$ \\
\rowcolor{mygray}                             &           & $\{4f\}$  &   $\{B_{2g}\}$ \\
\rowcolor{mygray}                             &           & $\{4e\}$ &   $\{B_{3g}\}$ \\
\rowcolor{mygray} 			   & $60$ & $\{8d, 4b, 4a\}$ & $\{B_{1g}, B_{3g}, B_{2g}\}$ \\
\rowcolor{mygray}                             &           & $\{4c\}$   &  $\{B_{2g}\}$ \\
\rowcolor{mygray} 			   & $61$ & $\{8c, 4b, 4a\}$ & $\{B_{1g}, B_{3g}, B_{2g}\}$ \\  
\rowcolor{mygray} 			   & $62$ & $\{8d, 4b, 4a\}$ & $\{B_{1g}, B_{3g}, B_{2g}\}$ \\  
\rowcolor{mygray}                             &           & $\{4c\}$   & $\{B_{2g}\}$ \\
\rowcolor{mygray} 			   & $63$ & $\{16h, 8f, 8d, 4b\}$ & $\{B_{1g}, B_{3g}, B_{2g}\}$ \\
\rowcolor{mygray}                             &           & $\{8g, 4c\}$  &  $\{B_{1g}\}$ \\
\rowcolor{mygray}                             &           & $\{8e, 4a\}$  & $\{B_{3g}\}$ \\
\rowcolor{mygray}                & $64$ & $\{16g, 8f, 8e, 8c\}$ & $\{B_{1g}, B_{3g}, B_{2g}\}$ \\
\rowcolor{mygray}                             &           & $\{8d, 4b, 4a\}$ & $\{B_{3g}\}$ \\
\rowcolor{mygray} 			   & $65$ & $\{16r, 8n, 8m\}$ & $\{B_{1g}, B_{3g}, B_{2g}\}$ \\
\rowcolor{mygray}                             &           & $\{8q, 8p, 4l, 4j, 4i, 4f, 4e\}$ & $\{B_{1g}\}$ \\
\rowcolor{mygray}                             &           & $\{8o\}$ & $\{B_{2g}\}$ \\

\rowcolor{mygray}  		   & $66$ & $\{16m, 8k, 8h\}$ & $\{B_{1g}, B_{3g}, B_{2g}\}$ \\
\rowcolor{mygray}                             &           & $\{8l, 8j, 8i, 4f, 4e, 4d, 4c, 4b\}$ & $\{B_{1g}\}$\\
\rowcolor{mygray}                             &           & $\{8g\}$  & $\{B_{3g}\}$ \\
\hline
\end{tabular}
 
\end{minipage}\hfill 
\begin{minipage}{.45\columnwidth}
\centering
\begin{tabular}{|c|ccc|}
\hline
\hline
\rowcolor{dgray}PG & SG & WP & $\Gamma_{\mathbf{N}}$ \\
\hline
 \hline
  \rowcolor{mygray}  $\mathbf{mmm}$	   & $67$ & $\{16o, 8n, 8m, 8k, 8j, 4g, 4f, 4e\}$ & $\{B_{1g}, B_{3g}, B_{2g}\}$ \\
\rowcolor{mygray}                             &           & $\{8l\}$  & $\{B_{1g}\}$ \\
\rowcolor{mygray}                             &           & $\{8i, 8h, 4d, 4c\}$ & $\{B_{3g}\}$ \\
\rowcolor{mygray}  			   & $68$ & $\{16i, 8g, 8f, 8e, 8d, 8c, 4b, 4a\}$ & $\{B_{1g}, B_{3g}, B_{2g}\}$ \\
\rowcolor{mygray}                             &           &  $\{8h\}$  &   $\{B_{1g}\}$ \\
\rowcolor{mygray}  			   & $69$ &  $\shortstack{$\{32p, 16o, 16n, 16m,$ \\ $16l, 16k, 16j, 8e, 8d, 8c\}$}$ &  $\{B_{1g}, B_{3g}, B_{2g}\}$ \\
\rowcolor{mygray}                             &           & $\{8i, 4b\}$  &    $\{B_{1g}\}$   \\
\rowcolor{mygray}                             &           & $\{8h\}$       & $\{B_{2g}\}$   \\
\rowcolor{mygray}                             &           & $\{8g\}$       & $\{B_{3g}\}$ \\
 \rowcolor{mygray} 			   & $70$ & $\{32h, 16g, 16f, 16e, 16d, 16c\}$ & $\{B_{1g}, B_{3g}, B_{2g}\}$ \\
\rowcolor{mygray}                             &           & $\{8b\}$  &  $\{B_{1g}\}$ \\
\rowcolor{mygray} 			   & $71$ & $\{16o, 8m, 8l, 8k, 4j, 4i, 4h, 2d, 2b\}$ & $\{B_{1g}, B_{3g}, B_{2g}\}$ \\
 \rowcolor{mygray}                            &           & $\{8n\}$  &   $\{B_{1g}\}$ \\
\rowcolor{mygray} 			   & $72$ & $\{16k, 8i, 8h, 8g, 8f, 8e, 4b, 4a\}$ & $\{B_{1g}, B_{3g}, B_{2g}\}$ \\
 \rowcolor{mygray}                            &           & $\{8j, 4d, 4c\}$  &   $\{B_{1g}\}$ \\
\rowcolor{mygray} 			   & $73$ & $\{16f, 8e, 8c, 8b, 8a\}$ & $\{B_{1g}, B_{3g}, B_{2g}\}$ \\
\rowcolor{mygray}                             &           & $\{8d\}$   &    $\{B_{2g}\}$ \\
\rowcolor{mygray} 			   & $74$ & $\{16j, 8i, 8h, 8g, 4e, 4d, 4c\}$ & $\{B_{1g}, B_{3g}, B_{2g}\}$ \\
\rowcolor{mygray}                             &           & $\{4b\}$      &   $\{B_{2g}\}$  \\
\rowcolor{mygray}                             &           & $\{8f, 4a\}$ & $\{B_{3g}\}$ \\
\hline
$\mathbf{4}$ & $75$ & $\{4d, 2c\}$ & $\{B\}$ \\
 & $76$ & $\{4a\}$ & $\{B\}$ \\ 
 & $77$ & $\{4d, 2c, 2b, 2a\}$ & $\{B\}$ \\ 
 & $78$ & $\{4a\}$ & $\{B\}$ \\ 
 & $79$ & $\{8c, 4b, 2a\}$ & $\{B\}$ \\ 
 & $80$ & $\{8b, 4a\}$ & $\{B\}$ \\ 
 \hline
\rowcolor{mygray}$\mathbf{\overline{4}}$ & $81$ & $\{4h, 2g, 2f, 2e\}$ & $\{B\}$ \\ 
\rowcolor{mygray}& $82$ & $\{8g, 4f, 4e, 2d, 2c, 2b\}$ & $\{B\}$ \\ 
\hline
$\mathbf{4/m}$ & $83$ & $\{8l, 4k, 4j, 4i, 2f, 2e\}$ & $\{B_{g}\}$ \\ 
& $84$ & $\{8k, 4j, 4i, 4h, 4g, 2d, 2c, 2b, 2a\}$ & $\{B_{g}\}$ \\ 
 & $85$ & $\{8g, 4f, 4e, 4d\}$ & $\{B_{g}\}$ \\ 
& $86$ & $\{8g, 4f, 4e, 4d, 4c\}$ & $\{B_{g}\}$ \\ 
& $87$ & $\{16i, 8h, 8g, 8f, 4e, 4d, 4c, 2b\}$ & $\{B_{g}\}$ \\ 
& $88$ & $\{16f, 8e, 8d, 8c, 4b, 4a\}$ & $\{B_{g}\}$ \\ 
\hline
\rowcolor{mygray}$\mathbf{422}$ & $89$ & $\{8p, 4i\}$ & $\{B_{1}, A_{2}, B_{2}\}$ \\
\rowcolor{mygray}			& 	    & $\{2h, 2g\}$ & $\{A_{2}\}$ \\
\rowcolor{mygray}			& 	    & $\{4o, 4n, 4m, 4l, 2f, 2e\}$ & $\{B_{1}\}$ \\
\rowcolor{mygray}			& 	    & $\{4k, 4j\}$  &  $\{B_{2}\}$ \\
\rowcolor{mygray}			& $90$ & $\{8g, 4d\}$ & $\{B_{1}, A_{2}, B_{2}\}$ \\
\rowcolor{mygray}			& 	    & $\{2c\}$       & $\{A_{2}\}$ \\
\rowcolor{mygray}			& 	    & $\{4f, 4e, 2b, 2a\}$ & $\{B_{2}\}$ \\
\rowcolor{mygray}			& $91$ & $\{8d\}$ & $\{B_{1}, A_{2}, B_{2}\}$ \\
\rowcolor{mygray}			& 	    & $\{4b, 4a\}$ & $\{B_{1}\}$ \\
\rowcolor{mygray}			& 	    & $\{4c\}$ &  $\{B_{2}\}$ \\
\rowcolor{mygray}			& $92$ & $\{8b\}$ & $\{B_{1}, A_{2}, B_{2}\}$ \\
\rowcolor{mygray}			& 	    & $\{4a\}$ & $\{B_{2}\}$ \\
\rowcolor{mygray}			& $93$ & $\{8p, 4i, 4h, 4g\}$ & $\{B_{1}, A_{2}, B_{2}\}$ \\
\rowcolor{mygray}			& 	    & $\{4m, 4l, 4k, 4j, 2d, 2c, 2b, 2a\}$ & $\{B_{1}\}$ \\
\rowcolor{mygray}			& 	    & $\{4o, 4n, 2f, 2e\}$ & $\{B_{2}\}$ \\
		\hline		
 \end{tabular}
 \end{minipage}
 \end{table*}

\begin{table*}
\begin{minipage}{.45\columnwidth}
\centering
\begin{tabular}{|c|ccc|}
\hline
\hline
\rowcolor{dgray}PG & SG & WP & $\Gamma_{\mathbf{N}}$ \\
\hline
\hline
\rowcolor{mygray}	$\mathbf{422}$	& $94$ & $\{8g, 4d, 4c\}$ & $\{B_{1}, A_{2}, B_{2}\}$ \\
\rowcolor{mygray}			& 	    &  $\{4f, 4e, 2b, 2a\}$  & $\{B_{2}\}$ \\
\rowcolor{mygray}			& $95$ & $\{8d\}$ & $\{B_{1}, A_{2}, B_{2}\}$ \\
\rowcolor{mygray}			& 	    &  $\{4b, 4a\}$ & $\{B_{1}\}$ \\
\rowcolor{mygray}			& 	    &  $\{4c\}$ & $\{B_{2}\}$ \\
\rowcolor{mygray}			& $96$ & $\{8b\}$ & $\{B_{1}, A_{2}, B_{2}\}$ \\
\rowcolor{mygray}			& 	    &  $\{4a\}$ & $\{B_{2}\}$ \\
\rowcolor{mygray}   		& $97$ & $\{16k, 8j, 8i, 8f, 4e, 4d\}$ & $\{B_{1}, A_{2}, B_{2}\}$ \\
\rowcolor{mygray}			& 	    &  $\{8h, 4c\}$ & $\{B_{1}\}$ \\
\rowcolor{mygray}			& 	    &  $\{8g, 2b\}$ & $\{B_{2}\}$ \\
\rowcolor{mygray}			& $98$ & $\{16g, 8f, 8c\}$ & $\{B_{1}, A_{2}, B_{2}\}$ \\
\rowcolor{mygray}			& 	    &  $\{8e, 8d, 4b, 4a\}$  & $\{B_{2}\}$ \\
\hline
 $\mathbf{4mm}$ & $99$ &$\{8g\}$ & $\{B_{1}, B_{2}, A_{2}\}$ \\
 			   &          & $\{4f, 4e, 2c\}$ & $\{B_{1}\}$ \\
 			   &          & $\{4d\}$  	     & $\{B_{2}\}$ \\
			   & $100$ & $\{8d\}$     & $\{B_{1}, B_{2}, A_{2}\}$ \\
 			   &          &  $\{2a\}$      & $\{A_{2}\}$ \\
 			   &          & $\{4c, 2b\}$ & $\{B_{2}\}$ \\	
  	   & $101$ &  $\{8e, 4c\}$ & $\{B_{1}, B_{2}, A_{2}\}$ \\
 			   &          & $\{4d, 2b, 2a\}$ & $\{B_{2}\}$ \\
			   & $102$ & $\{8d, 4b\}$ & $\{B_{1}, B_{2}, A_{2}\}$ \\
 			   &          & $\{4c, 2a\}$   & $\{B_{2}\}$ \\

 	   & $103$ & $\{8d, 4c\}$ & $\{B_{1}, B_{2}, A_{2}\}$ \\
 			   &          & $\{2b, 2a\}$ & $\{A_{2}\}$ \\
			   & $104$ & $\{8c, 4b\}$ & $\{B_{1}, B_{2}, A_{2}\}$ \\
 			   &          & $\{2a\}$         & $\{A_{2}\}$  \\
			   &$105$ & $\{8f\}$ & $\{B_{1}, B_{2}, A_{2}\}$ \\
 			   &          & $\{4e, 4d, 2c, 2b, 2a\}$ & $\{B_{1}\}$ \\
			   & $106$ & $\{8c, 4b, 4a\}$ & $\{B_{1}, B_{2}, A_{2}\}$ \\ 
			   & $107$ & $\{16e, 8d, 4b\}$ & $\{B_{1}, B_{2}, A_{2}\}$ \\
 			   &          & $\{8c, 2a\}$  	       & $\{B_{2}\}$ \\
		   & $108$ & $\{16d, 4a\}$ & $\{B_{1}, B_{2}, A_{2}\}$ \\
 			   &          & $\{8c, 4b\}$ & $\{B_{2}\}$ \\
			   & $109$ & $\{16c, 8b, 4a\}$ & $\{B_{1}, B_{2}, A_{2}\}$ \\ 
			   & $110$ & $\{16b, 8a\}$ & $\{B_{1}, B_{2}, A_{2}\}$ \\		
\hline
\rowcolor{mygray}$\mathbf{\overline{4}2m}$ & $111$ & $\{8o, 4m\}$ & $\{B_{1}, B_{2}, A_{2}\}$ \\
					\rowcolor{mygray} &           & $\{4l, 4k, 4j, 4i, 2f, 2e\}$ & $\{B_{1}\}$ \\
					\rowcolor{mygray} &           & $\{4n, 2h, 2g\}$ & $\{B_{2}\}$ \\
\rowcolor{mygray} & $112$ & $\{8n, 4m, 4l, 4k\}$ & $\{B_{1}, B_{2}, A_{2}\}$ \\
\rowcolor{mygray} & 		  & $\{2f, 2e\}$ & $\{A_{2}\}$ \\
\rowcolor{mygray} & 		  & $\{4j, 4i, 4h, 4g, 2d, 2c, 2b, 2a\}$ & $\{B_{1}\}$ \\
\rowcolor{mygray} & $113$ & $\{8f, 4d\}$ & $\{B_{1}, B_{2}, A_{2}\}$ \\
\rowcolor{mygray} & 		  & $\{2b, 2a\}$ & $\{A_{2}\}$ \\
\rowcolor{mygray} & 		  & $\{4e, 2c\}$ & $\{B_{2}\}$ \\
\rowcolor{mygray} & $114$ & $\{8e, 4d, 4c\}$ & $\{B_{1}, B_{2}, A_{2}\}$ \\
\rowcolor{mygray} & 		  & $\{2b, 2a\}$ 	     & $\{A_{2}\}$ \\
\hline
\end{tabular}
\end{minipage}\hfill 
\begin{minipage}{.45\columnwidth}
\centering
\begin{tabular}{|c|ccc|}
\hline
\hline
\rowcolor{dgray}PG & SG & WP & $\Gamma_{\mathbf{N}}$ \\
\hline
\hline
$\mathbf{\overline{4}m2}$ & $115$ & $\{8l\}$ & $\{B_{1}, B_{2}, A_{2}\}$ \\
					&            & $\{4i, 4h\}$ & $\{A_{2}\}$ \\
					&            & $\{4k, 4j, 2g, 2f, 2e\}$ & $\{B_{1}\}$ \\
					& $116$ & $\{8j, 4i, 4h, 4g\}$ & $\{B_{1}, B_{2}, A_{2}\}$ \\
					&            & $\{4f, 4e, 2b, 2a\}$ & $\{A_{2}\}$ \\
					&            & $\{2d, 2c\}$ & $\{B_{2}\}$ \\
					& $117$ & $\{8i, 4f, 4e\}$ & $\{B_{1}, B_{2}, A_{2}\}$ \\
					&            & $\{4h, 4g, 2d, 2c\}$ & $\{A_{2}\}$ \\
					&            & $\{2b, 2a\}$             & $\{B_{2}\}$ \\
& $118$ & $\{8i, 4h, 4e\}$ & $\{B_{1}, B_{2}, A_{2}\}$ \\
					&            & $\{4g, 4f, 2d, 2c\}$ & $\{A_{2}\}$ \\
					&            & $\{2b, 2a\}$  & $\{B_{2}\}$ \\
					& $119$ & $\{16j, 8i, 8h, 4f, 4e, 2d, 2c\}$ & $\{B_{1}, B_{2}, A_{2}\}$ \\
					&            & $\{8g, 2b\}$ & $\{A_{2}\}$ \\
					& $120$ & $\{16i, 8g, 8f, 8e, 4c, 4a\}$ & $\{B_{1}, B_{2}, A_{2}\}$ \\
					&            & $\{8h, 4d\}$ & $\{A_{2}\}$ \\
					&            & $\{4b\}$       & $\{B_{2}\}$ \\
				& $121$ & $\{16j, 8h, 8g, 4d\}$ & $\{B_{1}, B_{2}, A_{2}\}$ \\
					&            & $\{8f, 4c\}$ & $\{B_{1}\}$ \\
					&            & $\{8i, 4e, 2b\}$ & $\{B_{2}\}$ \\
 					& $122$ & $\{16e, 8d, 8c, 4b\}$ & $\{B_{1}, B_{2}, A_{2}\}$ \\
					&            &  $\{4a\}$ & $\{A_{2}\}$ \\
 \hline
\rowcolor{mygray}$\mathbf{4/mmm}$ & $123$ & $\{16u, 8q, 8p\}$ & $\{B_{1g}, A_{2g}, B_{2g}\}$ \\
\rowcolor{mygray}				 & 	        & $\{8t, 8s, 4o, 4n, 4m, 4l, 4i, 2f, 2e\}$ & $\{B_{1g}\}$\\
\rowcolor{mygray}				 & 	        & $\{8r, 4k, 4j\}$ & $\{B_{2g}\}$ \\
\rowcolor{mygray} 	 & $124$ & $\{16n, 8m, 8i, 4e\}$ & $\{B_{1g}, A_{2g}, B_{2g}\}$ \\
\rowcolor{mygray}	  			 & 	        & $\{4h, 4g, 2d, 2b\}$ & $\{A_{2g}\}$\\
\rowcolor{mygray}	  			 & 	        & $\{8l, 8k, 4f\}$ & $\{B_{1g}\}$ \\
\rowcolor{mygray}	  			 & 	        & $\{8j\}$  &  $\{B_{2g}\}$ \\
\rowcolor{mygray} 				 & $125$ & $\{16n\}$ & $\{B_{1g}, A_{2g}, B_{2g}\}$ \\
\rowcolor{mygray}	  			 & 	        & $\{4g\}$  & $\{A_{2g}\}$ \\
\rowcolor{mygray}	  			 & 	        & $\{8l, 8k\}$ & $\{B_{1g}\}$ \\
\rowcolor{mygray}	  			 & 	        & $\{8m, 8j, 8i, 4h, 4f, 4e\}$ & $\{B_{2g}\}$ \\
\rowcolor{mygray} 			 & $126$ &  $\{16k, 8g, 8f\}$ & $\{B_{1g}, A_{2g}, B_{2g}\}$ \\
\rowcolor{mygray}	  			 & 	        & $\{4e, 4d\}$ & $\{A_{2g}\}$ \\ 
\rowcolor{mygray}	  			 & 	        & $\{8j, 8i, 4c\}$ & $\{B_{1g}\}$ \\
\rowcolor{mygray}	  			 & 	        & $\{8h\}$ & $\{B_{2g}\}$ \\
\rowcolor{mygray} 		 & $127$ & $\{16l, 8j, 8i\}$ & $\{B_{1g}, A_{2g}, B_{2g}\}$ \\
\rowcolor{mygray}	  			 & 	        & $\{4e, 2b, 2a\}$ & $\{A_{2g}\}$ \\
\rowcolor{mygray}	  			 & 	        & $\{8k, 4h, 4g, 4f, 2d, 2c\}$ & $\{B_{2g}\}$ \\
\rowcolor{mygray} 			& $128$ & $\{16i, 8h, 8f, 4c\}$ & $\{B_{1g}, A_{2g}, B_{2g}\}$ \\
\rowcolor{mygray}	  			 & 	        & $\{4e, 2b, 2a\}$ & $\{A_{2g}\}$ \\
\rowcolor{mygray}	  			 & 	        & $\{8g, 4d\}$  & $\{B_{2g}\}$ \\
\rowcolor{mygray} 			& $129$ & $\{16k\}$ & $\{B_{1g}, A_{2g}, B_{2g}\}$ \\
\rowcolor{mygray}	  			 & 	        & $\{8i, 4f\}$ & $\{B_{1g}\}$ \\
\rowcolor{mygray}	  			 & 	        & $\{8j, 8h, 8g, 4e, 4d\}$ & $\{B_{2g}\}$ \\
\hline
\end{tabular}
\end{minipage}
\end{table*}

\begin{table*}
\begin{minipage}{.45\columnwidth}
\centering
 \begin{tabular}{|c|ccc|}
\hline
\hline
\rowcolor{dgray}PG & SG & WP & $\Gamma_{\mathbf{N}}$ \\
\hline
\hline
\rowcolor{mygray} 	$\mathbf{4/mmm}$		 & $130$ & $\{16g, 8e, 8d\}$ & $\{B_{1g}, A_{2g}, B_{2g}\}$ \\
\rowcolor{mygray}	  			 & 	        & $\{4c, 4b\}$ & $\{A_{2g}\}$ \\
\rowcolor{mygray}	  			 & 	        & $\{8f, 4a\}$  & $\{B_{2g}\}$ \\
\rowcolor{mygray} 		 & $131$ & $\{16r, 8q\}$ & $\{B_{1g}, A_{2g}, B_{2g}\}$ \\
\rowcolor{mygray}	  			 & 	        & $\shortstack{\{$8p, 8o, 4m, 4l, 4k, 4j, 4i,$\\ $4h, 4g, 2d, 2c, 2b, 2a\}$}$ & $\{B_{1g}\}$ \\
\rowcolor{mygray}	  			 & 	        & $\{8n\}$ & $\{B_{2g}\}$ \\
\rowcolor{mygray} 			& $132$ & $\{16p, 8n, 8k, 4f\}$ & $\{B_{1g}, A_{2g}, B_{2g}\}$ \\
\rowcolor{mygray}	  			 & 	        & $\{8m, 8l, 4e\}$ & $\{B_{1g}\}$ \\
\rowcolor{mygray}	  			 & 	        & $\{8o, 4j, 4i, 4h, 4g, 2c, 2a\}$ & $\{B_{2g}\}$ \\
\rowcolor{mygray} 			 & $133$ & $\{16k, 8g, 8f, 8e\}$ & $\{B_{1g}, A_{2g}, B_{2g}\}$ \\
\rowcolor{mygray}	  			 & 	        & $\{4d\}$ & $\{A_{2g}\}$ \\
\rowcolor{mygray}	  			 & 	        & $\{8i, 8h, 4b, 4a\}$ & $\{B_{1g}\}$ \\
\rowcolor{mygray}	  			 & 	        & $\{8j, 4c\}$ & $\{B_{2g}\}$ \\
\rowcolor{mygray} 				 & $134$ & $\{16n, 8h\}$ & $\{B_{1g}, A_{2g}, B_{2g}\}$ \\
\rowcolor{mygray}	  			 & 	        & $\{8j, 8i, 4c\}$ & $\{B_{1g}\}$ \\
\rowcolor{mygray}	  			 & 	        & $\{8m, 8l, 8k, 4g, 4f, 4e, 4d\}$ & $\{B_{2g}\}$ \\
\rowcolor{mygray} 				 & $135$ & $\{16i, 8h, 8f, 8e, 4c, 4a\}$ & $\{B_{1g}, A_{2g}, B_{2g}\}$ \\
\rowcolor{mygray}	  			 & 	        & $\{4b\}$ & $\{A_{2g}\}$ \\
\rowcolor{mygray}	  			 & 	        & $\{8g, 4d\}$ & $\{B_{2g}\}$ \\
\rowcolor{mygray} 				 & $136$ & $\{16k, 8i, 8h, 4c\}$ & $\{B_{1g}, A_{2g}, B_{2g}\}$ \\
\rowcolor{mygray}	  			 & 	        & $\{4d\}$ & $\{A_{2g}\}$ \\
\rowcolor{mygray}	  			 & 	        & $\{8j, 4g, 4f, 4e, 2b, 2a\}$ & $\{B_{2g}\}$ \\
\rowcolor{mygray} 		 & $137$ & $\{16h, 8e\}$ & $\{B_{1g}, A_{2g}, B_{2g}\}$ \\
\rowcolor{mygray}	  			 & 	        & $\{8g, 4d, 4c\}$ & $\{B_{1g}\}$ \\
\rowcolor{mygray}	  			 & 	        & $\{8f\}$ & $\{B_{2g}\}$ \\
\rowcolor{mygray} 				 & $138$ & $\{16j, 8f\}$ & $\{B_{1g}, A_{2g}, B_{2g}\}$ \\
\rowcolor{mygray}	  			 & 	        & $\{4b\}$ & $\{A_{2g}\}$ \\
\rowcolor{mygray}	  			 & 	        & $\{8i, 8h, 8g, 4e, 4d, 4c, 4a\}$ & $\{B_{2g}\}$ \\
\rowcolor{mygray} 				 & $139$ & $\{32o, 16n, 16l, 16k, 8g, 4d\}$ & $\{B_{1g}, A_{2g}, B_{2g}\}$ \\
\rowcolor{mygray}	  			 & 	        & $\{8j, 8i, 4c\}$ & $\{B_{1g}\}$ \\
\rowcolor{mygray}	  			 & 	        & $\{16m, 8h, 8f, 4e, 2b\}$ & $\{B_{2g}\}$ \\
\rowcolor{mygray} 				 & $140$ & $\{32m, 16k, 16j, 16i, 8f, 8e, 4a\}$ & $\{B_{1g}, A_{2g}, B_{2g}\}$ \\
\rowcolor{mygray}	  			 & 	        & $\{4c\}$ & $\{A_{2g}\}$ \\
\rowcolor{mygray}	  			 & 	        & $\{16l, 8h, 8g, 4d, 4b\}$ & $\{B_{2g}\}$ \\
\rowcolor{mygray} 				 & $141$ & $\{32i, 16h, 16g, 8e, 4b, 4a\}$ & $\{B_{1g}, A_{2g}, B_{2g}\}$ \\
\rowcolor{mygray}	  			 & 	        & $\{16f, 8d, 8c\}$ & $\{B_{1g}\}$ \\
\rowcolor{mygray}& $142$ & $\{32g, 16f, 16e, 16d, 16c, 8b, 8a\}$ & $\{B_{1g}, A_{2g}, B_{2g}\}$ \\ 
\hline
$\mathbf{32}$ & $149$ & $\{6l, 2i, 2h, 2g\}$ & $\{A_{2}\}$ \\ 
 & $150$ & $\{6g, 2d, 2c\}$ & $\{A_{2}\}$ \\ 
 & $151$ & $\{6c\}$ & $\{A_{2}\}$ \\ 
 & $152$ & $\{6c\}$ & $\{A_{2}\}$ \\ 
 & $153$ & $\{6c\}$ & $\{A_{2}\}$ \\ 
 & $154$ & $\{6c\}$ & $\{A_{2}\}$ \\ 
 & $155$ & $\{18f, 9e, 9d, 6c\}$ & $\{A_{2}\}$ \\ 
 \hline
\rowcolor{mygray}$\mathbf{3m}$ & $156$ & $\{6e\}$ & $\{A_{2}\}$ \\ 
\rowcolor{mygray}& $157$ & $\{6d, 2b\}$ & $\{A_{2}\}$ \\ 
\rowcolor{mygray}& $158$ & $\{6d, 2c, 2b, 2a\}$ & $\{A_{2}\}$ \\ 
\rowcolor{mygray}& $159$ & $\{6c, 2b, 2a\}$ & $\{A_{2}\}$ \\ 
\rowcolor{mygray}& $160$ & $\{18c\}$ & $\{A_{2}\}$ \\ 
\rowcolor{mygray}& $161$ & $\{18b, 6a\}$ & $\{A_{2}\}$ \\ 
\hline
 \end{tabular}
\end{minipage}\hfill 
\begin{minipage}{.45\columnwidth}
\centering
\begin{tabular}{|c|ccc|}
 \hline
\hline
\rowcolor{dgray}PG & SG & WP & $\Gamma_{\mathbf{N}}$ \\
\hline
\hline
 $\mathbf{\overline{3}m}$ & $162$ & $\{12l, 4h\}$ & $\{A_{2g}\}$ \\ 
& $163$ & $\{12i, 6g, 4f, 4e, 2b\}$ & $\{A_{2g}\}$ \\ 
& $164$ & $\{12j\}$ & $\{A_{2g}\}$ \\ 
& $165$ & $\{12g, 6e, 4d, 4c, 2b\}$ & $\{A_{2g}\}$ \\ 
& $166$ & $\{36i\}$ & $\{A_{2g}\}$ \\ 
& $167$ & $\{36f, 18e, 18d, 12c, 6b\}$ & $\{A_{2g}\}$ \\ 
\hline
\rowcolor{mygray}$\mathbf{6}$ & $168$ & $\{6d, 2b\}$ & $\{B\}$ \\ 
\rowcolor{mygray}& $169$ & $\{6a\}$ & $\{B\}$ \\ 
\rowcolor{mygray} & $170$ & $\{6a\}$ & $\{B\}$ \\ 
\rowcolor{mygray} & $171$ & $\{6c\}$ & $\{B\}$ \\ 
\rowcolor{mygray} & $172$ & $\{6c\}$ & $\{B\}$ \\ 
\rowcolor{mygray} & $173$ & $\{6c, 2b, 2a\}$ & $\{B\}$ \\ 
\hline
$\mathbf{\overline{6}}$ & $174$ & $\{6l, 2i, 2h, 2g\}$ & $\{A''\}$ \\ 
\hline
\rowcolor{mygray}$\mathbf{6/m}$ & $175$ & $\{12l, 4h\}$ & $\{B_{g}\}$ \\ 
\rowcolor{mygray}& $176$ & $\{12i, 6g, 4f, 4e, 2b\}$ & $\{B_{g}\}$ \\ 
\hline
$\mathbf{622}$ & $177$ & $\{12n, 4h\}$ & $\{A_{2}, B_{2}, B_{1}\}$ \\
			    &            & $\{6i, 2e\}$ & $\{A_{2}\}$ \\
			    &            & $\{6m, 6l, 2d, 2c\}$ & $\{B_{1}\}$\\
			    &            & $\{6k, 6j\}$ & $\{B_{2}\}$ \\
			    & $178$ & $\{12c\}$ & $\{A_{2}, B_{2}, B_{1}\}$ \\
			    &            & $\{6b\}$  & $\{B_{1}\}$ \\
			    &            & $\{6a\}$  & $\{B_{2}\}$ \\
			    & $179$ & $\{12c\}$ & $\{A_{2}, B_{2}, B_{1}\}$ \\
			    &            & $\{6b\}$ & $\{B_{1}\}$ \\
			    &            & $\{6a\}$ & $\{B_{2}\}$ \\
			    & $180$ & $\{12k\}$ & $\{A_{2}, B_{2}, B_{1}\}$ \\
			    &            & $\{6f, 6e\}$ & $\{A_{2}\}$\\
			    &            & $\{6j, 6i\}$  & $\{B_{1}\}$ \\
			    &            & $\{6h, 6g\}$ & $\{B_{2}\}$ \\
			    & $181$ & $\{12k\}$ & $\{A_{2}, B_{2}, B_{1}\}$ \\
			    &            & $\{6f, 6e\}$   & $\{A_{2}\}$\\
			    &            & $\{6j, 6i\}$    & $\{B_{1}\}$ \\
			    &            &  $\{6h, 6g\}$ & $\{B_{2}\}$ \\
			    & $182$ & $\{12i, 4f, 4e\}$ & $\{A_{2}, B_{2}, B_{1}\}$ \\
			    &            & $\{6h, 2d, 2c, 2b\}$ & $\{B_{1}\}$ \\
			    &            & $\{6g, 2a\}$ & $\{B_{2}\}$ \\
\hline
$\mathbf{6mm}$ & $183$ & $\{12f\}$ & $\{A_{2}, B_{2}, B_{1}\}$ \\
			  &		& $\{6e, 2b\}$ & $\{B_{1}\}$ \\
			  &		& $\{6d\}$  & $\{B_{2}\}$ \\
			  & $184$ & $\{12d, 4b\}$ & $\{A_{2}, B_{2}, B_{1}\}$ \\
			  &		& $\{6c, 2a\}$ & $\{A_{2}\}$ \\
    	  &  $185$ & $\{12d, 4b\}$ & $\{A_{2}, B_{2}, B_{1}\}$ \\
			  &		& $\{6c, 2a\}$ & $\{B_{2}\}$ \\
			  & $186$ & $\{12d\}$ & $\{A_{2}, B_{2}, B_{1}\}$ \\
			  &            & $\{6c, 2b, 2a\}$ & $\{B_{1}\}$ \\
\hline
\rowcolor{mygray}$\mathbf{\overline{6}m2}$ & $187$ & $\{12o\}$ & $\{A_{1}'', A_{2}'', A_{2}'\}$ \\
\rowcolor{mygray}					    & 		  & $\{6m, 6l\}$ & $\{A_{1}''\}$ \\
\rowcolor{mygray}					    & 		  &  $\{6n, 2i, 2h, 2g\}$ & $\{A_{2}''\}$ \\
\rowcolor{mygray}					    & $188$ & $\{12l, 4i, 4h, 4g\}$ & $\{A_{1}'', A_{2}'', A_{2}'\}$ \\
\rowcolor{mygray}					    & 		  & $\{6k, 2f, 2d, 2b\}$ & $\{A_{1}''\}$ \\
\rowcolor{mygray}					    & 		  & $\{6j, 2e, 2c, 2a\}$ & $\{A_{2}'\}$ \\
\hline
\end{tabular}
\end{minipage}
\end{table*}

\begin{table*}
\begin{minipage}{.45\columnwidth}
\begin{tabular}{|c|ccc|}
 \hline
\hline
\rowcolor{dgray}PG & SG & WP & $\Gamma_{\mathbf{N}}$ \\
\hline
\hline
$\mathbf{\overline{6}2m}$ & $189$ & $\{12l, 4h\}$ & $\{A_{1}'', A_{2}'', A_{2}'\}$ \\
	&	& $\{6k, 6j, 2d, 2c\}$ & $\{A_{2}'\}$ \\
	&	& $\{6i, 2e\}$ & $\{A_{2}''\}$ \\
 & $190$ & $\{12i, 4f, 4e\}$ & $\{A_{1}'', A_{2}'', A_{2}'\}$ \\
	&	  & $\{6g, 2a\}$ & $\{A_{1}''\}$ \\
	&	  & $\{6h, 2d, 2c, 2b\}$ & $\{A_{2}'\}$ \\
\hline
\rowcolor{mygray}$\mathbf{6/mmm}$ & $191$ & $\{24r\}$ & $\{A_{2g}, B_{2g}, B_{1g}\}$ \\
\rowcolor{mygray}				 & 		& $\{12q, 12p\}$ & $\{A_{2g}\}$ \\
\rowcolor{mygray}				 & 		& $\{12n\}$ & $\{B_{1g}\}$ \\
\rowcolor{mygray}				 & 		& $\{12o, 4h\}$ & $\{B_{2g}\}$ \\
\rowcolor{mygray} & $192$ & $\{24m, 8h\}$ & $\{A_{2g}, B_{2g}, B_{1g}\}$ \\
\rowcolor{mygray} &		  & $\{12l, 12i, 6g, 4e, 4d, 2b\}$ & $\{A_{2g}\}$ \\
\rowcolor{mygray} &		  & $\{12k, 4c\}$ & $\{B_{1g}\}$ \\
\rowcolor{mygray} &		  & $\{12j\}$ & $\{B_{2g}\}$ \\
\rowcolor{mygray} & $193$ & $\{24l, 8h\}$ & $\{A_{2g}, B_{2g}, B_{1g}\}$ \\
\rowcolor{mygray} & 		  & $\{12j, 4c\}$ & $\{A_{2g}\}$ \\
\rowcolor{mygray} & 		  & $\{12k, 12i, 6f, 4e, 4d, 2b\}$ & $\{B_{1g}\}$ \\
\rowcolor{mygray}& $194$ & $\{24l\}$ & $\{A_{2g}, B_{2g}, B_{1g}\}$ \\
\rowcolor{mygray} & 		  & $\{12j\}$ & $\{A_{2g}\}$ \\
\rowcolor{mygray} & 		  & $\{12k, 12i, 6g, 4f, 4e, 2a\}$ & $\{B_{2g}\}$ \\
\hline
$\mathbf{m\overline{3}m}$ & $221$ & $\{48n, 24l, 24k, 12h\}$ & $\{A_{2g}\}$ \\ 
& $222$ & $\{48i, 24g, 16f, 8c\}$ & $\{A_{2g}\}$ \\ 
& $223$ & $\{48l, 24k, 16i, 12h, 12g, 12f, 6b, 2a\}$ & $\{A_{2g}\}$ \\ 
& $224$ & $\{48l, 24h\}$ & $\{A_{2g}\}$ \\ 
& $225$ & $\{192l, 96j\}$ & $\{A_{2g}\}$ \\ 
 & $226$ & $\{192j, 96i, 96h, 64g, 48e, 24c, 8b\}$ & $\{A_{2g}\}$ \\ 
& $227$ & $\{192i\}$ & $\{A_{2g}\}$ \\ 
& $228$ & $\{192h, 96g, 96f, 64e, 48d, 32c, 16a\}$ & $\{A_{2g}\}$ \\ 
& $229$ & $\{96l, 48i\}$ & $\{A_{2g}\}$ \\ 
& $230$ & $\{96h, 48g, 48f, 32e, 24d, 24c, 16b, 16a\}$ & $\{A_{2g}\}$ \\ 
\hline
$\mathbf{432}$ & $207$ & $\{24k, 12h, 8g\}$ & $\{A_{2}\}$ \\ 
& $208$ & $\{24m, 12j, 12i, 12h, 8g, 6d, 2a\}$ & $\{A_{2}\}$ \\ 
& $209$ & $\{96j, 48i, 48h, 48g, 32f, 24d, 8c\}$ & $\{A_{2}\}$ \\ 
& $210$ & $\{96h, 48f, 32e, 8b, 8a\}$ & $\{A_{2}\}$ \\ 
& $211$ & $\{48j, 24i, 24h, 24g, 16f, 8c\}$ & $\{A_{2}\}$ \\ 
& $212$ & $\{24e, 8c\}$ & $\{A_{2}\}$ \\ 
& $213$ & $\{24e, 8c\}$ & $\{A_{2}\}$ \\ 
 & $214$ & $\{48i, 24h, 24g, 24f, 16e, 12d, 12c, 8b, 8a\}$ & $\{A_{2}\}$ \\ 
 \hline
\rowcolor{mygray}$\mathbf{\overline{4}3m}$ & $215$ & $\{24j, 12h\}$ & $\{A_{2}\}$ \\ 
\rowcolor{mygray}& $216$ & $\{96i\}$ & $\{A_{2}\}$ \\ 
\rowcolor{mygray}& $217$ & $\{48h, 24f\}$ & $\{A_{2}\}$ \\ 
\rowcolor{mygray}& $218$ & $\{24i, 12h, 12g, 12f, 8e, 6b, 2a\}$ & $\{A_{2}\}$ \\ 
\rowcolor{mygray}& $219$ & $\{96h, 48g, 48f, 32e, 24d, 24c, 8b, 8a\}$ & $\{A_{2}\}$ \\ 
\rowcolor{mygray}& $220$ & $\{48e, 24d, 16c, 12b, 12a\}$ & $\{A_{2}\}$ \\ 
\hline
\end{tabular}
\end{minipage}
\end{table*}

\FloatBarrier


\section{\label{app:socfreeMultipoleTables}Table of multipoles coupling to $\Gamma_{\mathbf{N}}$ in the SOC-free limit}

\begin{table}[htb!]
\begin{minipage}{.3\columnwidth}
\centering
\begin{tabular}{|ccccccc|}
\hline
\hline
\rowcolor{dgray}$\Gamma_{\mathbf{N}}$ & $V$ & $\left[V^{ 2}\right]$ & $\left[V^{ 3}\right]$ & $\left[V^{ 4}\right]$ & $\left[V^{ 5}\right]$ & $\left[V^{ 6}\right]$ \\
\hline
\hline
\rowcolor{mygray}\multicolumn{7}{|c|}{$\mathbf{2}$}\\
$B$ & $\text{\ding{51}}$ & $\text{\ding{51}}$ & $\text{\ding{51}}$ & $\text{\ding{51}}$ & $\text{\ding{51}}$ & $\text{\ding{51}}$ \\
\hline
\rowcolor{mygray}\multicolumn{7}{|c|}{$\mathbf{m}$}\\
$A"$ & $\text{\ding{51}}$ & $\text{\ding{51}}$ & $\text{\ding{51}}$ & $\text{\ding{51}}$ & $\text{\ding{51}}$ & $\text{\ding{51}}$ \\
\hline
\rowcolor{mygray}\multicolumn{7}{|c|}{$\mathbf{2/m}$}\\
$B_{g}$ & $ $ & $\text{\ding{51}}$ & $ $ & $\text{\ding{51}}$ & $ $ & $\text{\ding{51}}$ \\
\hline
\rowcolor{mygray}\multicolumn{7}{|c|}{$\mathbf{222}$}\\
$B_{1}$ & $\text{\ding{51}}$ & $\text{\ding{51}}$ & $\text{\ding{51}}$ & $\text{\ding{51}}$ & $\text{\ding{51}}$ & $\text{\ding{51}}$ \\
$B_{3}$ & $\text{\ding{51}}$ & $\text{\ding{51}}$ & $\text{\ding{51}}$ & $\text{\ding{51}}$ & $\text{\ding{51}}$ & $\text{\ding{51}}$ \\
$B_{2}$ & $\text{\ding{51}}$ & $\text{\ding{51}}$ & $\text{\ding{51}}$ & $\text{\ding{51}}$ & $\text{\ding{51}}$ & $\text{\ding{51}}$ \\
\hline
\rowcolor{mygray}\multicolumn{7}{|c|}{$\mathbf{mm2}$}\\
$A_{2}$ & $ $ & $\text{\ding{51}}$ & $\text{\ding{51}}$ & $\text{\ding{51}}$ & $\text{\ding{51}}$ & $\text{\ding{51}}$ \\
$B_{2}$ & $\text{\ding{51}}$ & $\text{\ding{51}}$ & $\text{\ding{51}}$ & $\text{\ding{51}}$ & $\text{\ding{51}}$ & $\text{\ding{51}}$ \\
$B_{1}$ & $\text{\ding{51}}$ & $\text{\ding{51}}$ & $\text{\ding{51}}$ & $\text{\ding{51}}$ & $\text{\ding{51}}$ & $\text{\ding{51}}$ \\
\hline
\rowcolor{mygray}\multicolumn{7}{|c|}{$\mathbf{mmm}$}\\
$B_{1g}$ & $ $ & $\text{\ding{51}}$ & $ $ & $\text{\ding{51}}$ & $ $ & $\text{\ding{51}}$ \\
$B_{3g}$ & $ $ & $\text{\ding{51}}$ & $ $ & $\text{\ding{51}}$ & $ $ & $\text{\ding{51}}$ \\
$B_{2g}$ & $ $ & $\text{\ding{51}}$ & $ $ & $\text{\ding{51}}$ & $ $ & $\text{\ding{51}}$ \\
\hline
\rowcolor{mygray}\multicolumn{7}{|c|}{$\mathbf{4}$}\\
$B$ & $ $ & $\text{\ding{51}}$ & $\text{\ding{51}}$ & $\text{\ding{51}}$ & $\text{\ding{51}}$ & $\text{\ding{51}}$ \\
\hline
\rowcolor{mygray}\multicolumn{7}{|c|}{$\mathbf{\overline{4}}$}\\
$B$ & $\text{\ding{51}}$ & $\text{\ding{51}}$ & $\text{\ding{51}}$ & $\text{\ding{51}}$ & $\text{\ding{51}}$ & $\text{\ding{51}}$ \\
\hline
\rowcolor{mygray}\multicolumn{7}{|c|}{$\mathbf{4/m}$}\\
$B_{g}$ & $ $ & $\text{\ding{51}}$ & $ $ & $\text{\ding{51}}$ & $ $ & $\text{\ding{51}}$ \\
\hline
\rowcolor{mygray}\multicolumn{7}{|c|}{$\mathbf{422}$}\\
$B_{1}$ & $ $ & $\text{\ding{51}}$ & $\text{\ding{51}}$ & $\text{\ding{51}}$ & $\text{\ding{51}}$ & $\text{\ding{51}}$ \\
$A_{2}$ & $\text{\ding{51}}$ & $ $ & $\text{\ding{51}}$ & $\text{\ding{51}}$ & $\text{\ding{51}}$ & $\text{\ding{51}}$ \\
$B_{2}$ & $ $ & $\text{\ding{51}}$ & $\text{\ding{51}}$ & $\text{\ding{51}}$ & $\text{\ding{51}}$ & $\text{\ding{51}}$ \\
\hline
\hline
\end{tabular}
\end{minipage}\hfill 
\begin{minipage}{.3\columnwidth}
\begin{tabular}{|ccccccc|}
\hline
\hline
\rowcolor{dgray}$\Gamma_{\mathbf{N}}$ & $V$ & $\left[V^{ 2}\right]$ & $\left[V^{ 3}\right]$ & $\left[V^{ 4}\right]$ & $\left[V^{ 5}\right]$ & $\left[V^{ 6}\right]$ \\
\hline
\hline
\rowcolor{mygray}\multicolumn{7}{|c|}{$\mathbf{4mm}$}\\
$B_{1}$ & $ $ & $\text{\ding{51}}$ & $\text{\ding{51}}$ & $\text{\ding{51}}$ & $\text{\ding{51}}$ & $\text{\ding{51}}$ \\
$B_{2}$ & $ $ & $\text{\ding{51}}$ & $\text{\ding{51}}$ & $\text{\ding{51}}$ & $\text{\ding{51}}$ & $\text{\ding{51}}$ \\
$A_{2}$ & $ $ & $ $ & $ $ & $\text{\ding{51}}$ & $\text{\ding{51}}$ & $\text{\ding{51}}$ \\
\hline
\rowcolor{mygray}\multicolumn{7}{|c|}{$\mathbf{\overline{4}2m}$}\\
$B_{1}$ & $ $ & $\text{\ding{51}}$ & $ $ & $\text{\ding{51}}$ & $\text{\ding{51}}$ & $\text{\ding{51}}$ \\
$B_{2}$ & $\text{\ding{51}}$ & $\text{\ding{51}}$ & $\text{\ding{51}}$ & $\text{\ding{51}}$ & $\text{\ding{51}}$ & $\text{\ding{51}}$ \\
$A_{2}$ & $ $ & $ $ & $\text{\ding{51}}$ & $\text{\ding{51}}$ & $\text{\ding{51}}$ & $\text{\ding{51}}$ \\
\hline
\rowcolor{mygray}\multicolumn{7}{|c|}{$\mathbf{\overline{4}m2}$}\\
$B_{1}$ & $\text{\ding{51}}$ & $\text{\ding{51}}$ & $\text{\ding{51}}$ & $\text{\ding{51}}$ & $\text{\ding{51}}$ & $\text{\ding{51}}$ \\
$B_{2}$ & $ $ & $ $ & $\text{\ding{51}}$ & $\text{\ding{51}}$ & $\text{\ding{51}}$ & $\text{\ding{51}}$ \\
$A_{2}$ & $ $ & $\text{\ding{51}}$ & $ $ & $\text{\ding{51}}$ & $\text{\ding{51}}$ & $\text{\ding{51}}$ \\
\hline
\rowcolor{mygray}\multicolumn{7}{|c|}{$\mathbf{4/mmm}$}\\
$B_{1g}$ & $ $ & $\text{\ding{51}}$ & $ $ & $\text{\ding{51}}$ & $ $ & $\text{\ding{51}}$ \\
$A_{2g}$ & $ $ & $ $ & $ $ & $\text{\ding{51}}$ & $ $ & $\text{\ding{51}}$ \\
$B_{2g}$ & $ $ & $\text{\ding{51}}$ & $ $ & $\text{\ding{51}}$ & $ $ & $\text{\ding{51}}$ \\
\hline
\rowcolor{mygray}\multicolumn{7}{|c|}{$\mathbf{32}$}\\
$A_{2}$ & $\text{\ding{51}}$ & $ $ & $\text{\ding{51}}$ & $\text{\ding{51}}$ & $\text{\ding{51}}$ & $\text{\ding{51}}$ \\
\hline
\rowcolor{mygray}\multicolumn{7}{|c|}{$\mathbf{3m}$}\\
$A_{2}$ & $ $ & $ $ & $\text{\ding{51}}$ & $\text{\ding{51}}$ & $\text{\ding{51}}$ & $\text{\ding{51}}$ \\
\hline
\rowcolor{mygray}\multicolumn{7}{|c|}{$\mathbf{\overline{3}m}$}\\
$A_{2g}$ & $ $ & $ $ & $ $ & $\text{\ding{51}}$ & $ $ & $\text{\ding{51}}$ \\
\hline
\rowcolor{mygray}\multicolumn{7}{|c|}{$\mathbf{6}$}\\
$B$ & $ $ & $ $ & $\text{\ding{51}}$ & $\text{\ding{51}}$ & $\text{\ding{51}}$ & $\text{\ding{51}}$ \\
\hline
\rowcolor{mygray}\multicolumn{7}{|c|}{$\mathbf{\overline{6}}$}\\
$A''$ & $\text{\ding{51}}$ & $ $ & $\text{\ding{51}}$ & $\text{\ding{51}}$ & $\text{\ding{51}}$ & $\text{\ding{51}}$ \\
\hline
\rowcolor{mygray}\multicolumn{7}{|c|}{$\mathbf{6/m}$}\\
$B_{g}$ & $ $ & $ $ & $ $ & $\text{\ding{51}}$ & $ $ & $\text{\ding{51}}$ \\
\hline
\hline
\end{tabular}
\end{minipage}\hfill 
\begin{minipage}{.3\columnwidth}
\begin{tabular}{|ccccccc|}
\hline
\hline
\rowcolor{dgray}$\Gamma_{\mathbf{N}}$ & $V$ & $\left[V^{ 2}\right]$ & $\left[V^{ 3}\right]$ & $\left[V^{ 4}\right]$ & $\left[V^{ 5}\right]$ & $\left[V^{ 6}\right]$ \\
\hline
\hline
\rowcolor{mygray}\multicolumn{7}{|c|}{$\mathbf{622}$}\\
$A_{2}$ & $\text{\ding{51}}$ & $ $ & $\text{\ding{51}}$ & $ $ & $\text{\ding{51}}$ & $\text{\ding{51}}$ \\
$B_{2}$ & $ $ & $ $ & $\text{\ding{51}}$ & $\text{\ding{51}}$ & $\text{\ding{51}}$ & $\text{\ding{51}}$ \\
$B_{1}$ & $ $ & $ $ & $\text{\ding{51}}$ & $\text{\ding{51}}$ & $\text{\ding{51}}$ & $\text{\ding{51}}$ \\
\hline
\rowcolor{mygray}\multicolumn{7}{|c|}{$\mathbf{6mm}$}\\
$A_{2}$ & $ $ & $ $ & $ $ & $ $ & $ $ & $\text{\ding{51}}$ \\
$B_{2}$ & $ $ & $ $ & $\text{\ding{51}}$ & $\text{\ding{51}}$ & $\text{\ding{51}}$ & $\text{\ding{51}}$ \\
$B_{1}$ & $ $ & $ $ & $\text{\ding{51}}$ & $\text{\ding{51}}$ & $\text{\ding{51}}$ & $\text{\ding{51}}$ \\
\hline
\rowcolor{mygray}\multicolumn{7}{|c|}{$\mathbf{\overline{6}2m}$}\\
$A_{1}''$ & $ $ & $ $ & $ $ & $\text{\ding{51}}$ & $ $ & $\text{\ding{51}}$ \\
$A_{2}''$ & $\text{\ding{51}}$ & $ $ & $\text{\ding{51}}$ & $\text{\ding{51}}$ & $\text{\ding{51}}$ & $\text{\ding{51}}$ \\
$A_{2}'$ & $ $ & $ $ & $\text{\ding{51}}$ & $ $ & $\text{\ding{51}}$ & $\text{\ding{51}}$ \\
\hline
\rowcolor{mygray}\multicolumn{7}{|c|}{$\mathbf{\overline{6}m2}$}\\
$A_{1}''$ & $ $ & $ $ & $\text{\ding{51}}$ & $ $ & $\text{\ding{51}}$ & $\text{\ding{51}}$ \\
$A_{2}''$ & $\text{\ding{51}}$ & $ $ & $\text{\ding{51}}$ & $\text{\ding{51}}$ & $\text{\ding{51}}$ & $\text{\ding{51}}$ \\
$A_{2}'$ & $ $ & $ $ & $ $ & $\text{\ding{51}}$ & $ $ & $\text{\ding{51}}$ \\
\hline
\rowcolor{mygray}\multicolumn{7}{|c|}{$\mathbf{6/mmm}$}\\
$A_{2g}$ & $ $ & $ $ & $ $ & $ $ & $ $ & $\text{\ding{51}}$ \\
$B_{2g}$ & $ $ & $ $ & $ $ & $\text{\ding{51}}$ & $ $ & $\text{\ding{51}}$ \\
$B_{1g}$ & $ $ & $ $ & $ $ & $\text{\ding{51}}$ & $ $ & $\text{\ding{51}}$ \\
\hline
\rowcolor{mygray}\multicolumn{7}{|c|}{$\mathbf{432}$}\\
$A_{2}$ & $ $ & $ $ & $\text{\ding{51}}$ & $ $ & $\text{\ding{51}}$ & $\text{\ding{51}}$ \\
\hline
\rowcolor{mygray}\multicolumn{7}{|c|}{$\mathbf{\overline{4}3m}$}\\
$A_{2}$ & $ $ & $ $ & $ $ & $ $ & $ $ & $\text{\ding{51}}$ \\
\hline
\rowcolor{mygray}\multicolumn{7}{|c|}{$\mathbf{m\overline{3}m}$}\\
$A_{2g}$ & $ $ & $ $ & $ $ & $ $ & $ $ & $\text{\ding{51}}$ \\
\hline
\hline
\end{tabular}
\end{minipage}
\caption{Table of the $(1,N)$ spatial part of the SOC-free multipoles coupling to the possible N\'eel vectors in each point group. Recall that in the spin space, the multipole has $M=1$, i.e. the true multipole coupling to $\mathbf{N}$ is the direct product of $\Gamma_{l=1}^{(s)}$ with the multipoles presented in this table.}\label{tab:socfreeMultipoles}
\end{table}
\FloatBarrier

\clearpage
\onecolumngrid

\section{\label{app:MultipoleCouplingComponents}Table of symmetry-allowed couplings with and without SOC}

\begin{table}[h!]
\setlength\tabcolsep{2pt}
\begin{tabular}{|ccccC|}
\hline
\hline
\rowcolor{dgray} \textbf{PG} & $\Gamma_{\mathbf{N}}$ & \textbf{SO-Free Components} & \multicolumn{2}{c|}{\textbf{Guaranteed \,\, SOC \,\, Coupling}} \tstrut\bstrut \\
\hline
\hline
\multirow{2}{*}{\parbox{1cm}{\vspace{0.65cm}\centering  $\mathbf{2}$}}  &      \multirow{2}{*}{  \parbox{1cm}{\vspace{0.65cm}\centering $B$}} &  \multirow{2}{*}{ \parbox{1cm}{\vspace{0.6cm}\centering $\{ z, x \}$}}       &  			 $aV$ 		&  $\{z N_y, \,\,y N_z, \,\,y N_x, \,\,x N_y\}$ \tstrut \\
     &						&		& \parbox{2cm}{\vspace{0.5cm}\centering$aeV^{2}$ }	 & \{ $z N_z R_z,\,\, z N_x R_z, \,\,z N_y R_y, \,\,z N_z R_x,\,\, z N_x R_x$, $y N_y R_z, \,\,y N_z R_y,\,\, y N_x R_y,\,\, y N_y R_x, \,\,x N_z R_z$, $x N_x R_z, \,\,x N_y R_y, \,\,x N_z R_x, \,\,x N_x
R_x$ \}\bstrut\\

\hline

\rowcolor{mygray}\multirow{-6}{*}{}						 &  	   \multirow{-2}{*}{} 	& 	 \multirow{-2}{*}{} &    		$aV$		 &	\{ $z N_z,\,\,z N_x,\,\,y N_y,\,\,x N_z,\,\,x N_x$ \}	\tstrut \\
 \rowcolor{mygray}    \parbox{1cm}{\vspace{0.2cm}\centering $\mathbf{m}$} &	\parbox{1cm}{\vspace{0.2cm}\centering $ A''$ }	& \parbox{1cm}{\vspace{0.22cm}\centering $\{ y \}$ }		      &\parbox{2cm}{\vspace{0.5cm}\centering$aeV^{2}$ } & \{ $z N_y R_z,\,\, z N_z R_y,\,\, z N_x R_y,\,\, z N_y R_x,\,\, y N_z R_z$, $y N_x R_z, \,\,y N_y R_y, \,\,y N_z R_x, \,\,y N_x R_x$, $x N_y R_z, \,\,x N_z R_y,\,\, x N_x R_y, \,\,x N_y R_x$ \} \bstrut \\

\hline

$\mathbf{2/m}$    &  $ B_{g}$ &$\{ x y,  y z  \}$  &  $aeV$  & \{ $N_y R_z,\,\,N_z R_y,\,\,N_x R_y,\,\,N_y R_x$ \} \tstrut\bstrut \\

\hline

\rowcolor{mygray} \multirow{-6}{*}{} &  \multirow{-2}{*}{}  &\multirow{-2}{*}{}  & $aV$  & \{ $y N_x, \,\, x N_y$ \}	\tstrut \\
 \rowcolor{mygray} 	 &  \parbox{1cm}{\vspace{-0.4cm}\centering$B_{1}$} & \parbox{1cm}{\vspace{-0.4cm}\centering $\{ z \}$ }&	\parbox{2cm}{\vspace{0.2cm}\centering$aeV^{2}$ } 	&  \{ $z N_z R_z,\,\,z N_y R_y,\,\,z N_x R_x,\,\,y N_y R_z$, $y N_z R_y,\,\,x N_x R_z,\,\,x N_z R_x$ \}  \bstrut   \\
 
\rowcolor{mygray} \multirow{-2}{*}{} & \multirow{-2}{*}{}  & \multirow{-2}{*}{}     &  $aV$  & $\{z N_y,\,\,y N_z\}$ \tstrut  \\
\rowcolor{mygray} \parbox{1cm}{  \vspace{-0.4cm}\centering$\mathbf{222}$ }	& 	\parbox{1cm}{\vspace{-0.4cm}\centering $B_{3}$}	&	\parbox{1cm}{\vspace{-0.4cm}\centering $\{  x\}$}	& 	\parbox{2cm}{\vspace{0.3cm}\centering $aeV^{2}$	 }& \{ $z N_x R_z,\,\,z N_z R_x, \,\,y N_x R_y,\,\,y N_y R_x,$ $x N_z R_z, \,\,x N_y R_y,\,\,x N_x R_x $\}	\bstrut  \\
\rowcolor{mygray} \multirow{-2}{*}{}  & \multirow{-2}{*}{} &  \multirow{-2}{*}{}    & $aV$  &  $\{z N_x,\,\,x N_z \}$	\tstrut \\
\rowcolor{mygray} &	\parbox{1cm}{\vspace{-0.4cm}\centering $ B_{2}$}	& \parbox{1cm}{\vspace{-0.4cm}\centering $\{ y\}$ } & \parbox{2cm}{\vspace{0.3cm}\centering $aeV^{2}$	 }	&  \{ $z N_y R_z, \,\,z N_z R_y, \,\,y N_z R_z, \,\,y N_y R_y$, $y N_x R_x,\,\,x N_x R_y,\,\,x N_y R_x$ \}	\bstrut \\

\hline

\multirow{5}{*}{} &  $A_{2}$ & $\{ x y \}$  &  $aeV$ & \{ $N_x R_y,N_y R_x$ \} \tstrut\bstrut\\
  &    \multirow{-2}{*}{}& \multirow{-2}{*}{ }  &  $aV$ & \{ $z N_x,\,\,x N_z$ \} \tstrut \\
  &	 \parbox{1cm}{\vspace{-0.4cm}\centering $ B_{2}$ }	&	\parbox{1cm}{\vspace{-0.4cm}\centering$\{y\}$}	&	\parbox{2cm}{\vspace{0.3cm}\centering $aeV^{2}$ } &  \{ $z N_y R_z,\,\,z N_z R_y,\,\,y N_z R_z,\,\,y N_y R_y$, $y N_x R_x,\,\,x N_x R_y,\,\,x N_y R_x$ \}\bstrut \\
   \parbox{1cm}{\vspace{-1cm}\centering $\mathbf{mm2}$ } &  \multirow{-2}{*}{ }   &   \multirow{-2}{*}{ }   &  $aV$ &\{ $z N_y\,\, y N_z$ \} \tstrut \\
   & \parbox{1cm}{\vspace{-0.4cm}\centering  $B_{1}$} 	&	\parbox{1cm}{\vspace{-0.4cm}\centering$\{ x\}$} &	\parbox{2cm}{\vspace{0.3cm}\centering$aeV^{2}$} &  \{ $z N_x R_z,\,\,z N_z R_x,\,\,y N_x R_y,\,\,y N_y R_x$, $x N_z R_z,\,\,x N_y R_y,\,\,x N_x R_x$ \}   \bstrut \\
\hline
\rowcolor{mygray}  &   $B_{1g}$ & $\{  x  y \}$  &  $aeV$ & \{ $N_x R_y,\,\,N_y R_x$ \} \tstrut\\
\rowcolor{mygray}  $\mathbf{mmm}$&    $ B_{3g}$ & $\{yz\}$   & $aeV$  & \{ $N_y R_z,\,\,N_z R_y$ \}\\
\rowcolor{mygray}   &    $ B_{2g}$ & $\{ xz\}$   & $aeV$  &  \{ $N_x R_z,\,\,N_z R_x$ \} \bstrut\\
\hline
$\mathbf{4}$  &   $B $ &$ \{ y^{2}- x^{2},  xy \}$  & $aeV$ & \{ $N_x R_x-N_y R_y,\,\,N_y R_x+N_x R_y$ \}\tstrut\bstrut \\
\hline
\rowcolor{mygray} \multirow{-2}{*}{}  &  \multirow{-2}{*}{}  &  \multirow{-2}{*}{}  & $aV$ & \{ $z N_z,\,\,x N_x+y N_y,\,\,x N_y-y N_x$ \} \tstrut\\
\rowcolor{mygray} \parbox{1cm}{\vspace{0.2cm}\centering$\mathbf{\overline{4}}$ }	&   \parbox{1cm}{\vspace{0.3cm}\centering $B $} &  \parbox{1cm}{\vspace{0.31cm}\centering$\{  z \}$} & \parbox{2cm}{\vspace{0.5cm}\centering$aeV^{2}$} & \{ $z N_z R_z,\,\,z \left(N_x R_x+N_y R_y \right),\,\,z \left(N_y R_x-N_x R_y \right)$, $\left(R_z \left(x N_x+y N_y \right)\right),\,\,R_z \left(x N_y-y N_x \right)$, $N_z \left(x R_x+y R_y \right),\,\,N_z \left(x R_y-y R_x \right)$ \}\bstrut\\
\hline
 $\mathbf{4/m}$ &      $ B_{g} $ & $\{  y^{2}- x^{2},   xy  \}$ & $aeV$ & \{ $N_x R_x-N_y R_y,\,\,N_y R_x+N_x R_y$ \} \tstrut\bstrut \\
\hline
\rowcolor{mygray}\multirow{-4}{*}{}  &    $ B_{1} $ & $\{   y^{2} - x^{2} \}$   &  $aeV$ & \{ $N_y R_y-N_x R_x$ \} \tstrut\bstrut\\
\rowcolor{mygray}    & \multirow{-2}{*}{}  &  \multirow{-2}{*}{}   &  $aV$ &\{ $y N_x-x N_y$ \} \tstrut \\
\rowcolor{mygray}  \parbox{1cm}{\vspace{-0.2cm}\centering$\mathbf{422}$} & \parbox{1cm}{\vspace{-0.2cm}\centering$A_{2} $} & \parbox{1cm}{\vspace{-0.2cm}\centering$\{ z\}$ } & \parbox{2cm}{\vspace{0.3cm}\centering$aeV^{2}$} & \{ $z N_z R_z,\,\,z \left(N_x R_x+N_y R_y \right)$, $R_z \left(x N_x+y N_y \right),\,\,N_z \left(x R_x+y R_y \right)$ \} \bstrut\\
\rowcolor{mygray}  &  $B_{2} $ & $\{xy\}$   & $aeV$  & \{ $N_y R_x+N_x R_y$ \} \tstrut\bstrut \\
\hline
\multirow{-4}{*}{}  &   $ B_{1} $ & $\{   y^{2}-x^{2} \}$   & $aeV$ &  \{ $N_y R_y-N_x R_x\} $ \tstrut\bstrut \\
\hline
\end{tabular}
\centering
\caption{For each point group and $\Gamma_{\mathbf{N}}$, the minimal SO-free multipole polynomial (see Eq.~\ref{eq:multipoleDef}) is given in the third column. The representation $\Gamma_{\xi}$ of the allowed tensor with SOC and its coupling to $N_{i}$ components appear in last two columns.
}\label{tab:CouplingComps}
\end{table}

\vspace{-3cm}
\begin{table*}[h!]
\setlength\tabcolsep{2pt}
\begin{tabular}{|ccccC|}
\hline
\hline
\rowcolor{dgray} \textbf{PG} & $\Gamma_{\mathbf{N}}$ & \textbf{SO-Free Components} & \multicolumn{2}{c|}{\textbf{Guaranteed \,\, SOC \,\, Coupling}} \tstrut\bstrut \\
\hline
\hline
\parbox{1cm}{\centering$\mathbf{4mm}$} & $ B_{2} $ & $\{ xy\}$  & $aeV$ &  \{ $N_y R_x+N_x R_y$ \} \tstrut\bstrut \\
   & \multirow{-2}{*}{}  \parbox{2cm}{\vspace{0.5cm}\centering$ A_{2}$}   & 	\multirow{-2}{*}{}  \parbox{2cm}{\vspace{0.5cm}\centering$\{xy(x^{2}-y^{2})\}$} &  \parbox{2cm}{\vspace{0.5cm}\centering $aeV[V^{2}]$} &  \{  $  z R_z \left(y N_x-x N_y\right),\,y^2 N_y R_x-x^2 N_x R_y$, $\, x y \left(N_x R_x-N_y R_y\right), \,x^2 N_y R_x-y^2 N_x R_y$, $\, z N_z \left(y R_x-x R_y\right),\,z^2 \left(N_y R_x-N_x R_y\right) $  \} \bstrut  \\ 
\hline
\rowcolor{mygray} \multirow{-5}{*}{}  &    $B_{1} $ & $\{  y^{2}- x^{2} $\}    &  $aeV$  & \{ $N_y R_y-N_x R_x$ \} \tstrut\bstrut \\
\rowcolor{mygray} &    &     &  $aV$ & \{ $y N_x-x N_y$ \}\tstrut \\
\rowcolor{mygray}   \parbox{1cm}{\vspace{0.1cm}\centering$\mathbf{\overline{4}2m}$} &  \parbox{1cm}{\vspace{-.4cm}\centering $ B_{2} $} &   \parbox{1cm}{\vspace{-.4cm}\centering$\{  z\}$ }   &  $aeV^{2}$ &  \{ $z N_z R_z,\,\,z \left(N_x R_x+N_y R_y \right)$, $R_z \left(x N_x+y N_y \right),\,\,N_z \left(x R_x+y R_y\right)$ \} \bstrut\\
\rowcolor{mygray}  &   $A_{2} $ & $\{   z(y^{2}-x^{2})\}$  & $aeV^{2}$ & \{ $z N_y R_y-z N_x R_x,\,\,R_z \left(y N_y-x N_x \right)$, $N_z \left(y R_y-x R_x\right)$ \} \tstrut\bstrut  \\
 \hline
 \multirow{-4}{*}{} &   \multirow{-2}{*}{}  &  \multirow{-2}{*}{} &  $aV$ & \{ $y N_x-x N_y$ \}  \tstrut\\
   & \parbox{1cm}{\vspace{-.8cm}\centering$ B_{1} $	}&	\parbox{1cm}{\vspace{-.4cm}\centering$\{   z \}$} &	$aeV^{2}$ & \{ $z N_z R_z,\,\,z \left(N_x R_x+N_y R_y \right)$, $R_z \left(x N_x+y N_y \right),\,\,N_z \left(x R_x+y R_y \right)$ \} \bstrut \\
 \parbox{1cm}{\vspace{-.8cm}\centering $\mathbf{\overline{4}m2}$ }&  $B_{2} $ & $\{   xyz\}$   & $aeV^{2}$  &  $\{z \left(N_y R_x+N_x R_y \right),\,\,R_z \left(y N_x+x N_y \right)$, $N_z \left(y R_x+x R_y \right)\}$\tstrut\bstrut   \\
  &   $ A_{2} $ & $\{   xy\}$   &  $aeV$   & \{ $N_y R_x+N_x R_y$ \}  \tstrut\bstrut  \\
\hline
\rowcolor{mygray} \multirow{-4}{*}{} &  $B_{1g} $ & $\{   y^{2}-x^{2}\}$   &  $aeV$    & \{ $N_y R_y-N_x R_x$ \}  \tstrut\bstrut  \\
\rowcolor{mygray} \parbox{1.5cm}{\vspace{.5cm}\centering$\mathbf{4/mmm}$}  &       \multirow{-2}{*}{}   \parbox{1cm}{\vspace{.5cm}\centering$ A_{2g} $ }  &  \multirow{-2}{*}{} \parbox{2cm}{\vspace{.5cm}\centering$\{   xy(x^{2}-y^{2}) \}$} & \parbox{2cm}{\vspace{.5cm}\centering$aeV[V^{2}]$  } &  \{ $  z R_z \left(y N_x-x N_y\right), y^2 N_y R_x-x^2 N_x R_y$, $ x y \left(N_x R_x-N_y R_y\right), x^2 N_y R_x-y^2 N_x R_y$, $ z N_z \left(y R_x-x R_y\right), z^2 \left(N_y R_x-N_x R_y\right) $ \}  \tstrut \bstrut\\  
\rowcolor{mygray}  &          $ B_{2g} $ & $\{  xy \}$   &  $aeV$  & $\{N_y R_x+N_x R_y\}$  \tstrut\bstrut  \\
\hline
\multirow{-2}{*}{}  &\multirow{-2}{*}{} & \multirow{-2}{*}{}  & $aV$ & \{ $y N_x-x N_y$ \}  \tstrut\\
  \parbox{1cm}{\vspace{.1cm}\centering$\mathbf{32}$}  & \parbox{1cm}{\vspace{.1cm}\centering$ A_{2}$} & \parbox{1cm}{\vspace{.1cm}\centering$\{    z \}$}  &  \parbox{2cm}{\vspace{.65cm}\centering$aeV^{2}$} & \{ $z N_z R_z,\,\,z \left(N_x R_x+N_y R_y \right)$, $R_z \left(x N_x+y N_y \right),\,\,N_z \left(x R_x+y R_y \right)$, $N_x \left(x R_x-y R_y \right)-N_y \left(y R_x+x R_y \right)$ \} \bstrut\\
 \hline
\rowcolor{mygray}   \parbox{1cm}{\vspace{.3cm}\centering$\mathbf{3m}$}  &      \parbox{1cm}{\vspace{.3cm}\centering$A_{2} $ }&   \parbox{2cm}{\vspace{.3cm}\centering$\{   y(y^{2} -3 x^{2})  \}$} &   \parbox{2cm}{\vspace{.25cm}\centering$aeV^{2}$}  & \{ $ z N_x R_y-z N_y R_x,R_z \left(y N_x-x N_y\right)$, $N_y \left(y \
R_y-x R_x\right)-N_x \left(y R_x+x R_y\right),N_z \left(y R_x-x R_y\right) $ \} \bstrut \\    
\hline \tstrut\bstrut
\parbox{1cm}{ \vspace{.9cm}\centering $\mathbf{\overline{3}m}$} &  \parbox{1cm}{\vspace{.9cm}\centering$ A_{2g} $} & \parbox{2cm}{\vspace{.9cm}\centering $\{   y ( y^{2} - 3x^{2})z  \}$}  &\parbox{1cm}{\vspace{.9cm}\centering $aeV[V^{2}]$}   &\vspace{-.6cm} \{ $  R_z \left(N_y \!\left(y^2-x^2\right)-4 x y N_x\right)\!,\,  z R_z \!\left(y N_x-x N_y\right)$, $N_z \left(R_y \left(y^2-x^2\right)-4 x y R_x \right),\left(x^2+y^2\right) \left(N_y R_x-N_x R_y\right)$, $ yN_{x}(2xR_{x}+yR_{y})-xN_{y}(xR_{x}+2yR_{y}),  z N_z \left(y R_x-x R_y\right)$, $ z \left(N_x \left(y R_x+x R_y\right)+N_y \left(x R_x-y R_y\right)\right)$, $ z^2 \left(N_y R_x-N_x R_y\right) $ \} \tstrut\bstrut \\ 
\hline
\rowcolor{mygray}\parbox{1cm}{\vspace{.3cm}\centering $\mathbf{6}$ }&     \parbox{1cm}{\vspace{.3cm}\centering $B $ }& \parbox{4cm}{\vspace{.3cm}\centering$\{    x (3y^{2}- x^{2}), y(y^{2} - 3x^{2} ) \}$}  & \parbox{2cm}{\vspace{.3cm}\centering  $aeV^{2}$ } &  \{ $N_y \left(y R_y-x R_x \right)-N_x \left(y R_x+x R_y\right)$, $N_y \left(y R_x+x R_y \right)+N_x \left(y R_y-x R_x \right)$ \} \tstrut\bstrut  \\
\hline
 \multirow{-2}{*}{} &  \multirow{-2}{*}{} &  \multirow{-2}{*}{}   & $aV$ & \{ $z N_z,x N_x+y N_y,x N_y-y N_x $ \} \tstrut\\
 \parbox{1cm}{\vspace{.2cm}\centering$\mathbf{\overline{6}}$ }&      \parbox{1cm}{\vspace{.2cm}\centering  $A'' $ }& \parbox{1cm}{\vspace{.3cm}\centering$\{   z  \}$ }  & \parbox{2cm}{\vspace{.5cm}\centering $aeV^{2}$ }& \{ $z N_z R_z,\,\,z \left(N_x R_x+N_y R_y \right),\,\,z \left(N_y R_x-N_x R_y \right)$, $R_z \left(x N_x+y N_y \right),\,\,R_z \left(x N_y-y N_x \right)$, $N_z \left(x R_x+y R_y \right),N_z \left(x R_y-y R_x \right)$ \}  \bstrut \\
\hline
\rowcolor{mygray} \parbox{1cm}{\vspace{1cm}\centering$\mathbf{6/m}$} &   \parbox{1cm}{\vspace{1cm}\centering$ B_{g} $} &   \parbox{4cm}{\vspace{1cm}\centering$\{   x z (x^{2} - 3y^{2}) ,  y z (y^{2}-3  x^{2})   \}$}   &  \parbox{2cm}{\vspace{1cm}\centering$aeV[V^{2}]$} &  \vspace{-.7cm}\{ $ R_z \left(N_y \left(y^2-x^2\right)-4 x y N_x\right)$, $R_z \left(N_x \left(y^2-x^2\right)+4 x y N_y\right)$, $N_z \left(R_y \left(y^2-x^2\right)-4 x y R_x\right)$, $N_z \left(R_x \left(x^2-y^2\right)-4 x y R_y\right)$, $ z \left(N_x \left(x R_x-y R_y\right)-N_y \left(y R_x+x R_y\right)\right)$, $ z \left(N_x \left(y R_x+x R_y\right)+N_y \left(x R_x-y R_y\right)\right) $ \} \tstrut\bstrut \\
\hline
\multirow{-2}{*}{}   &   \multirow{-2}{*}{}  &  \multirow{-2}{*}{}  & $aV$  & \{ $y N_x-x N_y$ \} \tstrut \\
   \parbox{1cm}{\vspace{0.6cm}\centering$\mathbf{622}$}  &    \parbox{1cm}{\vspace{-0.4cm}\centering$ A_{2} $} &  \parbox{1cm}{\vspace{-0.4cm}\centering$\{   z \}$} &   $aeV^{2}$ &  \{ $z N_z R_z,\,\,z \left(N_x R_x+N_y R_y \right)$, $R_z \left(x N_x+y N_y \right),\,\,N_z \left(x R_x+y R_y \right)$ \} \bstrut\\
 &          $ B_{2} $ & $\{  x(y^{2}-3x^{2})\}$    &  $aeV^{2}$ &  \{ $N_y \left(y R_x+x R_y \right)+N_x \left(y R_y-x R_x \right)$ \} \tstrut \bstrut \\
  &          $ B_{1} $ & $\{   y(y^{2}-3x^{2})\}$   & $aeV^{2}$  &  \{ $N_y \left(y R_y-x R_x \right)-N_x \left(y R_x+x R_y \right)$ \} \tstrut \bstrut  \\
\hline
\end{tabular}
\end{table*}

\vspace{-3cm}
\begin{table*}[h!]
\setlength\tabcolsep{2pt}
\begin{tabular}{|ccccC|}
\hline
\hline
\rowcolor{dgray} \textbf{PG} & $\Gamma_{\mathbf{N}}$ & \textbf{SO-Free Components} & \multicolumn{2}{c|}{\textbf{Guaranteed \,\, SOC \,\, Coupling}} \tstrut\bstrut \\
\hline
\hline
 
\rowcolor{mygray} \parbox{1cm}{\vspace{1cm}\centering$\mathbf{6mm}$} &    \parbox{1cm}{\centering$A_{2} $ } & \parbox{3.5cm}{\centering$\{   3x^{5}y - 10x^{3}y^{3} + 3xy^{5}\}$}    & \parbox{1cm}{\centering $aeV[V^{2}]$} & \vspace{-.7cm} \{ $ z R_z \left(y N_x-x N_y\right)$, $\left(x^2+y^2\right) \left(N_y R_x-N_x R_y\right)$, $ yN_{x}(2xR_{x}+yR_{y})-xN_{y}(xR_{x}+2yR_{y})$, $ z N_z \left(y R_x-x R_y\right), z^2 \left(N_y R_x-N_x R_y
\right) $ \} \tstrut \\
\rowcolor{mygray} &          $ B_{2} $ & $\{   x(y^{2}-3x^{2})\}$   & $aeV^{2}$ & \{ $N_y \left(y R_x+x R_y \right)+N_x \left(y R_y-x R_x \right)$ \} \tstrut \bstrut  \\
\rowcolor{mygray}  &          $ B_{1} $ & $\{   y(y^{2}-3x^{2})\}$   &  $aeV^{2}$ & \{ $N_y \left(y R_y-x R_x \right)-N_x \left(y R_x+x R_y \right)$ \} \tstrut \bstrut \\
\hline

\multirow{-5}{*}{} &  \parbox{1cm}{\vspace{0.3cm}\centering$A_{1}'' $} &  \parbox{2cm}{\vspace{0.3cm}\centering $\{ yz(y^{2}-3x^{2}\}$ } &  \parbox{2cm}{\vspace{0.3cm}\centering$aeV[V^{2}]$} & \{ $ R_z \left(N_y \left(y^2-x^2\right)-4 x y N_x\right)$, $N_z \left(R_y \left(y^2-x^2\right)-4 x y R_x\right)$, $ z \left(N_x \left(y R_x+x R_y\right)+N_y \left(x R_x-y R_y\right)\right) $ \} \tstrut \\
  &      &   & $aV$ & \{ $y N_x-x N_y$ \} \tstrut \\
\parbox{1cm}{\vspace{-1.6cm}\centering  $\mathbf{\overline{6}2m}$ }  &      \parbox{1cm}{\vspace{-0.4cm}\centering $ A_{2}'' $} &  \parbox{1cm}{\vspace{-0.4cm}\centering$\{   z\}$ }   &  \parbox{2cm}{\vspace{0.2cm}\centering$aeV^{2}$} &  \{ $z N_z R_z,\,\,z \left(N_x R_x+N_y R_y \right)$, $R_z \left(x N_x+y N_y \right),\,\,N_z \left(x R_x+y R_y \right)$ \} \bstrut \\
    
  &          $A_{2}' $ & $\{  y(y^{2}-3x^{2}))\}$  & $aeV^{2}$  & \{ $N_y \left(y R_y-x R_x \right)-N_x \left(y R_x+x R_y \right)$ \} \tstrut\bstrut  \\

\hline
\rowcolor{mygray} &  $ A_{1}''$ & $\{   x(3y^{2}-x^{2}) \}$   &$aeV^{2}$  & \{ $N_y \left(y R_x+x R_y \right)+N_x \left(y R_y-x R_x \right)$ \} \tstrut\bstrut\\

\rowcolor{mygray}  \multirow{-5}{*}{} &  \multirow{-2}{*}{} & \multirow{-2}{*}{} & $aV$ & \{ $y N_x-x N_y$ \}  \tstrut\\
\rowcolor{mygray}  &         \parbox{2cm}{\vspace{0.4cm}\centering  $A_{2}'' $} & \parbox{1cm}{\vspace{0.4cm}\centering $\{   z\}$}   &  \parbox{2cm}{\vspace{0.3cm}\centering$aeV^{2}$} &  \{ $z N_z R_z,z \left(N_x R_x+N_y R_y \right)$, $R_z \left(x N_x+y N_y \right),N_z \left(x R_x+y R_y \right)$ \} \bstrut\\

\rowcolor{mygray}\parbox{1cm}{\vspace{-1.6cm}\centering$\mathbf{\overline{6}m2}$ } & \parbox{1cm}{\vspace{0.4cm}\centering$A_{2}' $} & \parbox{2cm}{\vspace{0.4cm}\centering$\{   xz(x^{2}-3y^{2})\}$} &  \parbox{2cm}{\vspace{0.4cm}\centering$aeV[V^{2}]$} & \{ $ R_z \left(N_x \left(y^2-x^2\right)+4 x y N_y\right)$, $N_z \left(R_x \left(x^2-y^2\right)-4 x y R_y\right)$, $ z \left(N_x \left(x R_x-y R_y\right)-N_y \left(y R_x+x R_y\right)\right) $ \}  \tstrut\bstrut \\
\hline

\multirow{-6}{*}{}  &  \parbox{1cm}{\vspace{.4cm}\centering$ A_{2g} $}  & \parbox{3.5cm}{\vspace{.4cm}\centering$\{    3x^{5}y-10x^{3}y^{3} + 3xy^{5}\}$} &  \parbox{2cm}{\vspace{0.35cm}\centering$aeV[V^{2}]$} &  \{ $  z R_z \left(y N_x-x N_y\right),\left(x^2+y^2\right) \left(N_y R_x-N_x R_y\right)$, $ yN_{x}(2xR_{x}+yR_{y})-xN_{y}(xR_{x}+2yR_{y})$, $ z N_z \left(y R_x-x R_y\right),z^2 \left(N_y R_x-N_x R_y\right) $ \}  \\
\parbox{1.4cm}{\vspace{0.4cm}$\mathbf{6/mmm}$}  & \parbox{2cm}{\vspace{0.4cm} $ B_{2g} $ } & \parbox{2cm}{\vspace{0.4cm}$\{   yz(y^{2}-3x^{2})\}$} &  \parbox{2cm}{\vspace{0.4cm}\centering$aeV[V^{2}]$} & \{ $ R_z \left(N_y \left(y^2-x^2\right)-4 x y N_x\right)$, $N_z \left(R_y \left(y^2-x^2\right)-4 x y R_x\right)$, $ z \left(N_x \left(y R_x+x R_y\right)+N_y \left(x R_x-y R_y\right)\right) $ \} \tstrut \\
  
  & \parbox{2cm}{\vspace{0.4cm}\centering$ B_{1g}$} & \parbox{2cm}{\vspace{0.4cm}\centering$\{   xz(x^{2}-3y^{2})\}$} & \parbox{2cm}{\vspace{0.4cm}\centering$aeV[V^{2}]$} & \{ $ R_z \left(N_x \left(y^2-x^2\right)+4 x y N_y\right)$, $N_z \left(R_x \left(x^2-y^2\right)-4 x y R_y\right)$, $ z \left(N_x \left(x R_x-y R_y\right)-N_y \left(y R_x+x R_y\right)\right) $ \} \tstrut\bstrut \\
\hline
\rowcolor{mygray}\parbox{0.8cm}{\centering$\mathbf{432}$} &  \parbox{1cm}{\centering$ A_{2} $ }& \parbox{1cm}{\centering$\{   x y  z \}$ }  &  \parbox{2cm}{\centering$aeV^{2}$ }& \{ $N_z \left(y R_x+x R_y \right)+N_y \left(z R_x+x R_z \right)+N_x \left(z R_y+y R_z \right)$ \} \tstrut\bstrut  \\
\hline 
\parbox{1.6cm}{\vspace{0.4cm}\centering$\mathbf{\overline{4}3m}$}  &\parbox{1cm}{\vspace{0.4cm}\centering $A_{2} $} &\parbox{4cm}{\vspace{0.4cm}\centering$\{(x^{2}- y^{2}) (x^{2}- z^{2})(y^{2}- z^{2})  \}$}  & \parbox{2cm}{\vspace{0.4cm}\centering$aeV[V^{2}]$ } &  \{$ N_z R_z \left(y^2-x^2\right)+N_y R_y \left(x^2-z^2\right)+N_x R_x \left(z^2-y^2\right)$, $ z N_z \left(x R_x\!-\!y R_y\right)\!+\!x N_x \left(y R_y\!-\!z R_z\right)\!+\!N_y \left(y z R_z\!-\!x y R_x\right)$\} \tstrut\bstrut\\
\hline
\rowcolor{mygray} \parbox{0.8cm}{\vspace{0.3cm}\centering$\mathbf{m\overline{3}m}$} &  \parbox{1cm}{\vspace{0.3cm}\centering$A_{2g} $} & \parbox{4cm}{\vspace{0.3cm}\centering$\{ (x^{2}\!-\!y^{2})(x^{2}\!-\!z^{2})( y^{2}\!-\!z^{2})\}$}   &  \parbox{2cm}{\vspace{0.3cm}\centering$aeV[V^{2}]$} & \{ $ N_z R_z \left(y^2-x^2\right)+N_y R_y \left(x^2-z^2\right)+N_x R_x \left(z^2-y^2\right)$, $\left(z N_z \left(x R_x\!-\!y R_y\right)\!+\!x N_x \left(y R_y\!-\!z R_z\right)\!+\!N_y \left(y z R_z\!-\!x y R_x\right)\right) $\} \tstrut\bstrut\\
\hline
\end{tabular}
\end{table*}

\clearpage
\twocolumngrid


\bibliography{references_v2}

\end{document}